\documentclass[%
 reprint,
 superscriptaddress,
 amsmath,amssymb,
 prres,
]{revtex4-1}

\usepackage{graphicx}
\usepackage{dcolumn}
\usepackage{bm}
\usepackage[normalem]{ulem}
\usepackage{multirow}
\usepackage[hidelinks]{hyperref}
\usepackage{chngcntr}
\usepackage{wrapfig}
\usepackage{url}
\usepackage{color}
\usepackage[dvipsnames]{xcolor}

\DeclareMathAlphabet{\pazocal}{OMS}{zplm}{m}{n}

\usepackage[raggedright]{titlesec}

\begin{document}


\pagestyle{empty}

\title{Optical dispersions through intracellular inhomogeneities}

\author{Masaki Watabe}
\email{Corresponding author: masaki@riken.jp}
\affiliation{Laboratory for Biologically Inspired Computing, RIKEN Center for Biosystems Dynamics Research, Suita, Osaka 565-0874, Japan}
\author{Yasuhiro Hirano}
\affiliation{Nuclear Dynamics Group, Graduate School of Frontier Biosciences Osaka University, Suita, Osaka 565-0871, Japan}
\author{Atsuko Iwane}
\affiliation{Laboratory for Cell Field Structure, RIKEN Center for Biosystems Dynamics Research, Higashi-Hiroshima, Hiroshima 739-0046, Japan}
\author{Osamu Matoba}
\affiliation{The Graduate School of System Informatics, Kobe University, Kobe, Hyogo 657-8501, Japan}
\author{Koichi Takahashi}
\affiliation{Laboratory for Biologically Inspired Computing, RIKEN Center for Biosystems Dynamics Research, Suita, Osaka 565-0874, Japan}
\affiliation{Institute for Advanced Bioscience, Keio University, Fujisawa, Kanagawa 252-8520, Japan}

\date{\today}

\begin{abstract}
Transport of intensity equation (TIE) exhibits a non-interferometric correlation between intensity and phase variations of intermediate fields (e.g., light and electron) in biological imaging. Previous TIE formulations have generally assumed a free space propagation of monochromatic coherent field functions crossing phase distributions along a longitudinal direction. Here, we modify the TIE with fractal (or self-similar) organization models based on intracellular refractive index turbulence. We then implement the TIE simulation over a broad range of fractal dimensions and wavelengths. Simulation results show how the intensity propagation through the spatial fluctuation of intracellular refractive index interconnects fractal-dimensionality with intensity dispersion (or transmissivity) within the picometer to micrometer wavelength range. In addition, we provide a spatial-autocorrelation of phase derivatives which allows the direct measurement and reconstruction of intracellular fractal profiles from optical and electron microscopy imaging.
\end{abstract}

\maketitle

\textit{Introduction.}
Biological cells are the basic structural and functional unit of life. All cells consist of a cytoplasm enclosed within a membrane, which includes biomolecules (e.g., proteins and nucleic acids) and intracellular organelles (e.g., the endoplasmic reticulum, Golgi apparatus, and mitochondria). These complex intracellular structures can be visualized using optical imaging systems and electron microscopes. As visible light has a wavelength ranging from $360\ {\rm nm}$ to $760\ {\rm nm}$, which are up to five orders of magnitude longer than wavelength of electrons, optical imaging techniques tend to filter out the inner structures of smaller intracellular objects otherwise observable through electron microscopy.

%

The complex amplitude of a wave function describing optical fields is generally composed of two major physical quantities: amplitude and phase functions \cite{born1999}. Variations in amplitudes (or intensities) can be captured directly by photosensitive detectors such as the human eye and scientific CMOS cameras. Variations in phases, however, are barely visible using optical microscopy systems and are sensitive to light scattering and fluctuations due to invisible spatial distributions of intracellular optical properties, e.g., refractive index and optical thickness. These phase variations have been formalized into a partial differential equation, termed transport of intensity equation (TIE) \cite{teague1983, zuo2020, mitome2021}, which is dependent on wavelength and the intensity variations of intermediate fields (i.e., photon and electron). This equation represents, in particular, the free space propagation of a monochromatic coherent wave function to a given phase distribution along a longitudinal direction.


A key challenge to bridging the large resolution gap that lies between optical and electron microscopy techniques consists of finding meaningful variations in both intensity and phase functions. Over a broad range of wavelengths, intensity and phase variations of intermediate fields are inextricably linked via the TIE. Previous TIE formalizations have generally assumed that a complex plane wave function describing intermediate fields can be propagated through an arbitrary phase distribution in free space. Intensity transport of the wave functions interacting with the complex refractive aspects of cellular interiors, however, remains elusive. In this article, we modify the TIE with fractal (or self-similar) organization models for intracellular refractive index turbulence. We then implement the TIE simulation in the picometer to micrometer wavelength range, computing intensity dispersions and reductions caused by inhomogeneous spatial distributions of intracellular refractive index. Crucially, we show how the {\it a priori} lateral intensity distribution of a cell can be described as a function of its fractal dimensions. Such fractal media is optically thin and transparent through the visible wavelengths but exhibits the compression of fractal-dimensionality and the increase of intensity dispersions in the wavelength range of electron microscopy imaging. We also derive a spatial-autocorrelation function that interconnects phase derivatives with intracellular fractal profiles and demonstrate the direct measurements and reconstruction of optical properties from fluorescent cell imaging. Our work facilitates further extensions to bioimage simulation modules \cite{mahajan2022, weigert2018, girsault2016, linden2016, *linden2017, venkataramani2016, watabe2019, *watabe2015, angiolini2015, rezatofighi2013, sbalzarini2013, boulanger2009}, incorporating, in particular, light scattering and fluctuations caused by nonuniform spatial distributions of intracellular optical properties. Such modification of the simulation is of particular relevance to the numerical evaluation and verification of observational invariance (or symmetry) as postulated in data science, likely leading to more realistic simulations of biological imaging.

%
%

\textit{Transport of intensity equation.}
For a specific intermediate field wavelength $\lambda$ (e.g., photon or electron) we reformulate the TIE for intensity propagation by using the laws of conservation of energy (i.e., Helmholtz equation), parameterizing, in particular, a complex refractive index for the intracellular media,
\begin{equation}
\label{eqn;index}
n(r_{\bot},z) = n_0 [1 + \Delta n(r_{\bot},z) + i \kappa(r_{\bot},z)]
\end{equation}
where $n_0$ and $\Delta n(r_{\bot},z)$ are the mean and spatial fluctuation of the refractive index, respectively. $\kappa(r_{\bot},z)$ in the imaginary part denotes the spatial distribution of the attenuation index. In this parameterization, we assume that a monochromatic coherent wave function propagating through the refractive index distribution along the axial axis (see Figure \ref{fig01;tie}) can be written in the form of
\begin{equation}
\label{eqn;wave}
\psi(r_{\bot},z) = \sqrt{I(r_{\bot},z)}\ e^{i \phi(r_{\bot},z)} e^{ik n_0 z}
\end{equation}
where the square root of the intensity distribution $\sqrt{I}$ denotes the amplitude of the wave function; $k$ is the wave number in free space $k = 2\pi/\lambda$. Substituting Eq. (\ref{eqn;index}) and (\ref{eqn;wave}) into the Helmholtz equation, we can derive TIE as follows.
\begin{eqnarray}
\label{eqn;tie}
\frac{\partial I(r_{\bot},z)}{\partial z} & = & - \frac{1}{k n_0}\ \nabla_{\bot}\cdot \big[ I(r_{\bot},z)\ \nabla_{\bot}\phi(r_{\bot},z) \big] \nonumber\\
& & -\ 2\ \frac{\partial \alpha(r_{\bot},z)}{\partial z}\ I(r_{\bot},z)
\end{eqnarray}
where $\nabla_{\bot}$ is the two-dimensional Nabla operator in the transverse direction, i.e., $\nabla_{\bot} = \left(\partial/\partial x, \partial/\partial y \right)$. $\alpha(r_{\bot}, z)$ and $\beta(r_{\bot}, z)$ represent amplitude and phase variations arising from the spatial fluctuations of the refractive and attenuation index, 
\begin{eqnarray}
\label{eqn;alpha}
\alpha(r_{\bot}, z) & = & k n_0 \int [1 + \Delta n(r_{\bot},z)]\ \kappa(r_{\bot},z)\ dz\\
{\rm and} \nonumber\\
\label{eqn;beta}
\beta(r_{\bot}, z) & = & k n_0 \int \Big[ \Delta n(r_{\bot},z) - \frac{1}{2} \kappa(r_{\bot},z)^2 \Big]\ dz\ \ \ 
\end{eqnarray}
where $\beta$-factor vanishes in the TIE modification. This factor may be more relevant to phase changes in transport of phase equation (TPE) \cite{zuo2020}. Furthermore, the derivation details are discussed in the section A.1 of Supplementary Material (SM).

The left-hand side of Eq. (\ref{eqn;tie}) represents the axial intensity differentiation that guides the intensity propagation along the z-axis. The right-hand side of the equation exhibits the total energy variation in a lateral intensity distribution. Such energy variations are composed of two parts. (1) The first part $-\nabla_{\bot} \cdot [I \nabla_{\bot} \phi]/(k n_0)$ represents the intensity and phase variations arising from the spatial fluctuation of the refractive and attenuation index. There are two terms in these variations: $I \nabla_{\bot}^2 \phi$ and $\nabla_{\bot} I \cdot \nabla_{\bot} \phi$. A scalar product of the intensity distribution and the Laplacian operator acting on the phase distribution can be interpreted as the intensity variation caused by convergence or divergence of the intensity the radius of local phase curvature of which is inversely proportional to $\nabla_{\bot}^2 \phi$. The dot product of the gradient of intensity and phase distributions represents a measure of the translational effects of the phase gradient in the direction of the intensity gradient. (2) The second part $- 2 (\partial \alpha/\partial z) I$ relates to Beer-Lambert's law of intensity reduction in an inhomogeneous medium, exhibiting, in particular, the energy absorption and scattering of the intermediate fields traveling through an intracellular medium. 

\begin{figure}
\centering
\includegraphics[width=0.99\linewidth]{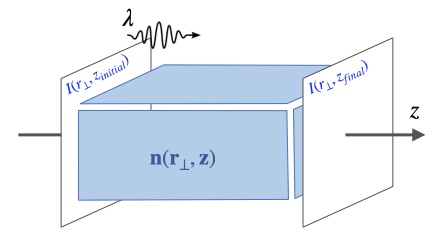}

\caption{For a given wavelength $\lambda$, the initial intensity distribution of a complex plane wave $I(r_{\bot},z_{initial})$ propagates through intracellular refractive index $n(r_{\bot},z)$, exhibiting the final intensity distribution $I(r_{\bot},z_{finial})$.}
\label{fig01;tie}
\end{figure}

\textit{Fractal cell modeling.} A wave function of intermediate fields propagating through a biological medium can encounter structures with dimensions ranging from the size of a protein macromolecule ($50$-$100\ {\rm nm}$) and intracellular compartments: for example, membranes ($10\ {\rm nm}$), nucleus ($5$-$10\ {\rm \mu m}$) and mitochondria ($0.2$-$2.0\ {\rm \mu m}$). Cells also contain a cytoskeleton made of filaments ($7$-$25\ {\rm nm}$), as well as nucleoli ($0.5$-$1.0\ {\rm \mu m}$) and DNA in the form of chromatin in nucleus. These complex structures are often modeled under the assumption that the constituents of cellular interiors are tightly filled with various discrete particles, with their surfaces pushed together to form contiguous biological components. Such cell modeling, when combined with Mie scattering theory, provides a homogeneous medium in a limited range of wavelengths and voxel size of the cell simulations \cite{schmitt1998, kaizu2019, chew2019, *chew2018, *arjunan2010}. Because of these constraints, the spatial fluctuations arising from the interactions between incoherent wave functions and discrete particles cannot be made to line up at the microscopic levels of intensity transport theory without breaking modeling assumptions.

Interactions of a wave function with intracellular inhomogeneities are associated with a turbulence in the refractive index distribution for which the spatial fluctuations are fractal in nature \cite{glaser2016, rogers2013, wax2010, *rogers2010}. For sake of model simplicity, we consider intracellular constituents as a continuous fractal media rather than aggregation of various discrete particles. For a given fractal dimension $D_f$, which usually represents a mathematical index for characterizing fractal patterns or regularities by quantifying their complexity, continuous random fields of refractive index fluctuation $\Delta n$ can be modeled by the Whittle-Mat\'ern (WM) covariance (or correlation) function,
\begin{eqnarray}
B_n(\rho) & = & \frac{\sigma^2_n 2^{1-\nu}}{|\Gamma(\nu)|} \left( \frac{\rho}{l_{c}} \right)^{\nu} K_{\nu} \left( \frac{\rho}{l_{c}} \right)
\label{eqn;Bn}
\end{eqnarray}
where $l_c$ are $\sigma_n^2$ are correlation length and the variance of excessive refractive index, respectively. $K_{\nu}(\cdot)$ denotes the $\nu$-th index of the modified Bessel function of second kind $\nu = (D_f - 3)/2$. The shape of the WM covariance function can be exhibited in a wide range of plausible fractal dimensions including: power law for $D_f < 3$, Henyey-Greenstein for $D_f = 3$, stretched exponential for $3 < D_f < 4$, Kolmogorov/von Karman for $D_f = 3.67$, exponential for $D_f = 4$ and Gaussian as $D_f \to \infty$. Figure \ref{fig02;fractal}(a) shows the WM covariance function for four different fractals models (see the SM section A.2.1 for model parameterizations).

\begin{figure}
\leftline{\bf ({\bf a}) \hspace{0.43\linewidth} ({\bf b})}
\centering
\includegraphics[width=0.49\linewidth]{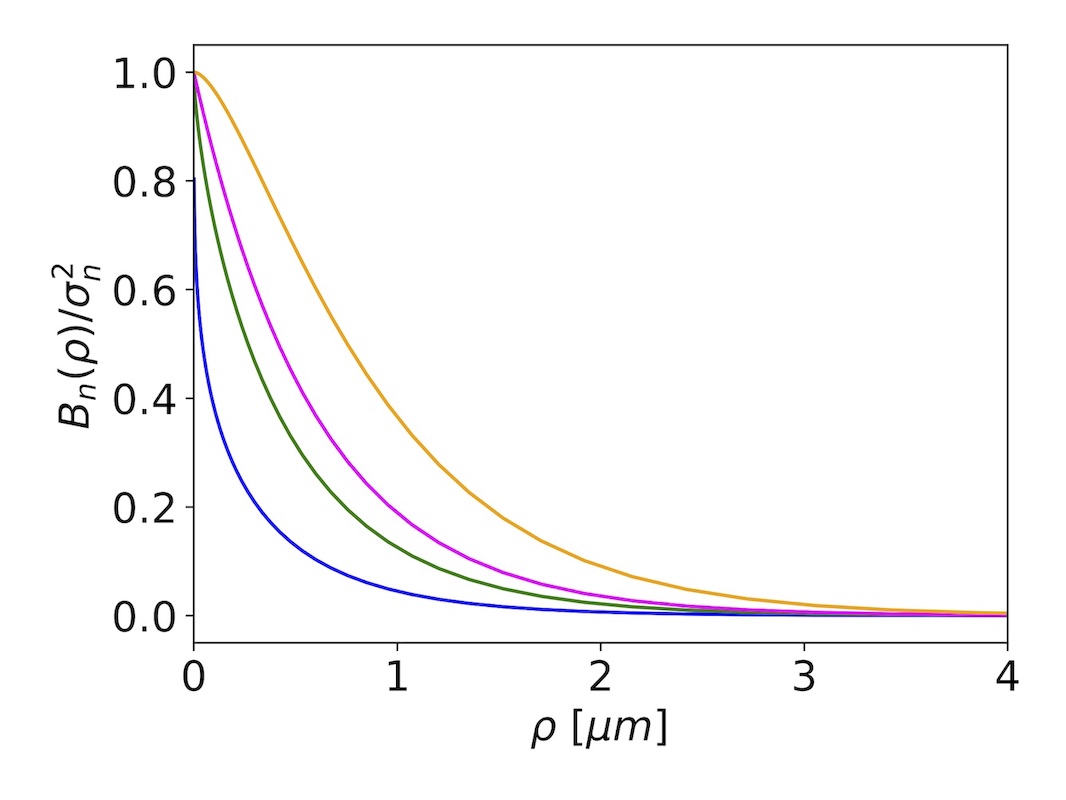}
\includegraphics[width=0.49\linewidth]{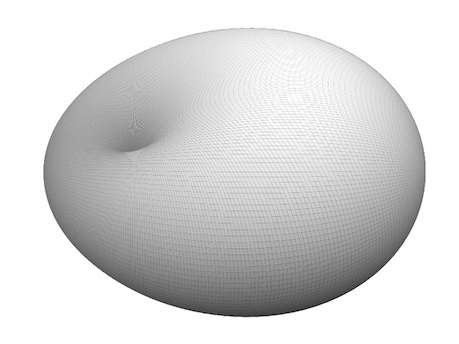}

\caption{Fractal model of intracellular refractive index turbulence. (a) The WM covariance as a function of distance between two different pixel positions $\rho$. Each colored lines denote the four representative models of fractal media: medium I (blue), medium II (green), medium III (magenta) and medium IV (orange).  (b) An example of the mean differential scattering cross-section per unit volume plotted in spherical coordinates ($l_c = 30\ {\rm nm}$). An incident wave function polarized in the vertical plane can propagate from left to right along the longitudinal direction. The dimple is located at the origin.}
\label{fig02;fractal}
\end{figure}

While the attenuation function in the fractal cell model can be represented by the intensity absorption coefficient $\mu_a$ and scattering coefficient $\mu_s$, most cells are transparent, weakly absorbing intermediate fields passing through their interiors. In a limiting case where $\mu_a \ll \mu_s$, we consider four following assumptions: (1) The attenuation function is independent of the spatial variation of scattering coefficient. (2) No backward scattering of the wave function. (3) No depolarization effects. (4) The attenuation function can be written in the form
\begin{eqnarray}
\kappa(r_{\bot},z) & = & \frac{\big<\mu_s(k n_0)\big> }{k n_0} = \frac{1}{k n_0} \iint_{\Omega}\ \big<\sigma(\hat{k}_o,\hat{k}_i) \big>\ d\Omega\ \ \ 
\label{eqn;kappa}
\end{eqnarray}
where $\Omega$ and $k n_0$ are solid angle and the effective wave number of intermediate fields, respectively. $\left<\mu_s(k n_0)\right>$ represents the mean scattering coefficient derived from an integration of the mean differential scattering cross-section $\big<\sigma(\hat{k}_o,\hat{k}_i)\big>$ over all angles. Figure \ref{fig02;fractal}(b) shows an example of the scattering cross-section in spherical coordinates, exhibiting, in particular, forward-directed scattering along the longitudinal direction (see the SM section A.2.2 for the $\sigma$ definition and more figures for the four representative fractal media).

\textit{TIE simulations via fractal cell modeling.}
Fundamental characteristics of the intensity transport through continuous fractal media can be extracted from numerical simulations and analyses of the TIE. Here we assume that phase functions in the TIE are given by Eq. (\ref{eqn;beta}), i.e., $\phi(r_{\bot},z) = -\beta(r_{\bot},z)$, and then implement TIE simulations over a wide range of fractal dimension and wavelength, computing the propagation and variation of lateral intensity distribution through the fractal cell models along the axial direction (see Figure \ref{fig01;tie}). The finite difference approximation of the intensity propagation for the $n$-th image frame can be represented in the form
\begin{eqnarray}
I^{n+1}_{ij} & = & I^{n}_{ij} + \delta z \left( \frac{\partial I}{\partial z}\right)^{n}_{ij} \nonumber\\
& = & I^{n}_{ij} + \delta z \bigg\{ \frac{1}{k n_0} \big[ \nabla_{\bot}\cdot \big( I\ \nabla_{\bot}\beta \big) \big] \nonumber\\
& & \hspace{1.2cm}  \ - 2 [1 + \Delta n(r_{\bot},z)] \big<\mu_s(k n_0)\big> I\ \bigg\}^{n}_{ij}
\label{eqn;fde}
\end{eqnarray}
where $i$ and $j$ represent the index of pixel position for a given image frame; with convergence condition for TIE simulations being $\delta z < k n_0 \delta x \delta y$. 

To analyze biophysical effects arising from intensity propagation via fractal cell modeling, we run the TIE simulations for the following two initial intensity conditions $I(r_{\bot},z_{initial})$: a uniform intensity distribution $1,000\ {\rm counts/pixel}$; and standard image in an intensity range from $500$ to $1,000\ {\rm counts/pixel}$. Our simulation results for a thicker cell sample ($\sim 10\ {\rm \mu m}$) are summarized in Figure \ref{fig03;results} (see the SM section B for more simulation results). For $\lambda = 507\ {\rm nm}$, the intensity propagation of the standard image through the fractal medium III ($D_f = 4.00$) exhibits intensity attenuation and intensity dispersion in the final intensity image [see Figure \ref{fig03;results}(b)]. Figure \ref{fig03;results}(d) shows a comparison between initial and final intensity histograms, thereby dispersing and reducing the intensity distribution of the initial standard image. However, for the fractal medium I ($D_f = 3.25$), there is significant transmissivity in the final intensity distribution [see Figure \ref{fig03;results}(a)]; with almost complete overlap between initial and final intensity histograms is shown in Figure \ref{fig03;results}(a).

\begin{figure}
\leftline{\bf ({\bf a}) \hspace{0.41\linewidth} ({\bf b})}
\centering
\includegraphics[width=0.49\linewidth]{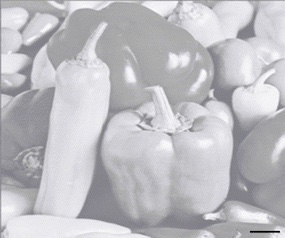}
\includegraphics[width=0.49\linewidth]{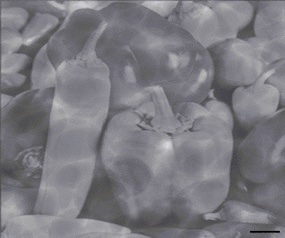}

\leftline{\bf ({\bf c}) \hspace{0.41\linewidth} ({\bf d})}
\centering
\includegraphics[width=0.49\linewidth]{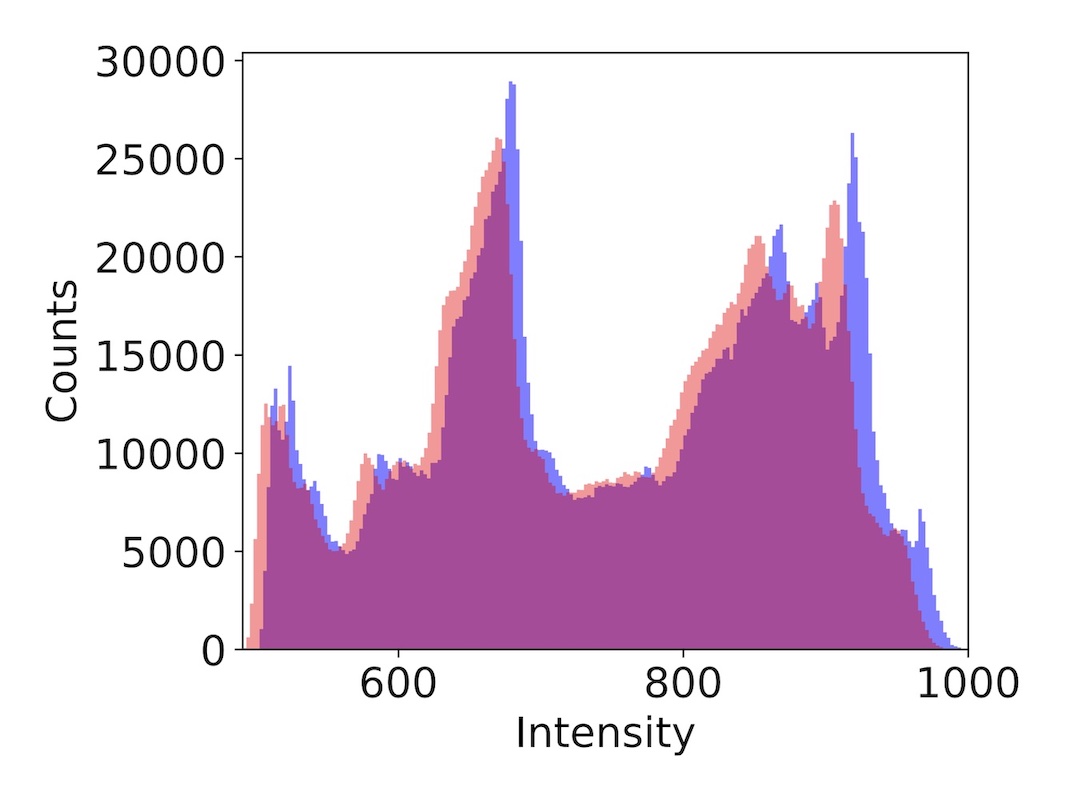}
\includegraphics[width=0.49\linewidth]{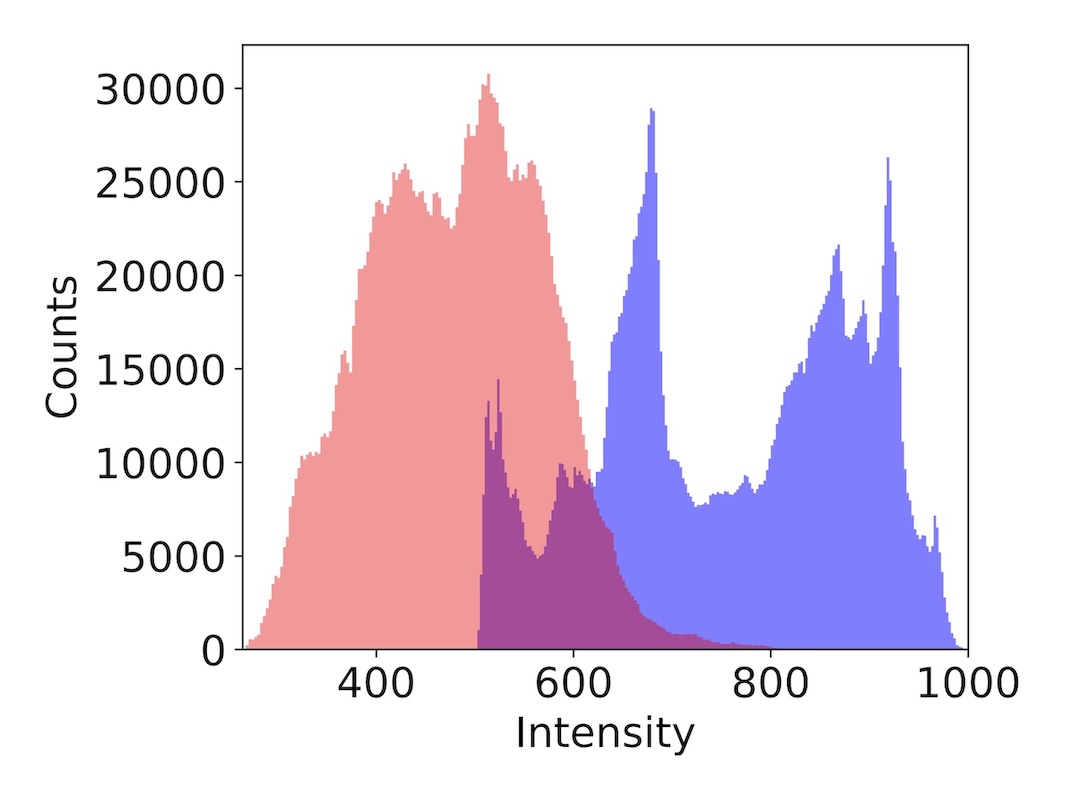}

\leftline{\bf ({\bf e}) \hspace{0.41\linewidth} ({\bf f})}
\centering
\includegraphics[width=0.49\linewidth]{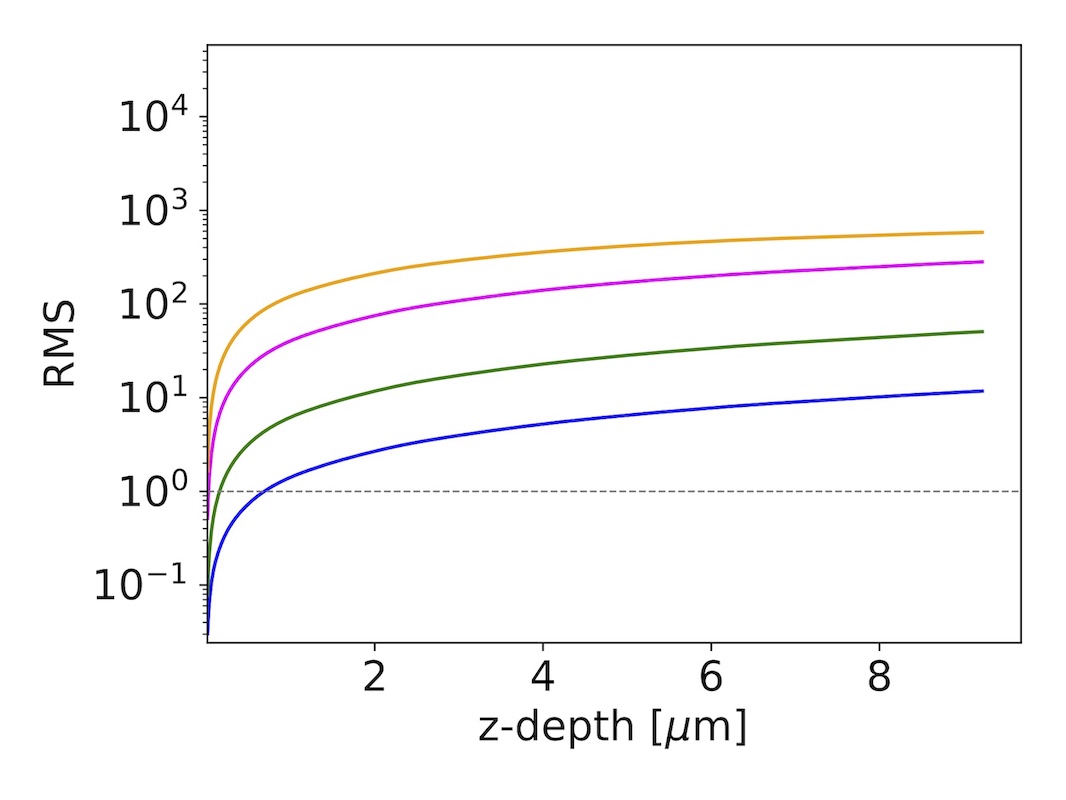}
\includegraphics[width=0.49\linewidth]{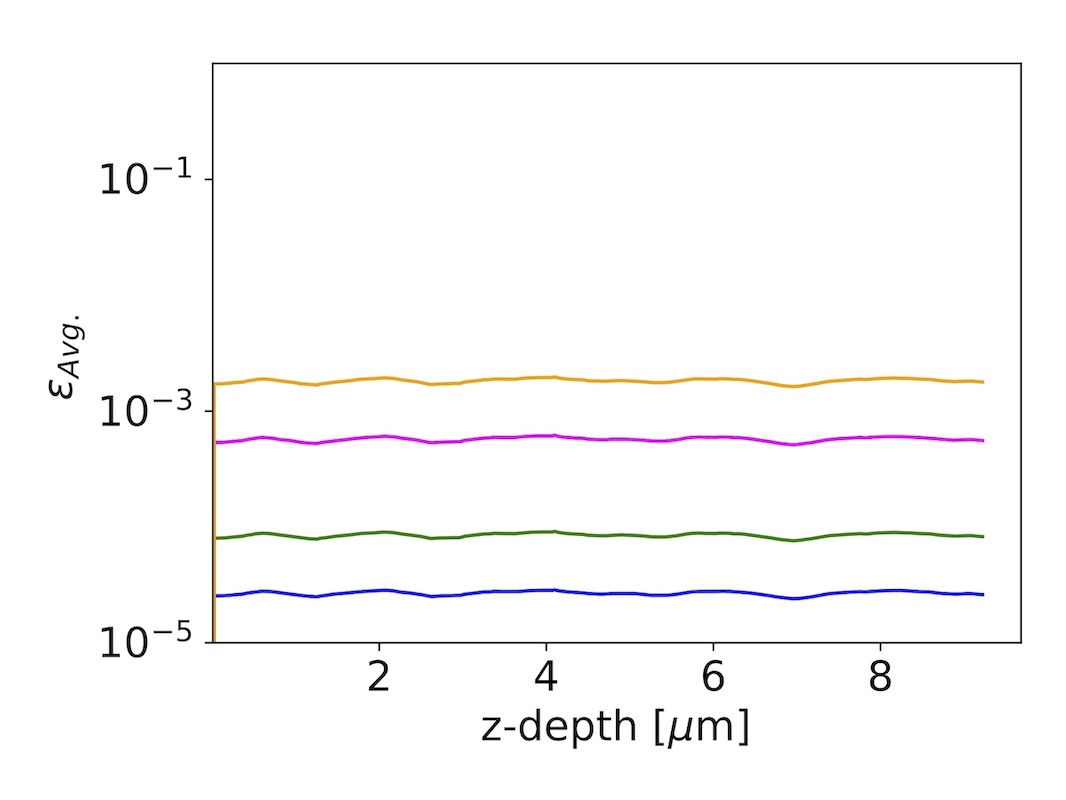}

\caption{The simulation results for thicker cell sample ($\sim 10\ {\rm \mu m}$). (a) Final intensity distribution for medium I ($D_f = 3.25$). Scale bar: $1.00\ {\rm \mu m}$. (b) Final intensity distribution for medium III ($D_f = 4.00$). (c) Histogram comparison between initial and final intensity distributions for medium I ($D_f = 3.25$). Blue and red colored areas represent histograms for initial and final intensity distributions, respectively. (d) Comparison of intensity histograms for medium III ($D_f = 4.00$). (e) RMS variations arising from intensity propagation shown as a function of the z-axis. Each colored lines denote the four different fractal models: medium I (blue), medium II (green), medium III (magenta) and medium IV (orange). (f) Averaged fractional errors shown as a function of the z-axis.}
\label{fig03;results}
\end{figure}

Convergence and stability of the TIE simulation can be seen in the size and fractional error of the intensity dispersion which varies with axial distance. Figure \ref{fig03;results}(e) exhibits relatively fast convergence of the variation of root-mean-square (RMS) values along the axial direction ($z > 2$-$3\ {\rm \mu m}$). The TIE simulation for fractal medium I ($D_f = 3.25$) and medium II ($D_f = 3.67$) converges to steady state at relatively lower RMS values, roughly $50$ (green line) and $11$ (red line). However, at higher  fractal dimensions, i.e., $D_f = 4.00$ and $5.00$, the RMS value that gives rise to convergence of intensity dispersion is about $250$ (magenta line) and $500$ (orange line). In addition, Figures \ref{fig03;results}(f) shows propagation of averaged fractional errors along z-axis, $\epsilon(z_{p}) = |I(z_{p}) - I(z_{p-1})|/I(z_{p})$, thus implying no divergence of numerical errors.

\begin{figure}
\centering
\includegraphics[width=0.99\linewidth]{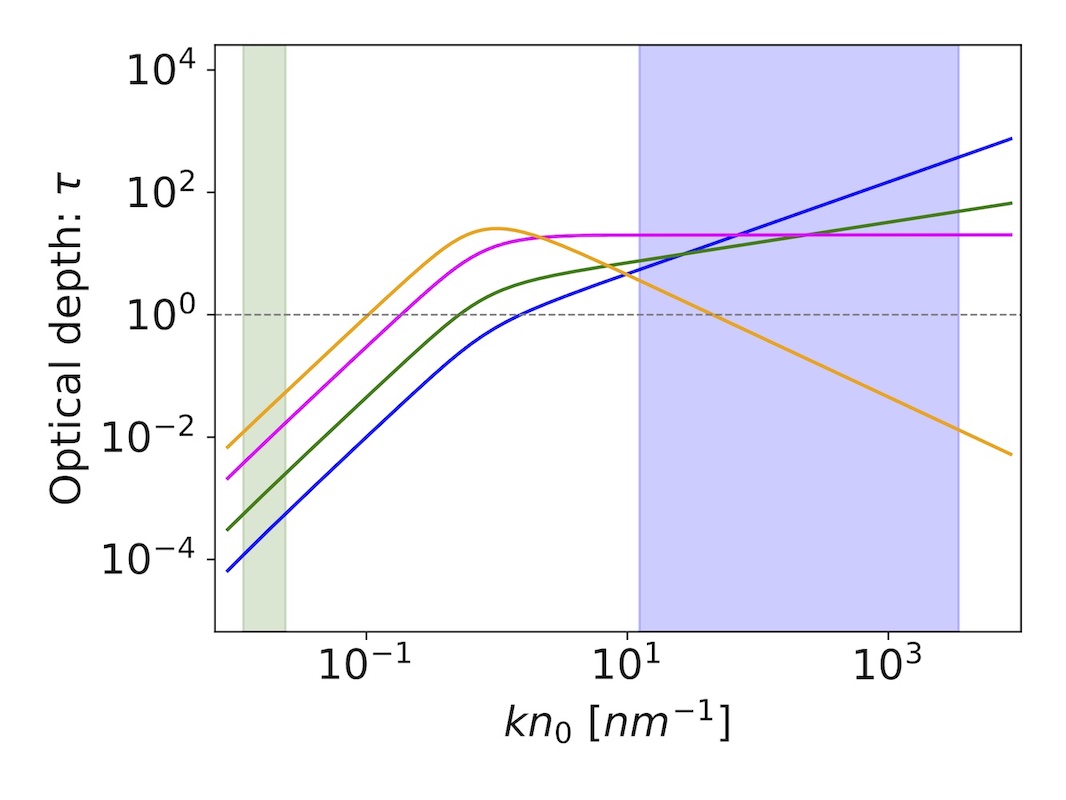}

\caption{Optical depth for sample thickness ($L = 110\ {\rm nm}$) represented over a wide range of effective wave number $k n_0$ values, from visible light (green) to electron beam (blue). Color lines denote the four different fractal media. If $\tau \gg 1$, then the sample is optically thick. If $\tau \ll 1$, then the sample is optically thin. Dashed line represents $\tau = 1$.}
\label{fig04;depth}
\end{figure}

Electron microscopy is capable of imaging thinner cell sample sizes less than $150\ {\rm nm}$ \cite{em}. In the wavelength range of electron beams, we consider optical depth (or thickness) to quantify a specific level of transparency of the four fractal media. Optical depth is dimensionless factor that generally represents a measure of scattering and absorption up to a specific ``depth" of intracellular media when intermediate fields (e.g., light and electron) travel through the structure inside the sample. In fractal cell modeling, optical depth can be defined as follows.
\begin{equation}
\tau = \int^{L}_{0} \left<\mu_s(k n_0)\right>dz
\end{equation}
where $L$ is physical thickness of biological samples. For $L = 110\ {\rm nm}$, Figure \ref{fig04;depth} shows optical depth over a broad range of wavelength from electron beams (blue region, $\lambda = 1.00\ {\rm pm} \sim 1.00\ {\rm nm}$) to visible light (green region, $\lambda = 360 \sim 760\ {\rm nm}$). The fractal media are optically thin ($\tau \ll 1$) and transparent through the visible light wavelength, weakly dispersing the light intensities passing through their interiors. Such intensity dispersion thus provides approximately-invariant transport between initial and final intensity distributions, $I(z_{final}) \sim I(z_{initial})$ [see Figures S10(a) in the SM section B2]. However, in a wavelength range of electron microscopy imaging, there is a significant tradeoff between fractal-dimensionality and optical depth. Fractal medium I (blue), II (green) and III (magenta) are optically thick ($\tau \gg 1$), exhibiting, in particular, intensity attenuation as well as intensity dispersion in the final standard image [see Figure S10(d)-(f)]. Mathematically speaking, as an increase of fractal dimensions, optical depth becomes thinner and the transmissivity thereof can be increased in the fractal medium IV (orange).

%

\textit{Applications.}
A further motivation to our work is the direct measurement and reconstruction of the refractive index profiles of cellular interiors. Recent experimental studies using the TIE via bioimaging \cite{mitome2021, rajput2021}, for example, have reconstructed intracellular phase distributions without modeling the refractive index turbulence. Likewise, J. D. Rogers and his colleagues have verified the apparent similarity in power spectral densities between the WM covariance function and scanning electron microscopy images \cite{rogers2013}. These qualitative verifications, however, were made without applying intensity transport theory and so cannot be valid as direct measurements of the refractive index profiles. For a proper comparison and measurement in bioimaging, a mathematical relation between the phase factor and the WM covariance model can be derived from the spatial-autocorrelation of Eq. (\ref{eqn;beta}) and then written in the form
\begin{eqnarray}
\left< \frac{\partial \beta(r)}{\partial z}\frac{\partial \beta(r + \rho)}{\partial z} \right> & = & (k n_0)^2 B_n(\rho) + \bigg[ \frac{\left<\mu_s(k n_0)\right>}{\sqrt{2 k n_0}} \bigg]^4\ \ \ \ 
\label{eqn;acorr}
\end{eqnarray}
when satisfying the four assumptions declared in the cell modeling section (see the SM section C.1 for derivation details). In this formula, the WM covariance function $B_n(\rho)$ and the mean scattering coefficient $\left<\mu(k n_0)\right>$ represent the shape and offset of the phase-differential autocorrelation along axial directions. Figure S11 shows no significant shape differences over a broad wavelength range from electron beams (dashed lines) to visible lights (solid lines); a large variation of the offset is found at $\rho > 10\ {\rm \mu m}$ due to the dependence on the effective wave number. Moreover, we use this mathematical relation to reconstruct intracellular optical properties (e.g., fractal dimension and scattering coefficient) from fluorescent cell imaging. In our demonstration, the covariance function is directly compared and fitted to the observed spatial-autocorrelation curves of phase derivatives. Table S4 and Figures S14 not only represent the best estimates in fractal model parameters but also their statistical uncertainties which gives a numerical indication of the level of validity and confidence in our fitting results (see the SM section C.2 for analysis details).


\textit{Conclusion.}
The TIE plays a key role in bridging the resolution gap between optical and electron microscopy techniques, providing, in particular, a non-interferometric correlation between intensity and phase functions of intermediate fields (e.g., electrons and light) over a broad range of wavelengths. These microscopy techniques are capable of directly imaging the intensity variations of intermediate fields passing through the interior of cells. Variations in phase functions, however, are barely visible with bioimaging systems, and sensitive to light scattering and fluctuation through invisible (or unobservable) spatial distributions of intracellular refractive and attenuation indices. In this paper, we reformulated the TIE through fractal modeling of these two indices. Our results from these TIE simulations revealed that intensity propagations through the refractive index fluctuation can lead to nonintuitive interconnections of fractal-dimensionality and intensity dispersion (or transmissivity) in the picometer to micrometer wavelength range. We also derived the spatial-correlation of phase derivatives that enables the direct measurements of the WM covariance model parameters from optical and electron microscopy imaging. 


Of further significance to our work is a numerical evaluation and verification of observational invariance (or symmetry) as postulated in data science. Most subcellular observations made via optical imaging are dedicated to an empirical (or data-driven) approach, its main function being to allow biophysicists to extract the regularities and patterns of the apparent intracellular properties captured with photosensitive devices \cite{meijering2020, moen2019, jordan2015, marx2013, danuser2011}. These regularities are often formalized into network models, mainly as a function of the observable intracellular components. However, in actual intracellular imaging, propagations and variations of light through inhomogeneous intracellular structures are dependent on the wavelength of light, as well as intracellular optical properties  such as the refractive index, optical thickness, and attenuation coefficients. These optical dependencies raise questions regarding whether the subcellular model representations can be conserved or violated through observational processes. Our work sheds light on these questions from the perspective of fundamental optics, and also suggests new developments and extensions to bioimage simulation modules \cite{mahajan2022, weigert2018, girsault2016, linden2016, *linden2017, venkataramani2016, watabe2019, angiolini2015, rezatofighi2013, sbalzarini2013, boulanger2009}, including, in particular, light scattering and fluctuations caused by the nonuniform spatial distributions of intracellular optical properties. Such an implementation is of broad relevance beyond just the TIE simulations presented here, and will likely lead to more realistic simulations of biological imaging.

\textit{Acknowledgments.}
We would like to thank Hideaki Yoshimura, Takahiro Nishimura, Naru Yoneda, Kozo Nishida and all the other members of the laboratory for biologically inspired computing in RIKEN, for their guidance and support throughout this research work. The research work is supported by JSPS (Japanese Society for the Promotion of Science) KAKENHI Grant Numbers JP20H05886, JP20H05891, JP20K21836 and JP21H05605.

%

\onecolumngrid

\newpage

\leftline{\bf\Large Supporting Material}

\maketitle

\newcolumntype{C}[1]{>{\centering\arraybackslash}p{#1}}

\setcounter{section}{0} \renewcommand{\thesection}{\Alph{section}}
\setcounter{subsection}{0} \renewcommand{\thesubsection}{\Alph{section}.\arabic{subsection}}
\setcounter{subsubsection}{0} \renewcommand{\thesubsubsection}{\Alph{section}.\arabic{subsection}.\arabic{subsubsection}}
\setcounter{figure}{0} \renewcommand{\thefigure}{S\arabic{figure}}
\setcounter{table}{0} \renewcommand{\thetable}{S\arabic{table}}
\setcounter{equation}{0} \renewcommand{\theequation}{S\arabic{equation}}

\titleformat*{\section}{\large\bfseries}
\titleformat*{\subsection}{\normalsize\bfseries}
\titleformat*{\subsubsection}{\normalsize\bfseries}



\tableofcontents
\vspace{1.0cm}
\addtocontents{toc}{\protect\rule{\textwidth}{.2pt}\par}

\newpage

\section{Intensity transport via fractal cell modeling}
\subsection{Modification of the TIE}
A plane wave function propagating through the refractive index distribution along the z-axis (see Figure 1 in the main text) obeys Helmholz equation,
\begin{equation}
\left[ \nabla^2 + k^2 n(r_{\bot},z)^2 \right] A(r_{\bot},z)\ e^{ik n_{0} z} = 0
\label{eqn;helm}
\end{equation}
where $\nabla^2$ denotes Laplacian operator in three dimensions; $A(r_{\bot},z)$ is a complex amplitude of the wave function. Substituting Eq. (1) in the main text into Eq. (\ref{eqn;helm}), we can deduce a paraxial wave equation as follows,
\begin{eqnarray}
\bigg\{ \nabla_{\bot}^2 + 2 k n_0 i \frac{\partial }{\partial z} + 2 (k n_0)^2 \Big[ \Delta n(r_{\bot},z) - \frac{1}{2} \kappa(r_{\bot},z)^2 \Big] + i\ 2 (k n_0)^2 [1 + \Delta n(r_{\bot},z)]\ \kappa(r_{\bot},z) \bigg\} A(r_{\bot},z) = 0
\label{eqn;pwe0}
\end{eqnarray}
where $\nabla_{\bot}$ is the two-dimensional Nabla operator in the transverse direction.
The amplitude propagator of the paraxial wave equation can be represented in the finite difference form of
\begin{eqnarray}
U(r_{\bot}, z + \delta z) & \cong & \frac{A(r_{\bot}, z + \delta z)}{A(r_{\bot}, z)} \nonumber\\
\nonumber\\
& = & \exp \bigg\{ \frac{i\delta z}{2k n_0} \nabla^2_{\bot} + i k n_0 \Big[ \Delta n(r_{\bot},z) - \frac{1}{2} \kappa(r_{\bot},z)^2 \Big] \delta z - k n_0 [1 + \Delta n(r_{\bot},z)]\ \kappa(r_{\bot},z) \delta z \bigg\}
\label{eqn;amp0}
\end{eqnarray}
where $\delta z$ denotes small changes in axial axis. The first term in the exponential factor represents the kinetic energy in amplitude propagation. The second term is coupled with an imaginary part of the wave function, representing, in particular, phase variation arising from the spatial distribution of refractive and attenuation index ($\delta \beta$). The third part exhibits the amplitude variation in a real part of wave function ($\delta \alpha$). Using Eq. (4) and (5) in the main text, Eq. (\ref{eqn;pwe0}) can be rewritten in the form of
\begin{eqnarray}
\left[ \nabla_{\bot}^2 + 2 k n_0 i \frac{\partial }{\partial z} + 2 k n_0 \frac{\partial}{\partial z} \bigg( \beta(r_{\bot},z) + i \alpha(r_{\bot},z) \bigg) \right] A(r_{\bot},z) = 0.
\label{eqn;pwe1}
\end{eqnarray}
\
\\


\noindent Here we assume that the scalar complex amplitude is given by 
\begin{eqnarray}
A(r_{\bot},z) & = & \sqrt{I(r_{\bot},z)}\ e^{i \phi(r_{\bot},z)}
\end{eqnarray}
where $I(r_{\bot}, z)$ and $\phi(r_{\bot}, z)$ are intensity and phase factor of the wave function, respectively. We then modify Eq. (\ref{eqn;pwe1}) as follows.\\

\noindent $A^{\ast} \times $ Eq. (\ref{eqn;pwe0}):
\begin{eqnarray}
A^{\ast} \nabla_{\bot}^2 A + 2 k n_0 i A^{\ast} \frac{\partial A}{\partial z} + 2 k n_0 A^{\ast} \frac{\partial}{\partial z} \bigg( \beta(r_{\bot},z) + i \alpha(r_{\bot},z) \bigg) A = 0
\label{eqn;pweA}
\end{eqnarray}

\noindent $A\ \times $ Eq. (\ref{eqn;pwe0})*:
\begin{eqnarray}
A\ \nabla_{\bot}^2 A^{\ast} - 2 k n_0 i A\ \frac{\partial A^{\ast}}{\partial z} + 2 k n_0 A\ \frac{\partial}{\partial z} \bigg( \beta(r_{\bot},z) - i \alpha(r_{\bot},z) \bigg) A^{\ast} = 0
\label{eqn;pweB}
\end{eqnarray} 

\noindent Subtracting Eq. (\ref{eqn;pweB}) from Eq. (\ref{eqn;pweA}), we obtain
\begin{eqnarray}
\nabla_{\bot} \left( A^{\ast} \nabla_{\bot} A - A \nabla_{\bot} A^{\ast} \right)
+ 2 k n_0 i \left( A^{\ast} \frac{\partial A}{\partial z} + A \frac{\partial A^{\ast}}{\partial z}\right)
+ 2 k n_0 \frac{\partial}{\partial z} \bigg( 2 i \alpha(r_{\bot},z) \bigg) A^{\ast}A = 0
\label{eqn;pwe2}
\end{eqnarray}
where $A^{\ast}A = I$ and $A^{\ast} \nabla_{\bot} A - A \nabla_{\bot} A^{\ast} = 2 i I \nabla_{\bot} \phi$. Substituting these amplitude properties into Eq. (\ref{eqn;pwe2}), we can deduce TIE as follows.
\begin{eqnarray}
\nabla_{\bot}\cdot \big[ I(r_{\bot},z)\ \nabla_{\bot} \phi(r_{\bot},z) \big] = - 
k n_0 \bigg[ \frac{\partial I(r_{\bot},z)}{\partial z} + 2 \frac{\partial \alpha(r_{\bot},z)}{\partial z} I(r_{\bot},z) \bigg].
\end{eqnarray}
where the $\beta$-factor  vanishes through this modification.


\newpage

\subsection{Fractal cell modeling}
We construct fractal cell models based on the Whittle-Mat\'ern (WM) covariance family, which is often used to describe continuous random fluctuation of intracellular refractive index \cite{glaser2016sm, rogers2013sm, wax2010sm, rogers2010sm}. 

\subsubsection{The index of refraction}
We construct four different fractal cell models from focused ion beam scanning electron microscopy (FIBSEM) images of {\it Cyanidioschyzon merolae} cells \cite{yudistira2020sm, ichinose2017sm}. Model parameterizations are shown in Table \ref{tabA2;models}. Our procedure for model construction is given as follows.

\begin{figure*}[!h]
\leftline{\bf \hspace{0.02\linewidth} (a) \hspace{5.2cm} (b) \hspace{5.2cm} (c)}
\centering
\includegraphics[width= 0.32\linewidth]{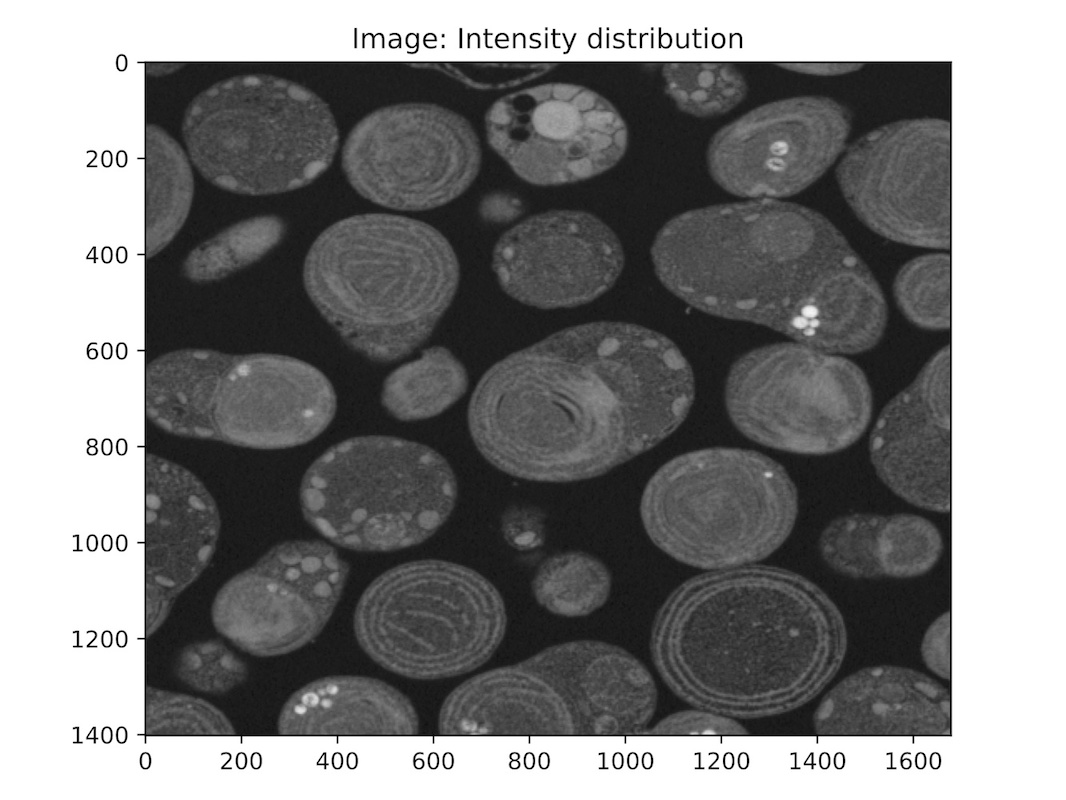}
\includegraphics[width= 0.32\linewidth]{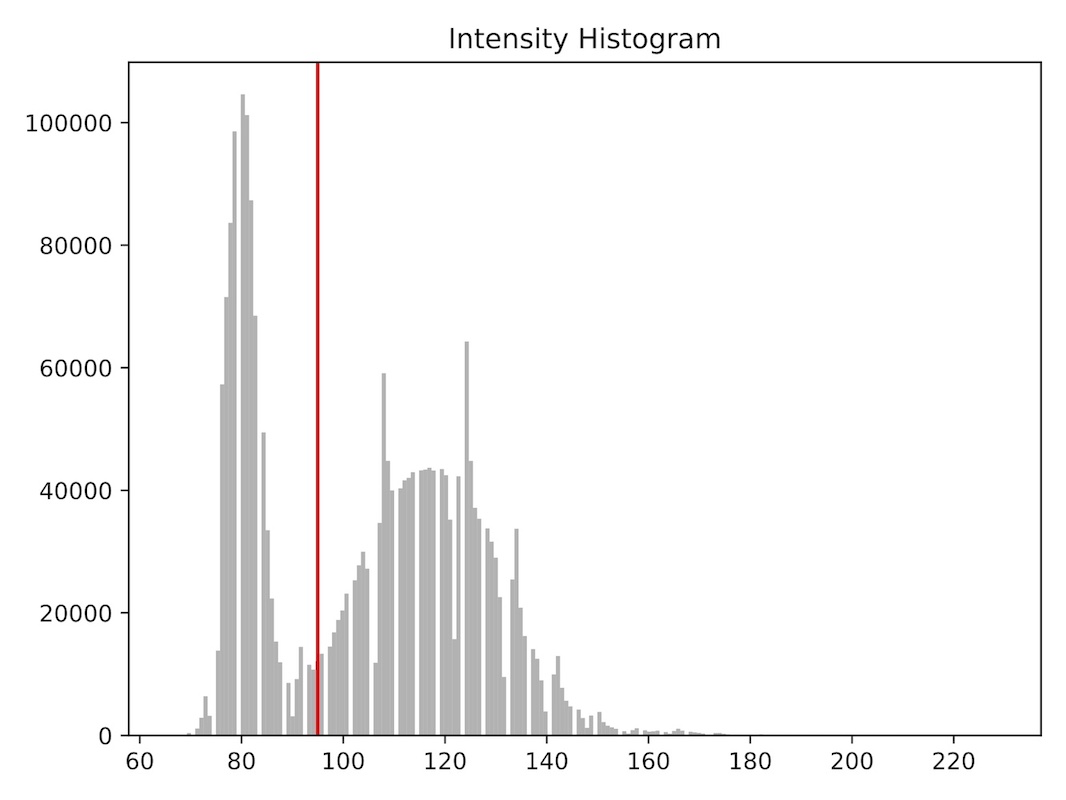}
\includegraphics[width= 0.32\linewidth]{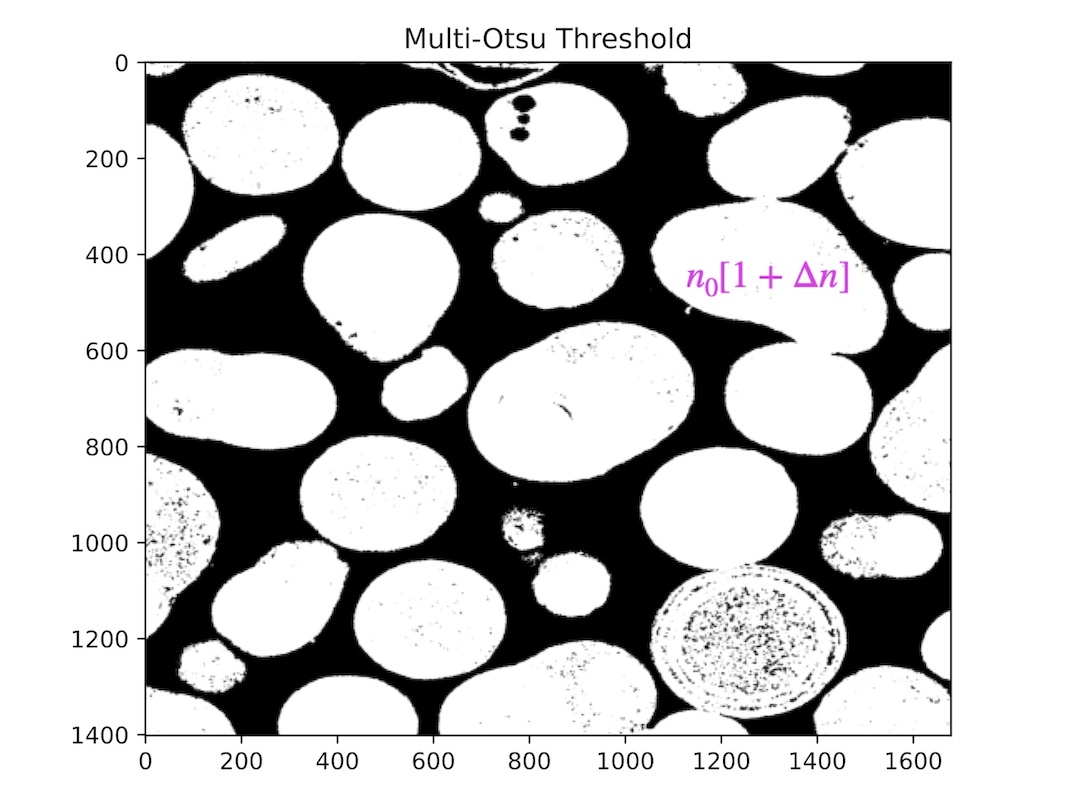}

\caption{(a) An example snapshot of FIBSEM cell imaging. (b) Intensity histogram of the FIBSEM cell image. Red line represents the threshold to separate intracellular regions ($I > I_{cut}$) and non-cellular background regions ($I < I_{cut}$) (c) FIBSEM image is converted into binary image: intracellular region (white) and background (black).}
\label{figA2;fibsem}
\end{figure*}

\begin{itemize}
\item[(1)] 3D cell imaging:
{\it C. merolae} is a unicellular haploid red alga that inhabits sulphate-rich environments such as acidic hot spring water (${\rm pH}\ 1.5$, $45^{\circ}C$) \cite{miyagishima2021sm}. The cell has a size of about $2$-$5\ {\rm \mu m}$ with relatively simple cellular architecture, containing only $4775$ protein-coding genes in its nucleus and the minimum number of membrane-enclosed organelles (i.e., a single nucleus, a single mitochondrion and a single chloroplast), lacking a vacuole and cell wall. Furthermore, cellular and organelle divisions can be highly synchronized through cell cycle progression. Therefore, the {\it C. merolae} cells is an excellent model system in which to study cellular and organelle division processes as well as structural biology.

\hspace{\parindent} FIBSEM imaging techniques involve combining serial gallium-ion beam etching of a resin block-face of cell samples with electron scanning of the exposed block surface. Such combinatorial operation allows accurate visualization of the 3D intracellular structure of {\it C. merolae} at the whole cell level. Parameters for the FIBSEM cell imaging are configured as follows: gallium-ion beam current, $1.9\ {\rm pA}$; electron wavelength, $\lambda = 4.00\ {\rm pm}$; defocus distance, $\delta z = 11.7\ {\rm nm}$; image pixel length, $\delta x = \delta y = 5.6\ {\rm nm}$; total number of images, $796$ frames; and each image size, $1679 \times 1402\ {\rm pixels}$. Figure \ref{figA2;fibsem}(a) shows an example snapshot of the intensity distribution for FIBSEM images of a {\it C. merolae} cell block.
\item[(2)] In the FIBSEM images, multi-Otsu thresholding method \cite{skimage} can be applied to split between intracellular region and non-cellular background region. The cell segmentation results for the $100$-th image frame are shown in Figures \ref{figA2;fibsem}(b) and (c).
\item[(3)] In binary images, we use a spatial statistical tool such as GSTools \cite{gstool} to generate spatial random fields of intracellular refractive index, following, in particular, the WM covariance function $B_n(\rho)$ [i.e., Eq. (6) in the main text].
\item[(4)] Lateral distributions and their histograms of refractive index $n_0 [1 + \Delta n]$ for the four different fractal media are shown in the left and middle columns of Figures \ref{figA2;results00}.
\end{itemize}

\vspace*{\fill}

\newpage

\vspace*{\fill}

\begin{table*}[!h]
    \centering
    \begin{tabular}{|l|l|}
    \hline
     & \hspace{5.0cm} Fractal model parameters \hspace{5.0cm} \\ \hline
    \hspace{1.0cm} Medium I \hspace{1.0cm} & \hspace{1.0cm} $n_{bg} = 1.33$, $n_0 = 1.38$, $\sigma^2_n = 0.25 \times 10^{-4}$, $l_c = 0.6\ {\rm \mu m}$, $D_f = 3.25$ (Stretched exp.) \\ \hline
    \hspace{1.0cm} Medium II & \hspace{1.0cm} $n_{bg} = 1.33$, $n_0 = 1.38$, $\sigma^2_n = 0.50 \times 10^{-4}$, $l_c = 0.6\ {\rm \mu m}$, $D_f = 3.67$ (von Karman) \\ \hline
    \hspace{1.0cm} Medium III & \hspace{1.0cm} $n_{bg} = 1.33$, $n_0 = 1.38$, $\sigma^2_n = 0.25 \times 10^{-3}$, $l_c = 0.6\ {\rm \mu m}$, $D_f = 4.00$ (Exponential) \\ \hline
    \hspace{1.0cm} Medium IV & \hspace{1.0cm} $n_{bg} = 1.33$, $n_0 = 1.38$, $\sigma^2_n = 0.50 \times 10^{-3}$, $l_c = 0.6\ {\rm \mu m}$, $D_f = 5.00$ (Gaussian) \\ \hline
    \end{tabular}
    \caption{Model parameters for four different fractal media. $n_{bg}$ represents the index of refraction in non-cellular background regions. The other model parameters are described in the main text.}
    \label{tabA2;models}
\end{table*}

\subsubsection{Differential scattering cross-section}

\begin{wrapfigure}[16]{r}{0.40\linewidth}
\centering
\includegraphics[width= 0.98\linewidth]{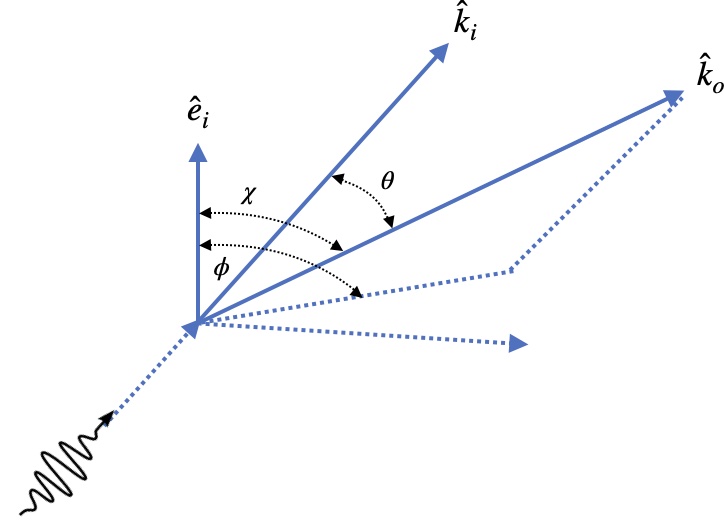}
\caption{The schematic view of a monochromatic plane wave function scattering from a continuous fractal medium.}
\label{figA2;scheme}
\end{wrapfigure}

\begin{figure*}[!h]
\leftline{\bf \hspace{0.02\linewidth} (a) \hspace{5.2cm} (b) \hspace{5.2cm} (c)}
\centering
\includegraphics[width= 0.32\linewidth]{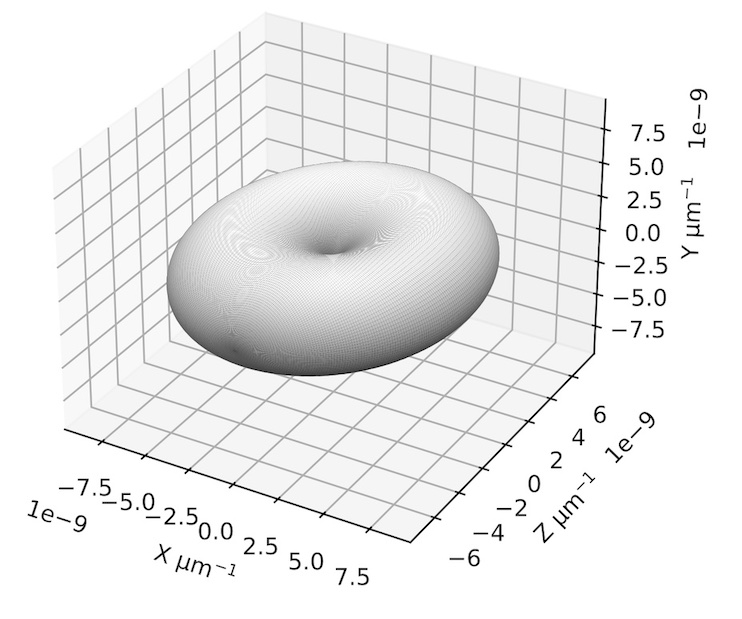}
\includegraphics[width= 0.32\linewidth]{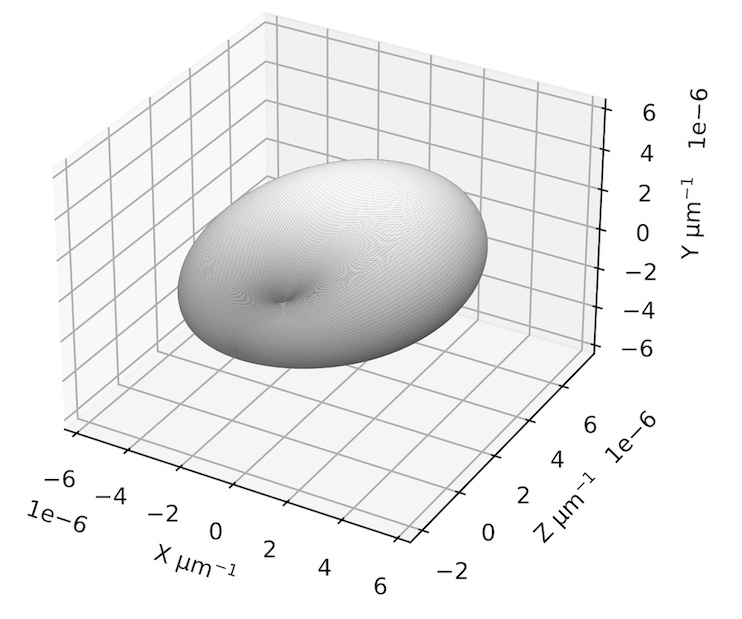}
\includegraphics[width= 0.32\linewidth]{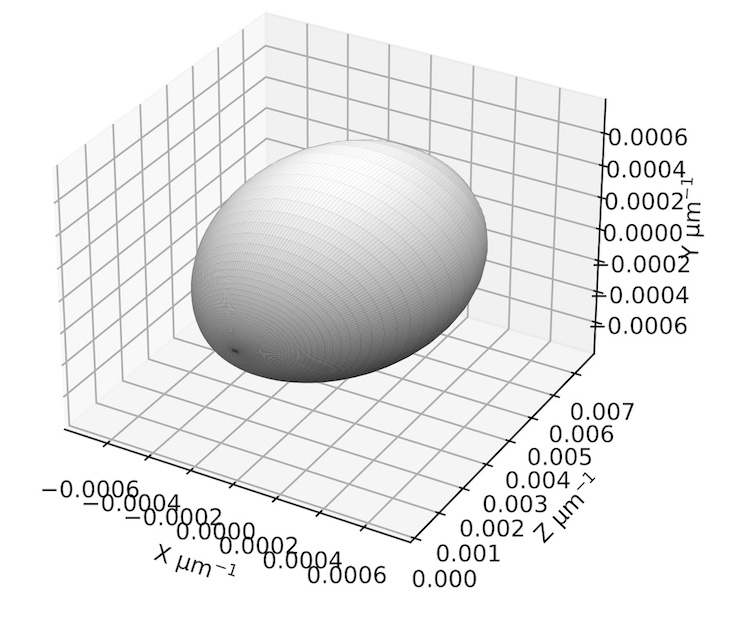}

\caption{Several examples of the mean differential scattering cross sections per unit volume plotted in spherical coordinate. An incident wave function propagates along the axial direction, and polarized in the y-axis. (a) The result for the correlation length $l_c = 3.00\ {\rm nm}$. The dimple is located in the origin. The other model parameters in these examples are configured to $n_0 = 1.38$, $\sigma^2_n = 0.25 \times 10^{-4}$ and $D_f = 3.25$. (b) The result for the correlation length $l_c = 30.0\ {\rm nm}$. (c) The result for the correlation length $l_c = 300\ {\rm nm}$.}
\label{figA2;xsec}
\end{figure*}

\begin{figure*}[!h]
\leftline{\bf \hspace{0.02\linewidth} (a) Medium I}
\centering
\includegraphics[width= 0.30\linewidth]{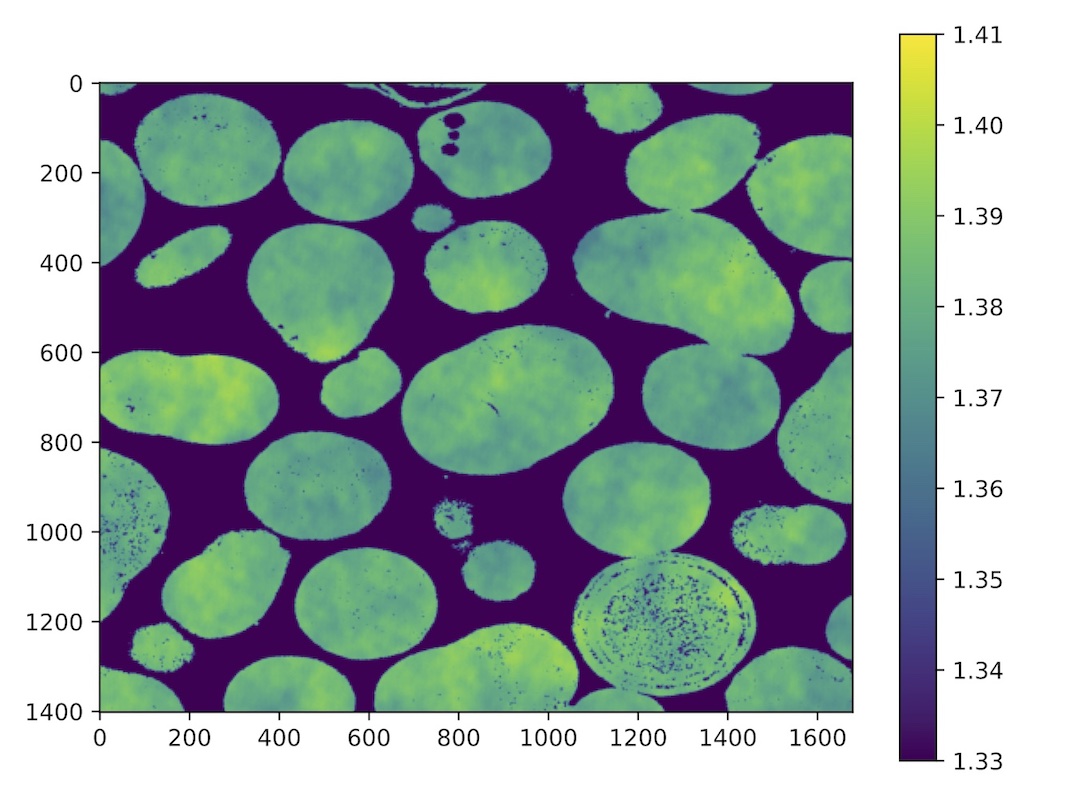}
\includegraphics[width= 0.30\linewidth]{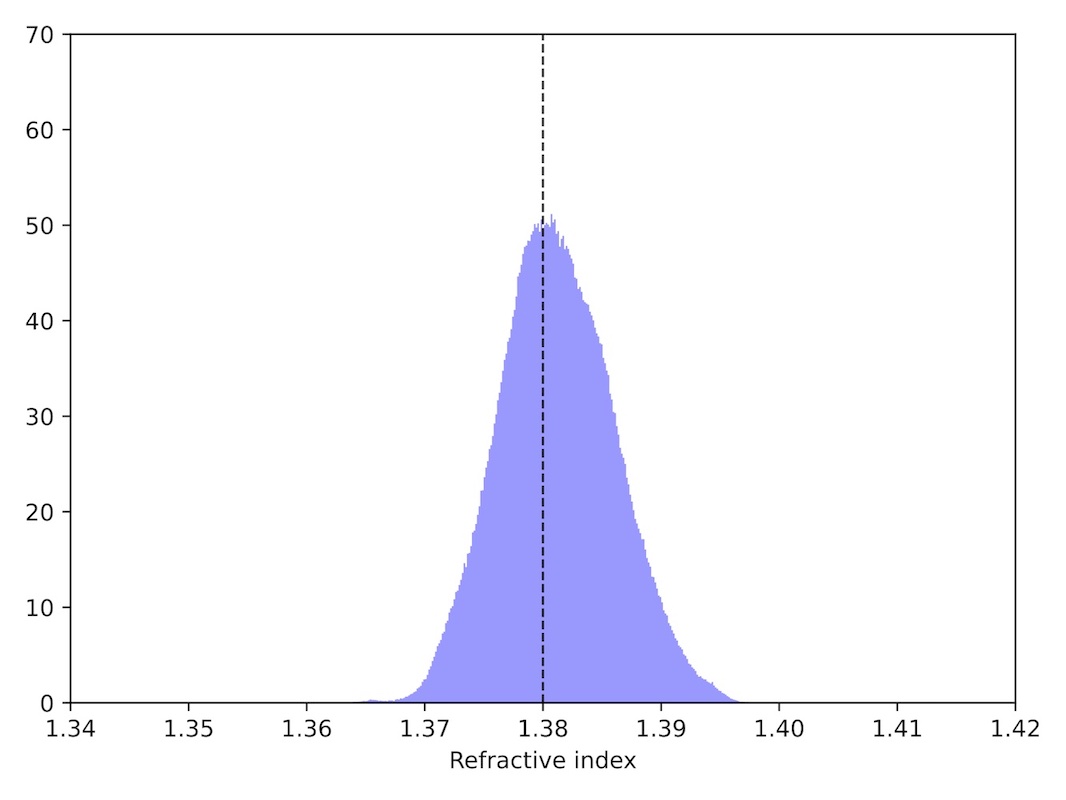}
\includegraphics[width= 0.28\linewidth]{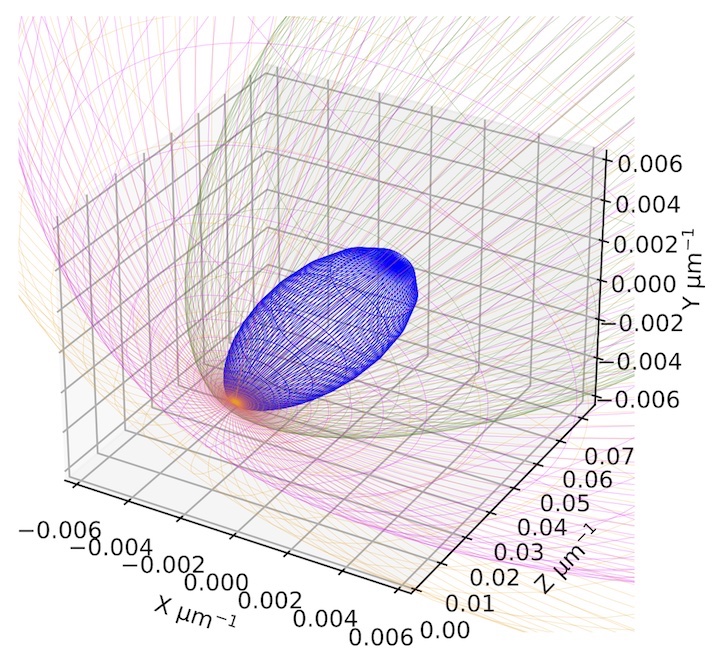}

\leftline{\bf \hspace{0.02\linewidth} (b) Medium II}
\centering
\includegraphics[width= 0.30\linewidth]{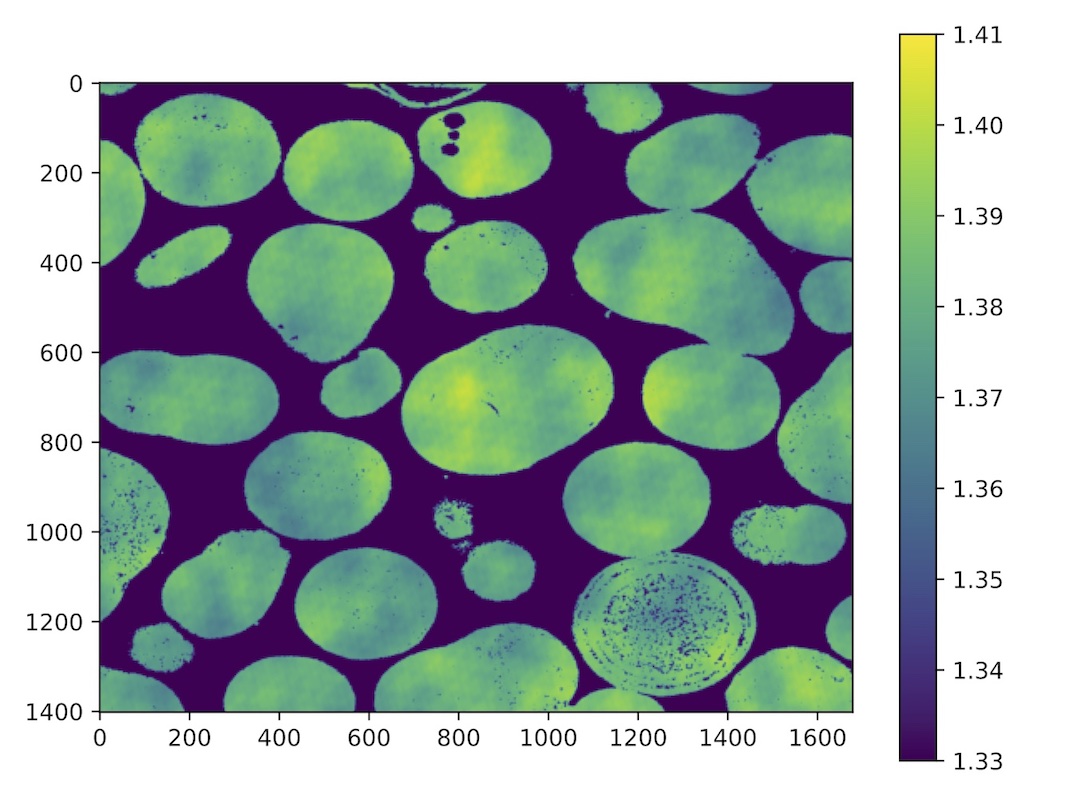}
\includegraphics[width= 0.30\linewidth]{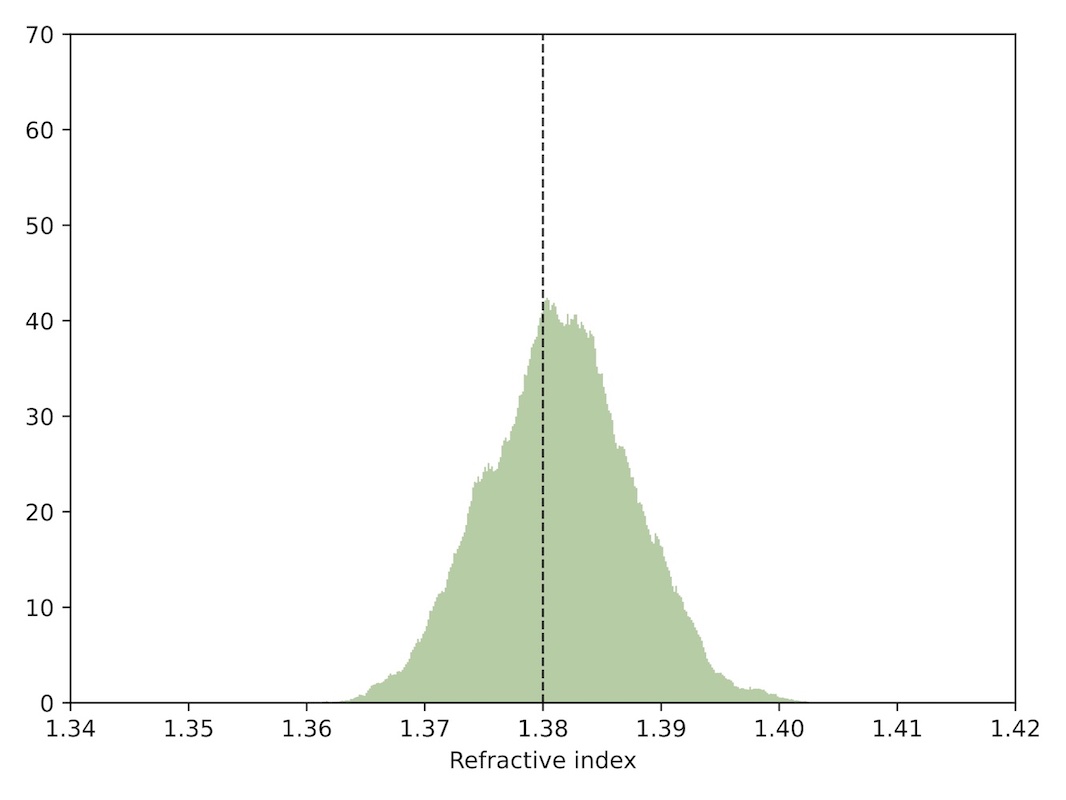}
\includegraphics[width= 0.28\linewidth]{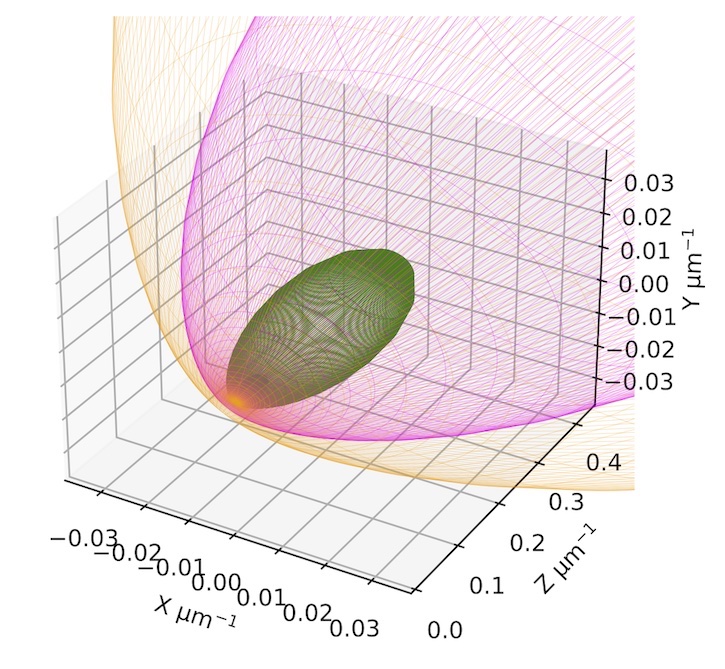}

\leftline{\bf \hspace{0.02\linewidth} (c) Medium III}
\centering
\includegraphics[width= 0.30\linewidth]{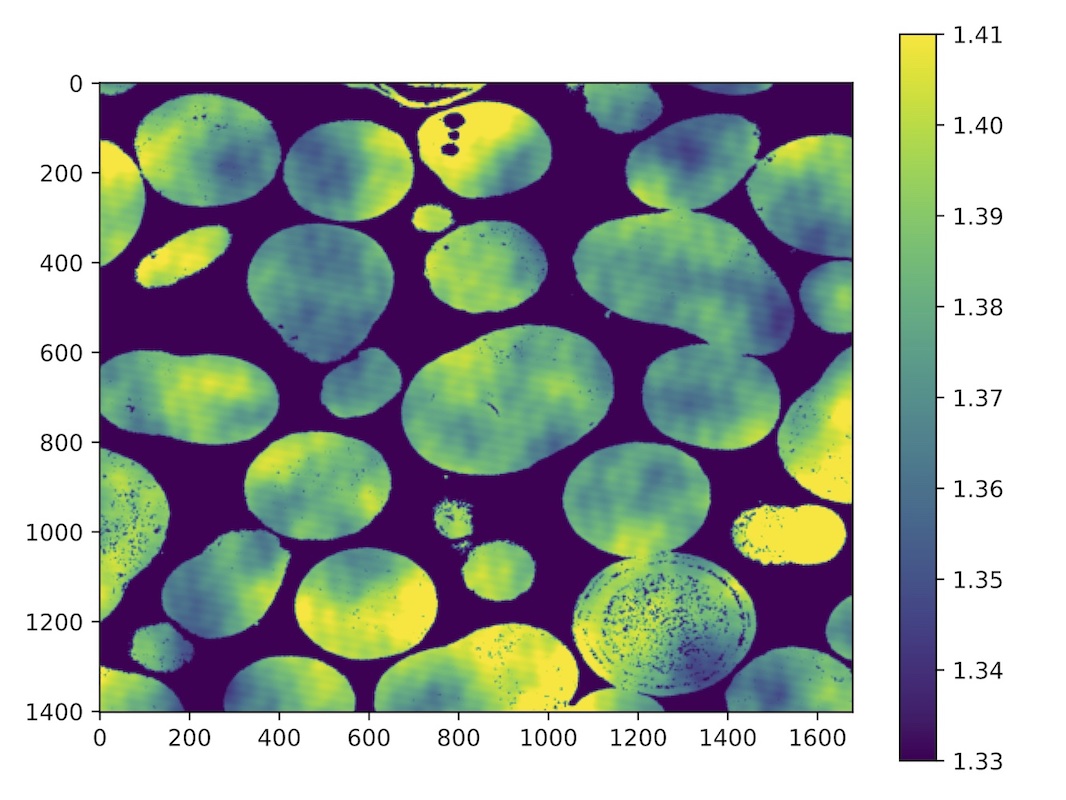}
\includegraphics[width= 0.30\linewidth]{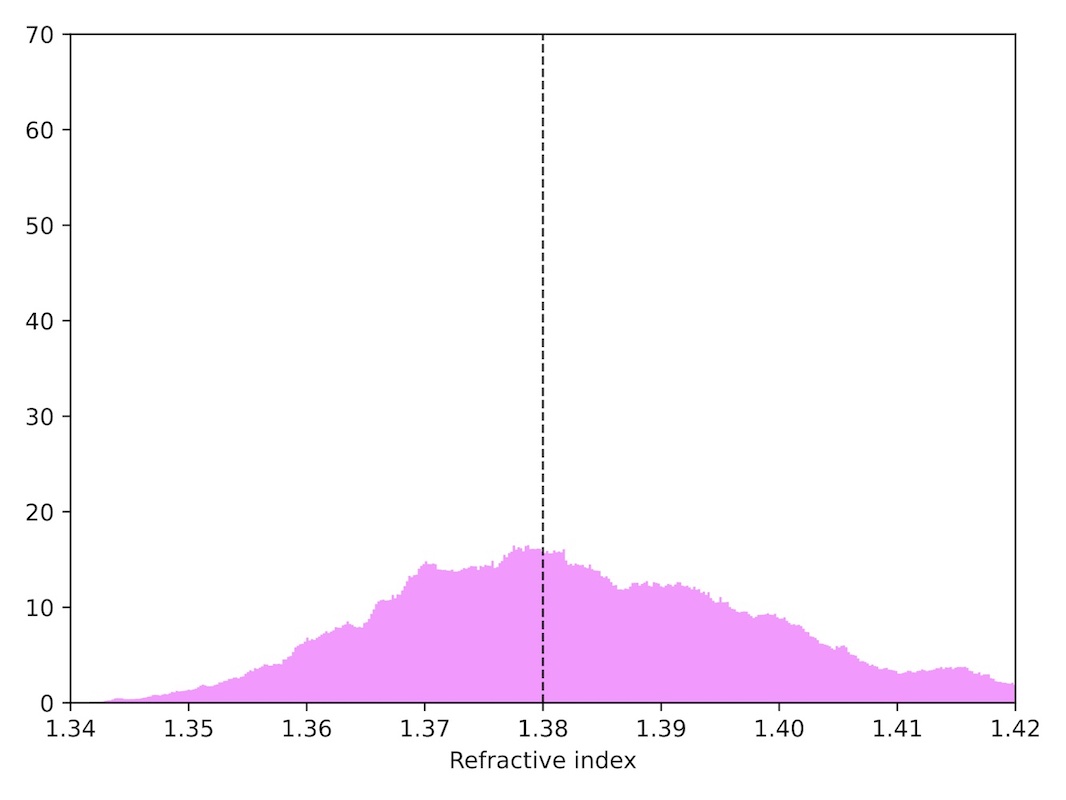}
\includegraphics[width= 0.28\linewidth]{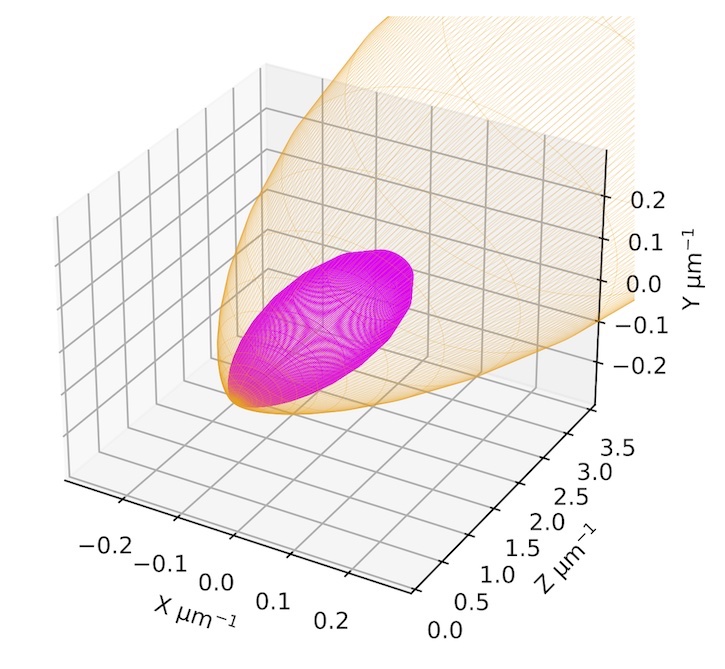}

\leftline{\bf \hspace{0.02\linewidth} (d) Medium IV}
\centering
\includegraphics[width= 0.30\linewidth]{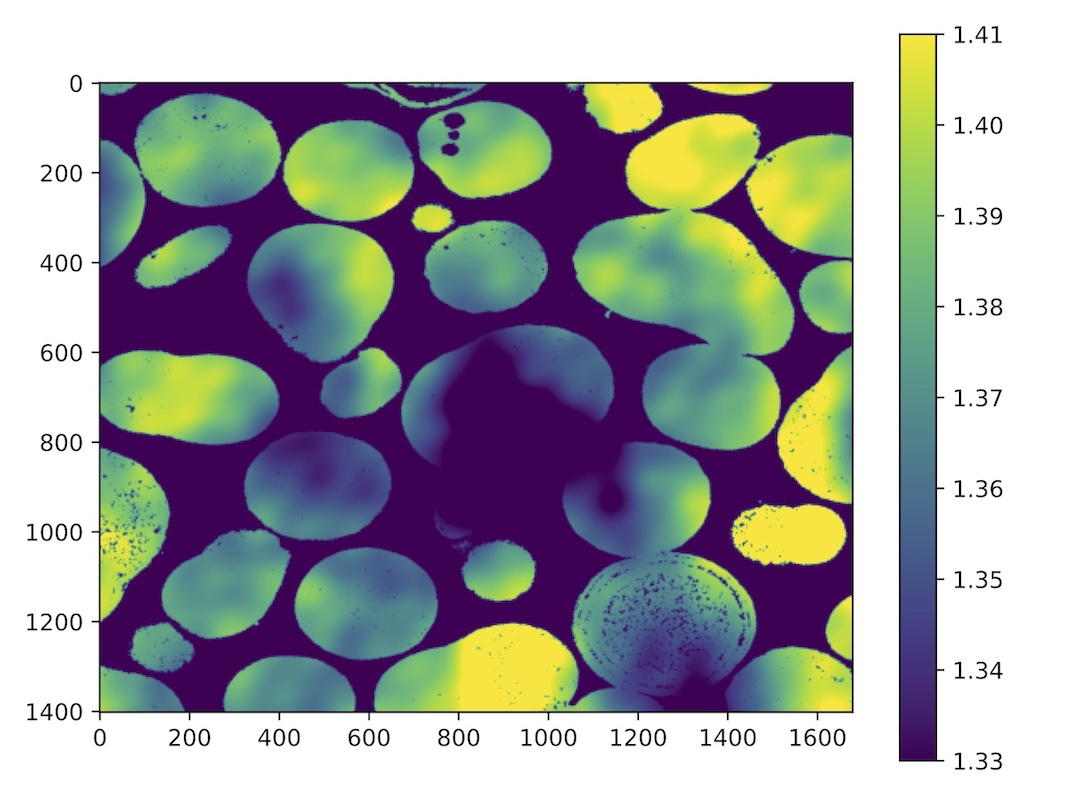}
\includegraphics[width= 0.30\linewidth]{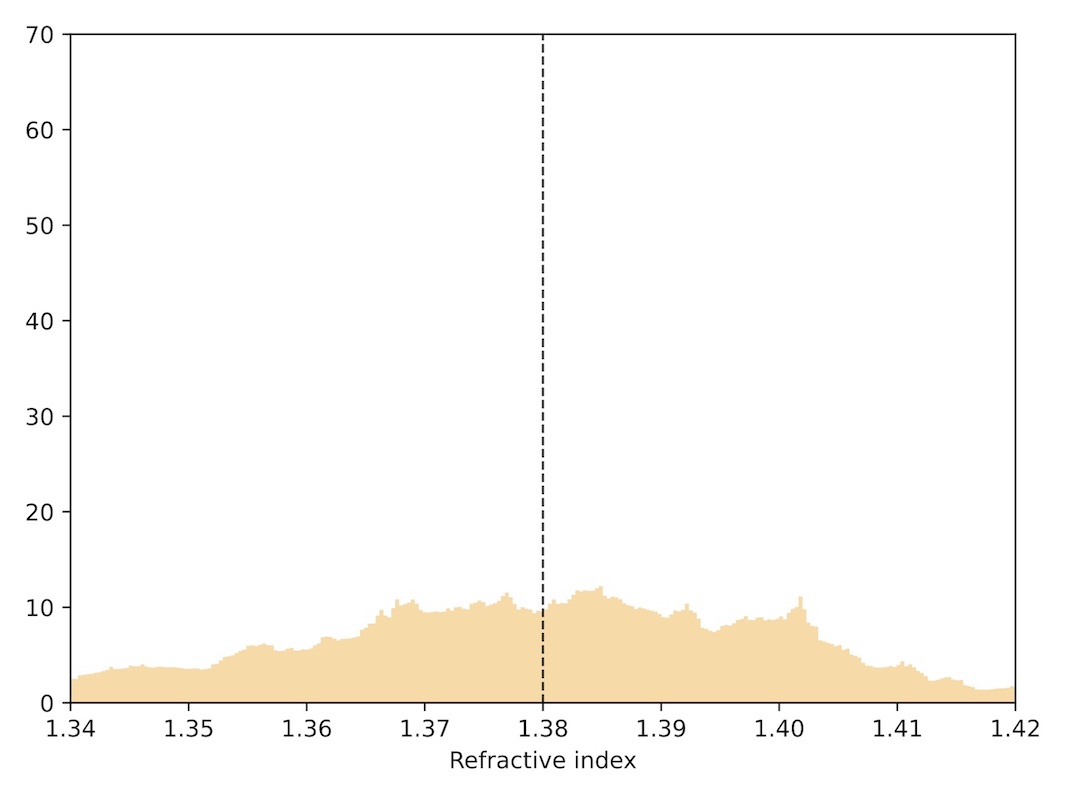}
\includegraphics[width= 0.28\linewidth]{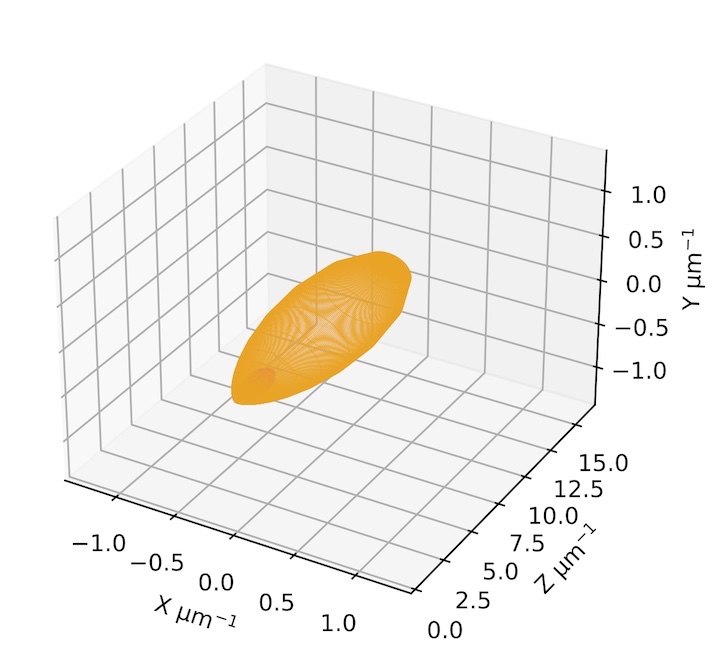}

\caption{Left and middle column figures show spatial distributions of refractive index and its histograms for the $100$-th image frame, respectively. In the right column figures, the mean differential scattering cross-section per unit volume are plotted in a spherical coordinate and then compared between the four different fractal media: medium I (blue), medium II (green), medium III (magenta) and medium IV (orange). The dimple of the cross section cannot be clearly seen in these column figures but definitely located at the origin.}
\label{figA2;results00}
\end{figure*}

In fractal cell modeling, an incident monochromatic wave function polarized in the direction $\hat{e}_i$ propagates along the longitudinal direction $\hat{k}_i$. The wave function can be scattered through continuous fractal media and then observed in the far field region at the direction $\hat{k}_o$ [see Figure \ref{figA2;scheme}]. The mean differential scattering cross-section per unit volume is defined as the averaged power per unit solid angle at direction $\hat{k}_o$ scattered by unit volume of the fractal medium and then can be written in the form of
\begin{eqnarray}
\big<\sigma(\hat{k}_o,\hat{k}_i)\big> & = & 2 \pi (k n_0)^4 \sin^2\chi \Phi_n(k_s) \\
& = & \frac{2 \sigma^2_n (k n_0)^4 l_c^3 \Gamma(D_f/2)}{\sqrt{\pi} \left| \Gamma((D_f-3)/2)\right|} \frac{(1 - \sin^2\theta\cos^2\phi)}{(1 + [2 k n_0 l_c \sin(\theta/2)]^2)^{D_f/2}} \nonumber
\end{eqnarray}
where $\sin^2\chi = 1 - \sin^2\theta\cos^2\phi$ and $k_s = 2 k n_0 \sin(\theta/2)$. The power spectral density of refractive index fluctuation $\Phi_n(k_s)$ can be derived from Fourier transformation of the WM covariance function $B_n(\rho)$. Figures \ref{figA2;xsec} show the examples of the mean differential cross-section plotted in a spherical coordinate. In Figure \ref{figA2;xsec}(a), the dimple is located right on the middle of an ellipsoidal-like shape of the scattering cross-section, representing, in particular, uniform scattering in all direction. Figures \ref{figA2;xsec}(b) and (c), however, exhibit forward scattering along the longitudinal direction (or z-axis). The dimple cannot be clearly seen in Figure \ref{figA2;xsec}(c) but definitely located at the origin. Furthermore, the mean differential cross-section for the four representative models of fractal media  (see Table \ref{tabA2;models} for model parameterization) are compared in the right column of Figures \ref{figA2;results00}.

%
%

\newpage

\newpage

\section{Results from the TIE simulations}
\subsection{Uniform intensity distribution}

\begin{figure*}[!h]
\leftline{\bf \hspace{0.02\linewidth} (a) Medium I}
\centering
\includegraphics[width= 0.30\linewidth]{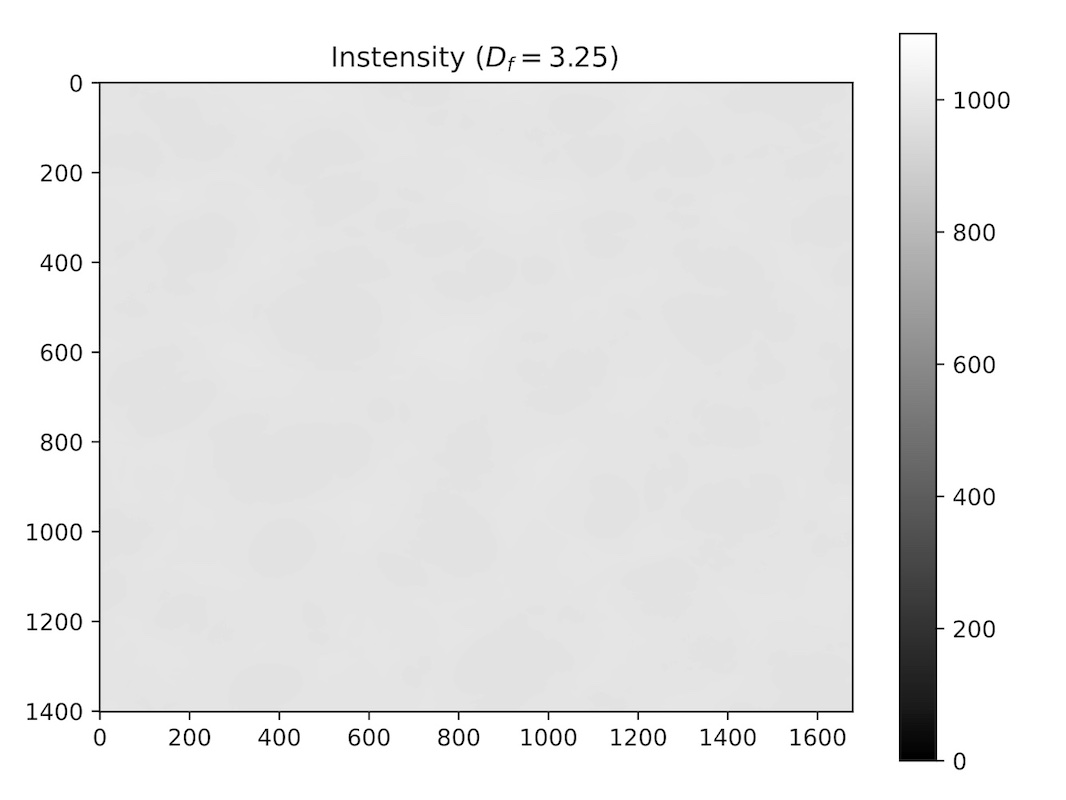}
\includegraphics[width= 0.30\linewidth]{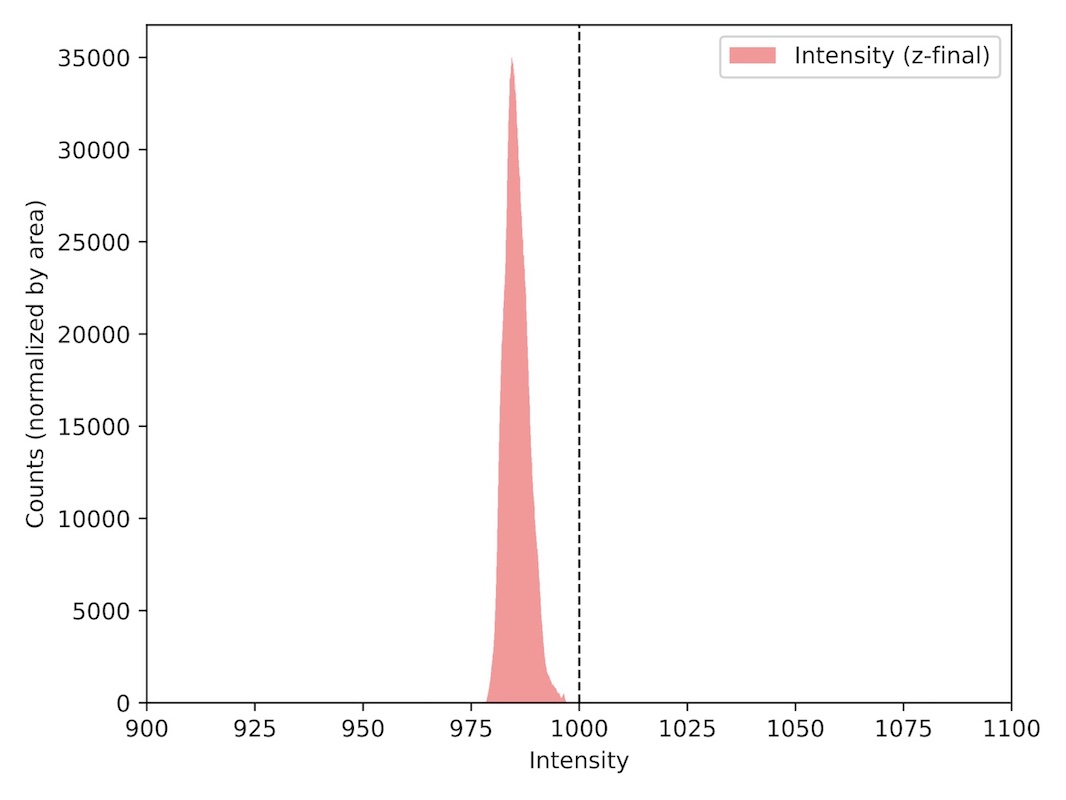}
\includegraphics[width= 0.30\linewidth]{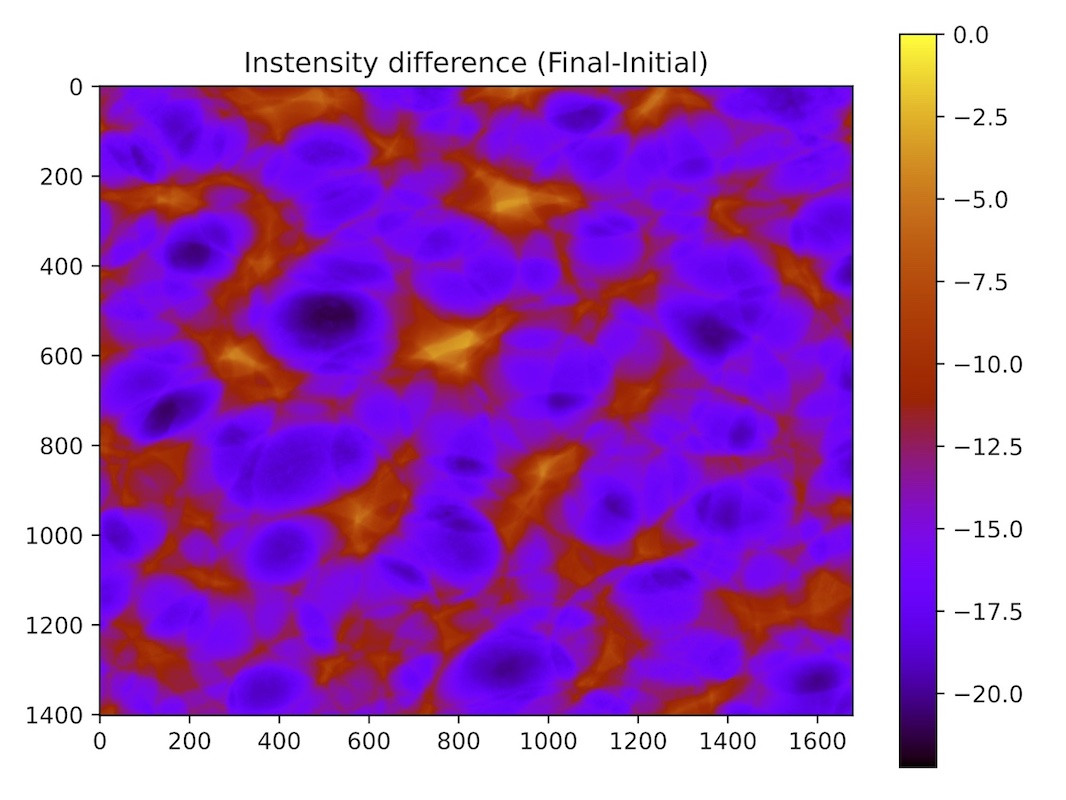}

\leftline{\bf \hspace{0.02\linewidth} (b) Medium II}
\centering
\includegraphics[width= 0.30\linewidth]{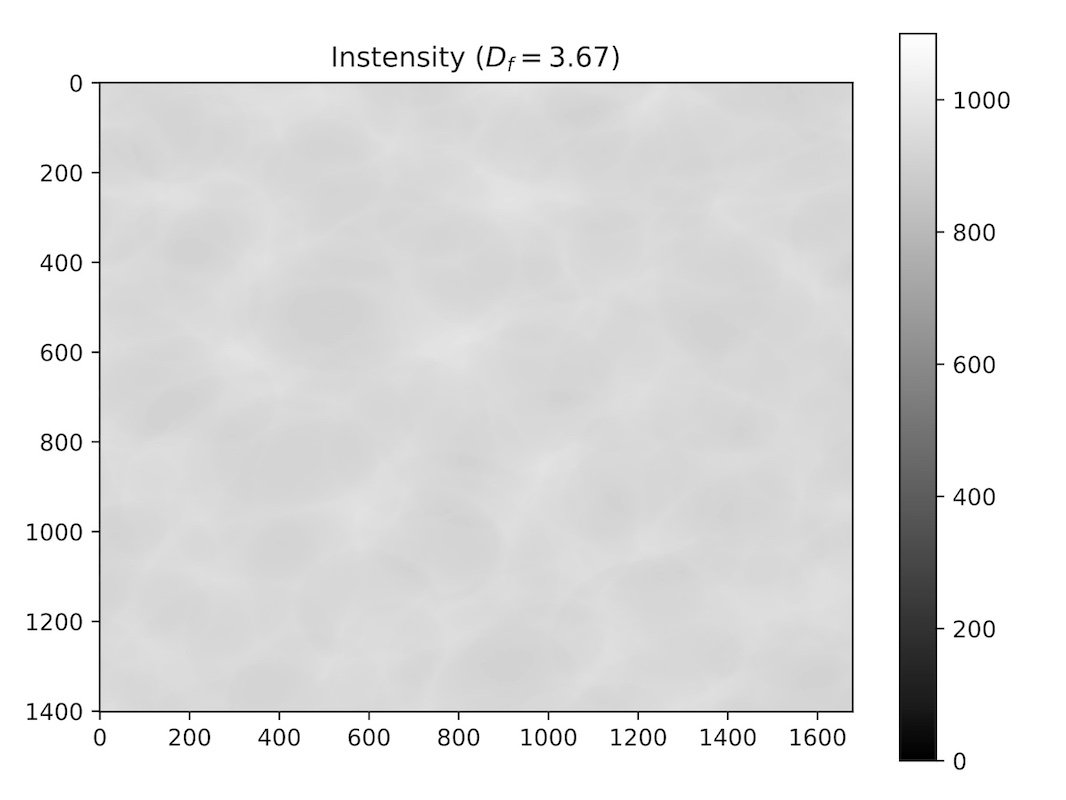}
\includegraphics[width= 0.30\linewidth]{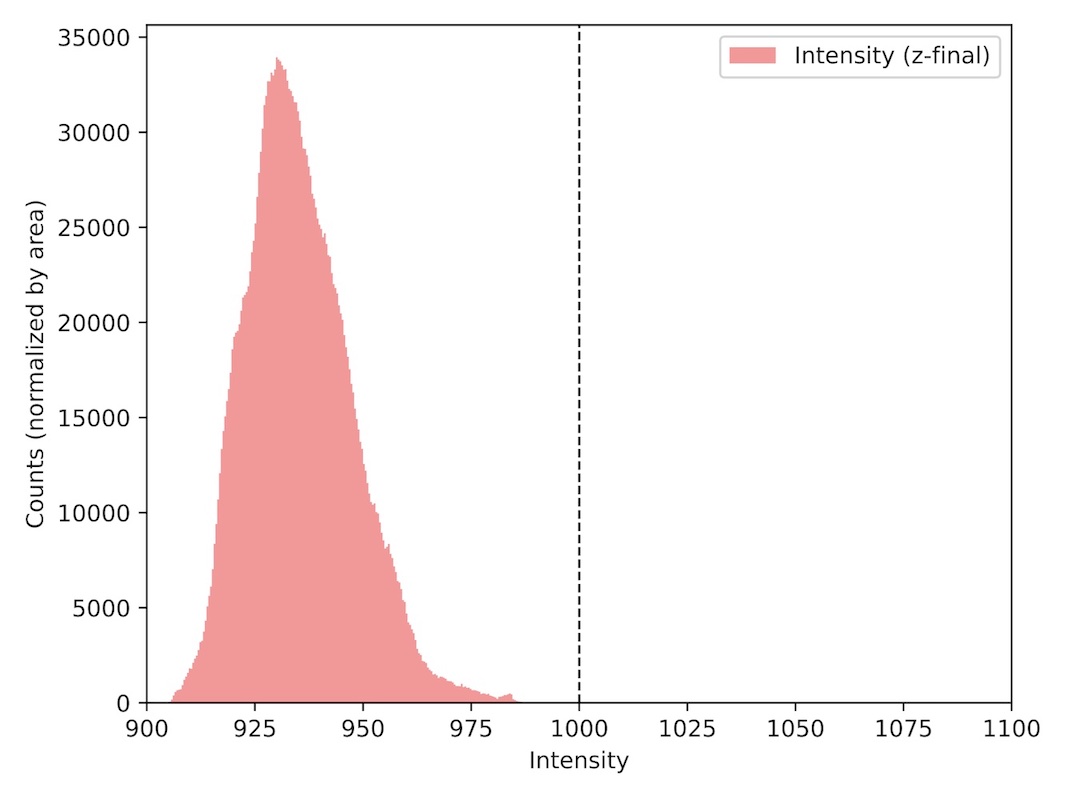}
\includegraphics[width= 0.30\linewidth]{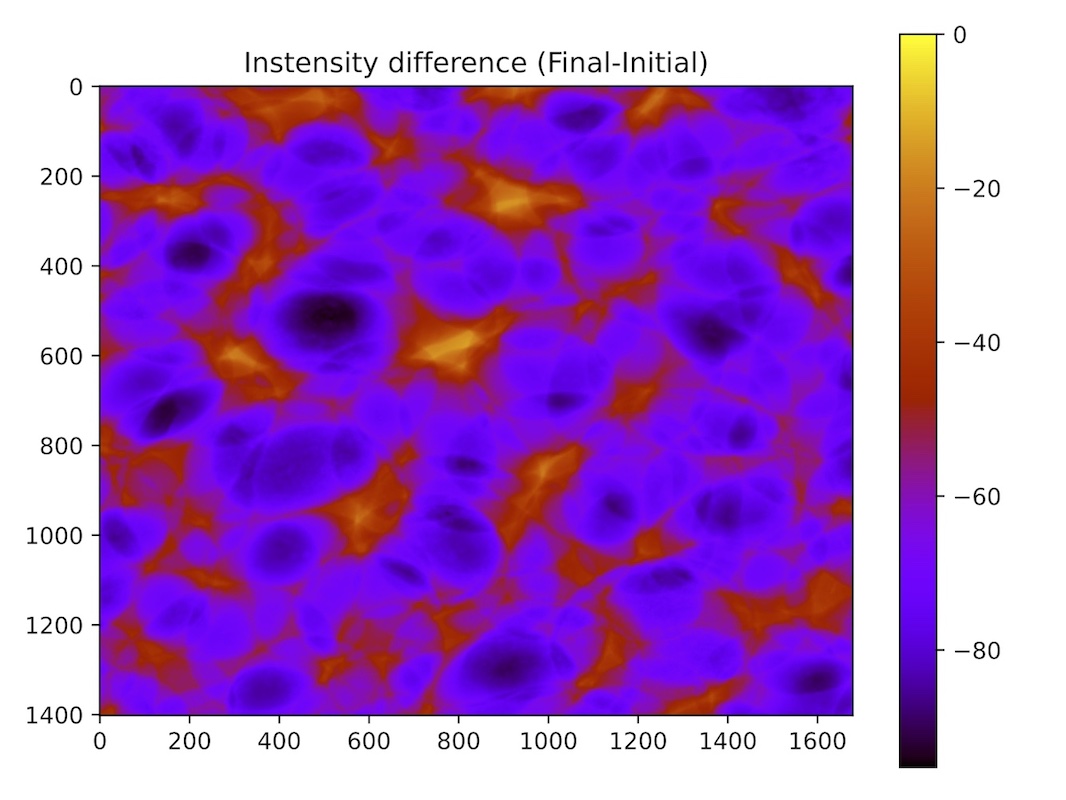}

\leftline{\bf \hspace{0.02\linewidth} (c) Medium III}
\centering
\includegraphics[width= 0.30\linewidth]{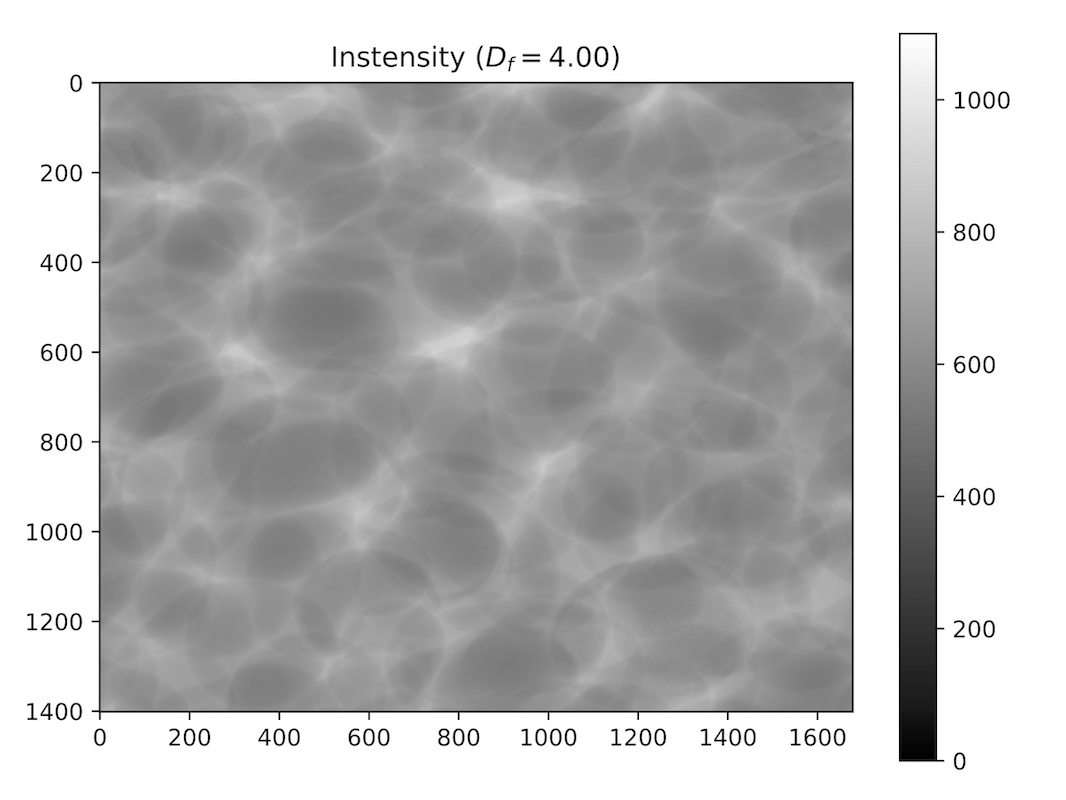}
\includegraphics[width= 0.30\linewidth]{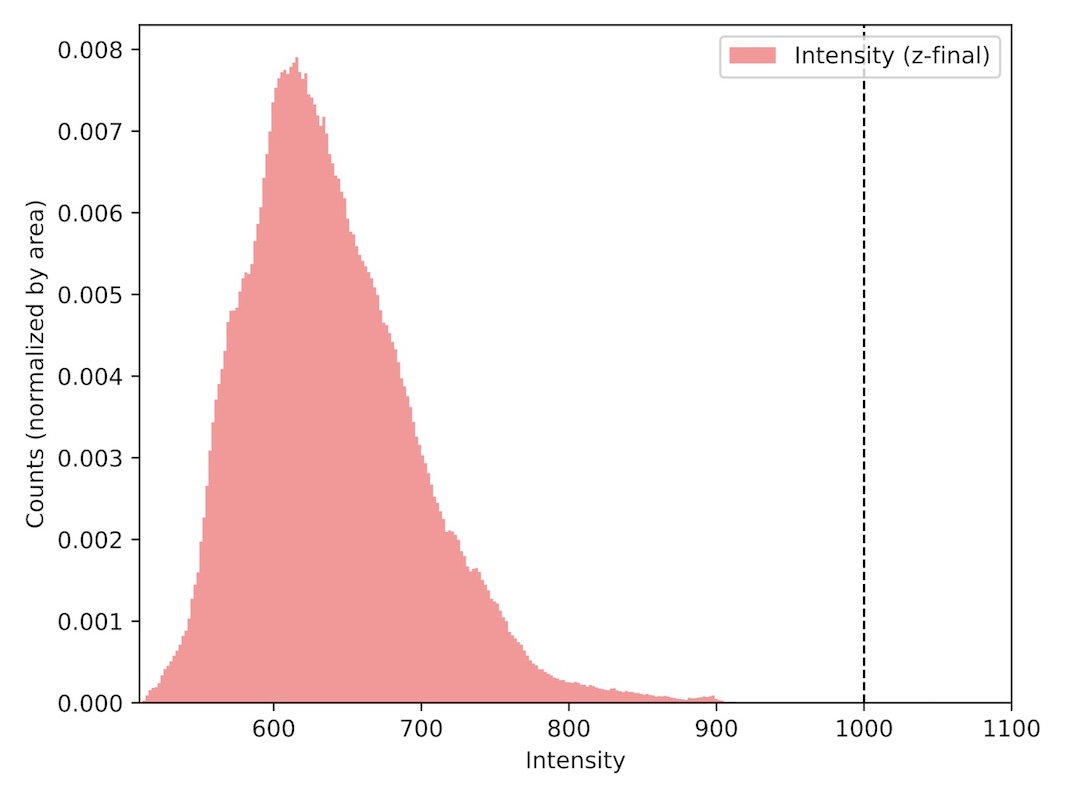}
\includegraphics[width= 0.30\linewidth]{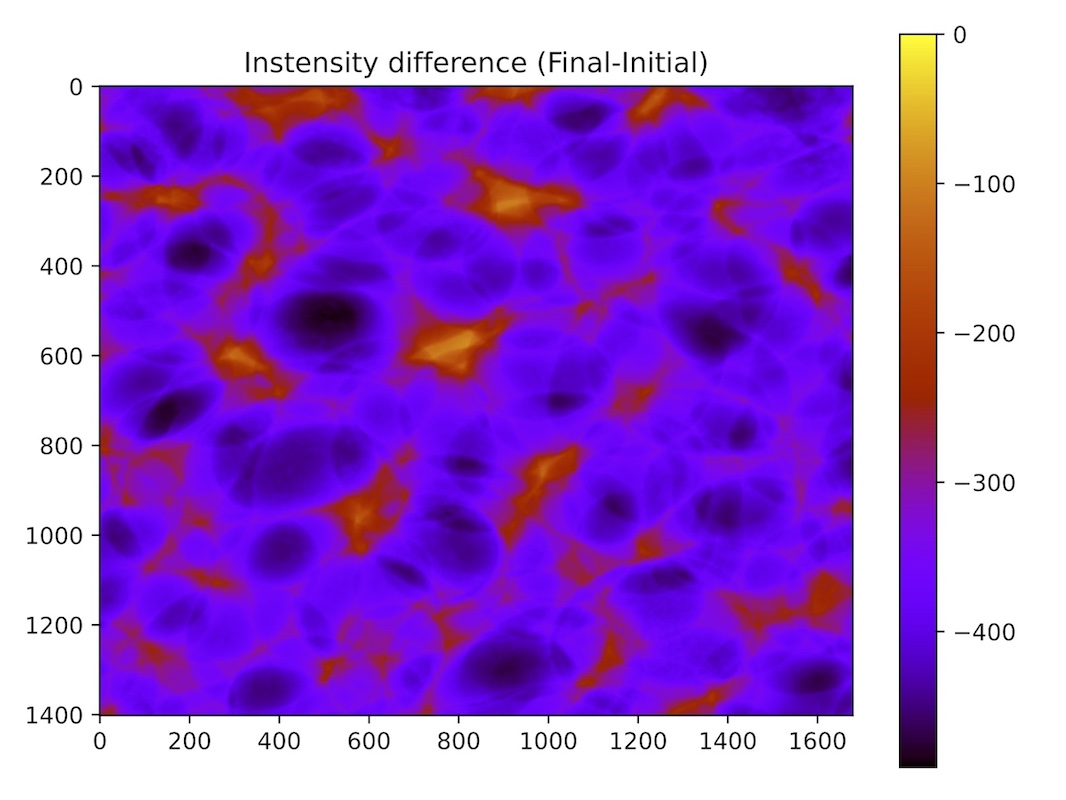}

\leftline{\bf \hspace{0.02\linewidth} (d) Medium IV}
\centering
\includegraphics[width= 0.30\linewidth]{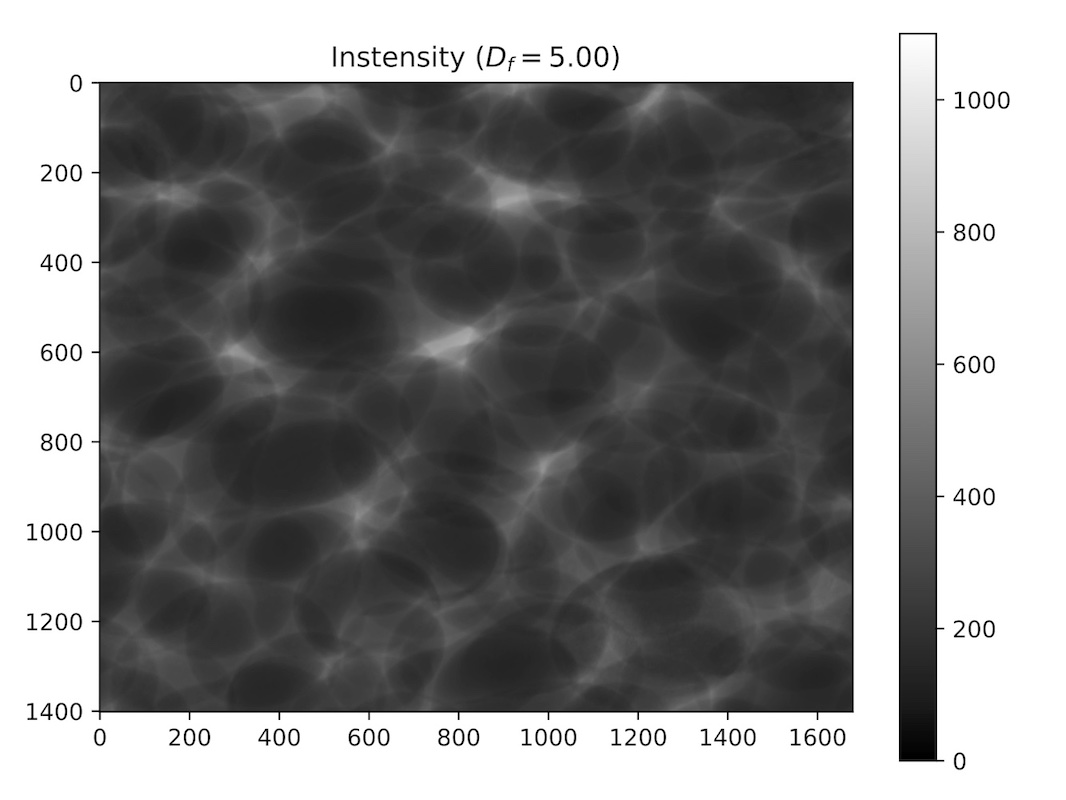}
\includegraphics[width= 0.30\linewidth]{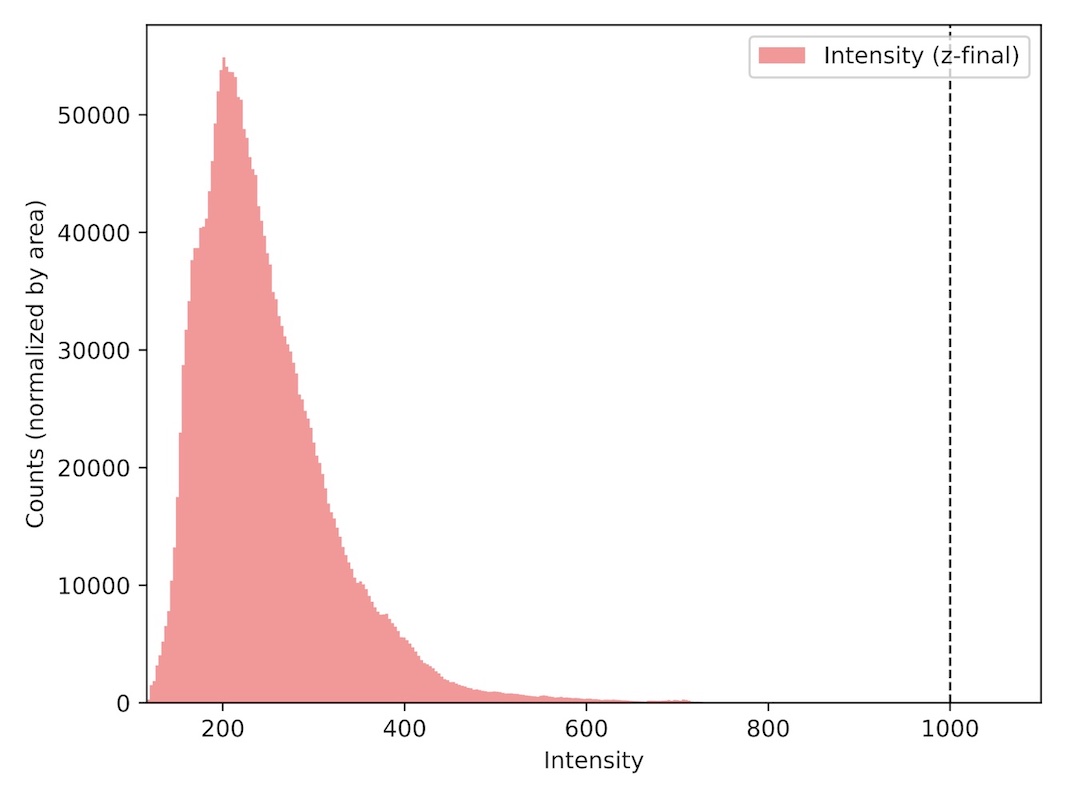}
\includegraphics[width= 0.30\linewidth]{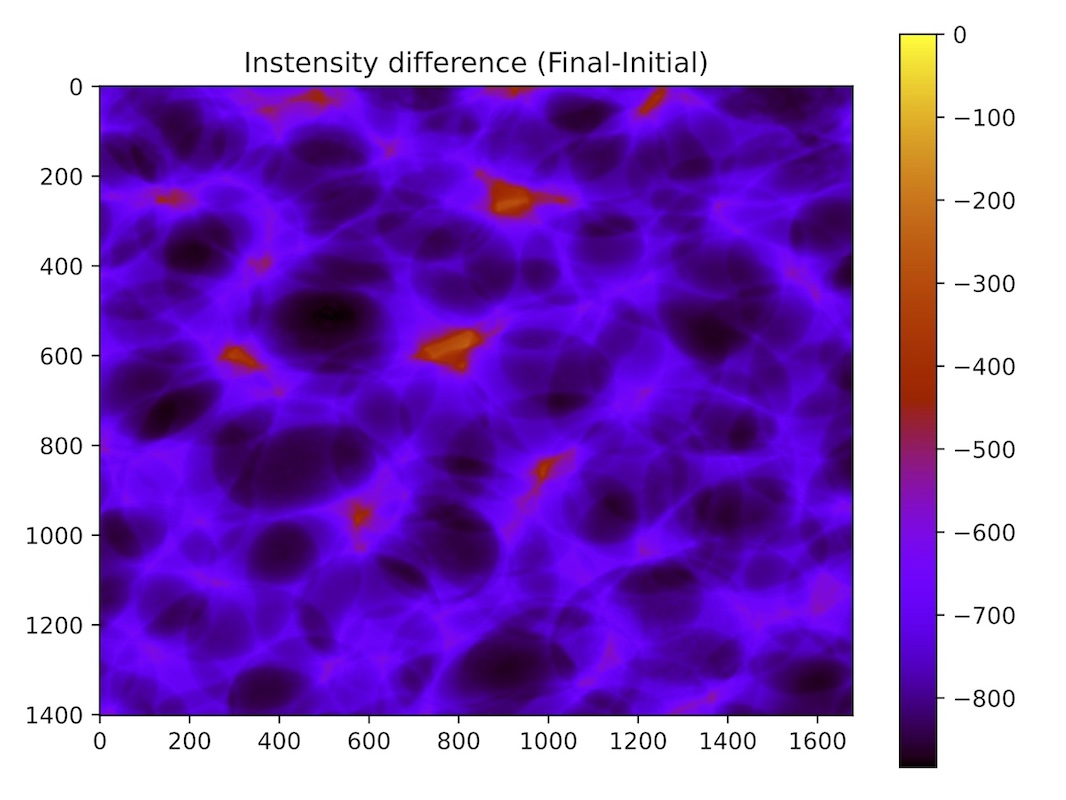}

\caption{Simulation results for propagation of uniform intensity distribution [$I(r_{\bot},z_{initial}) = 1,000\ {\rm counts/pixel}$] through the four different fractal media. Left, middle and right column figures represent final intensity distribution, histogram of final intensity distribution, and intensity difference $I(r_{\bot}, z_{initial}) - I(r_{\bot}, z_{final})$, respectively.}
\label{figB1;results01}
\end{figure*}

\newpage

Root-mean-square (RMS) and its relative values of intensity distributions between the $p$-th and $q$-th index of image frames are given by
\begin{equation}
\centering
RMS_{p,q} = \sqrt{\frac{\sum_{i,j}^{N} \left|I_{i,j}(z_{q}) - I_{i,j}(z_{p})\right|^2}{N}}\ \ {\rm and}\ \ Relative\ RMS_{p,q} = \sqrt{\frac{\sum_{i,j}^{N} \left|I_{i,j}(z_{q}) - I_{i,j}(z_{p})\right|^2}{\sum^{N}_{i,j} |I_{i,j}(z_{p})|^2}}
\end{equation}
where $N$ is the total number of image pixels.

\begin{table*}[!h]
    \centering
    \begin{tabular}{|l|c|c|}
    \hline
     & \hspace{2.0cm} RMS \hspace{2.0cm} &\hspace{2.0cm} Relative RMS ($\%$)\hspace{2.0cm} \\ \hline
    \hspace{1.0cm} Medium I\ \ \ (blue) \hspace{1.0cm} & $14.7039$ & $1.47$ \\ \hline
    \hspace{1.0cm} Medium II\ \ (green) & $65.7219$ & $6.57$ \\ \hline
    \hspace{1.0cm} Medium III (magenta) & $365.3491$ & $36.53$ \\ \hline
    \hspace{1.0cm} Medium IV (orange) & $756.3070$ & $75.63$ \\ \hline
    \end{tabular}
    \caption{RMS and its relative values between initial and final intensity distributions for wavelength $507\ {\rm nm}$.}
    \label{tabB1;rms}
\end{table*}

\begin{figure*}[!h]
\leftline{\bf \hspace{0.06\linewidth} (a) \hspace{0.42\linewidth} (b)}
\centering
\includegraphics[width=0.46\linewidth]{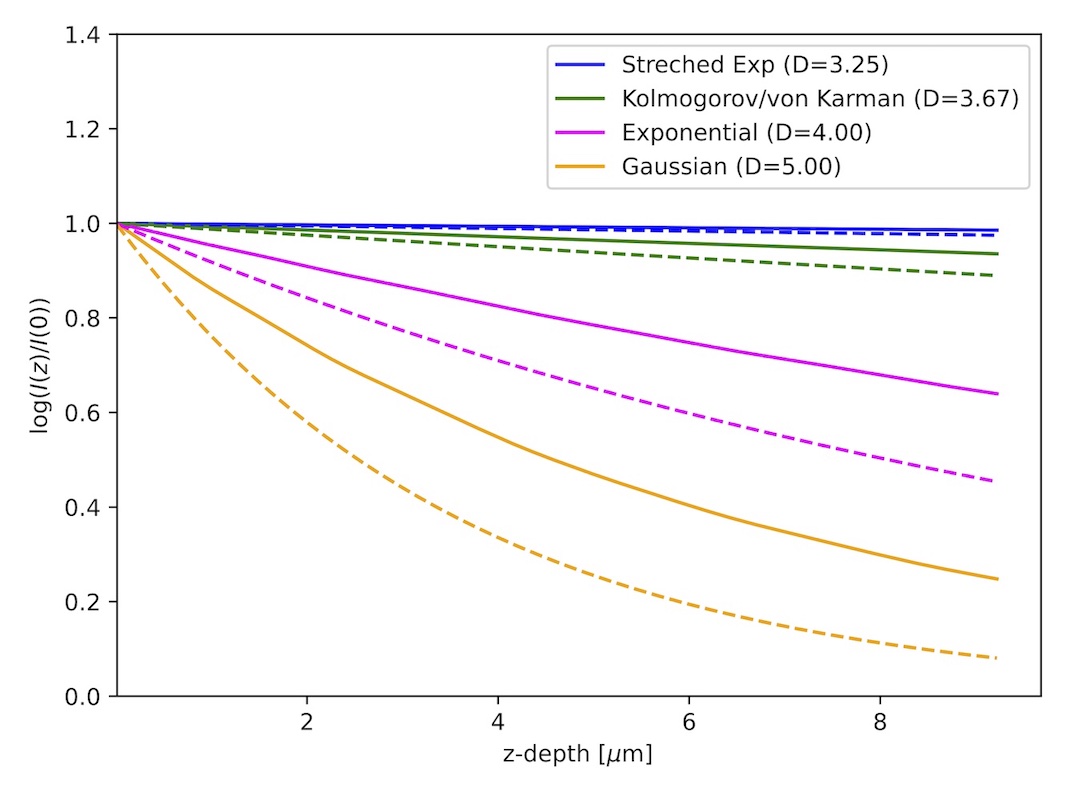}
\includegraphics[width=0.46\linewidth]{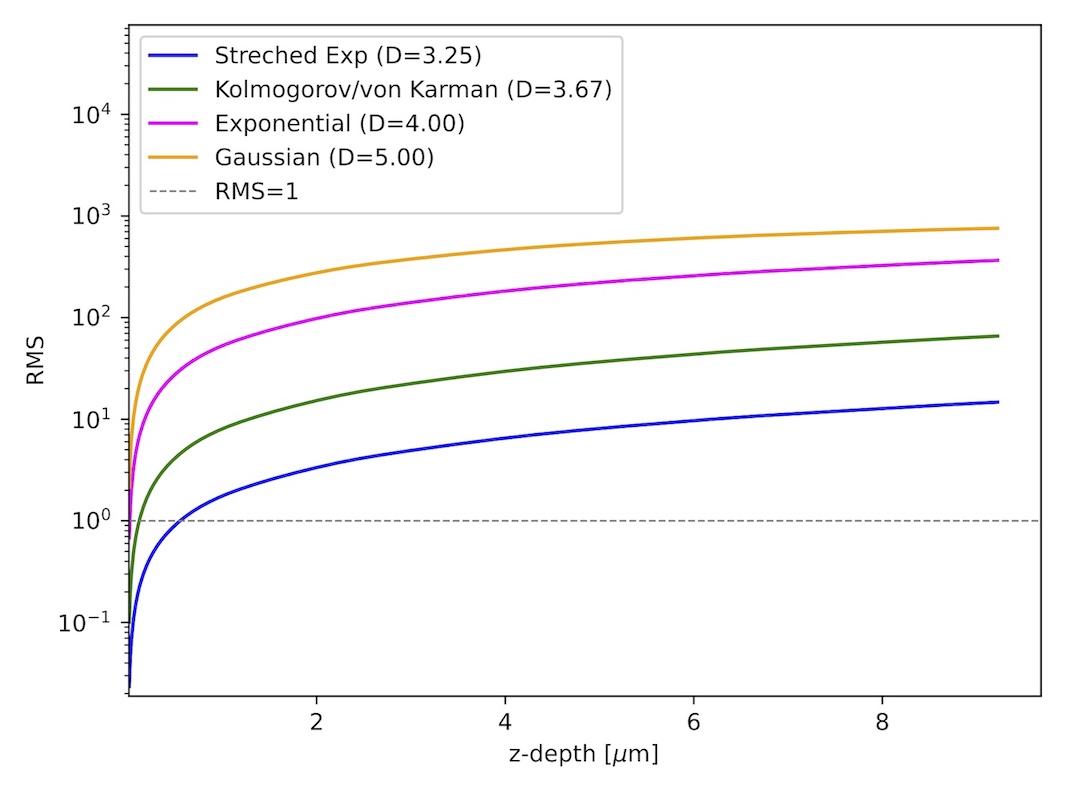}

\leftline{\bf \hspace{0.06\linewidth} (c) \hspace{0.42\linewidth} (d)}
\centering
\includegraphics[width=0.46\linewidth]{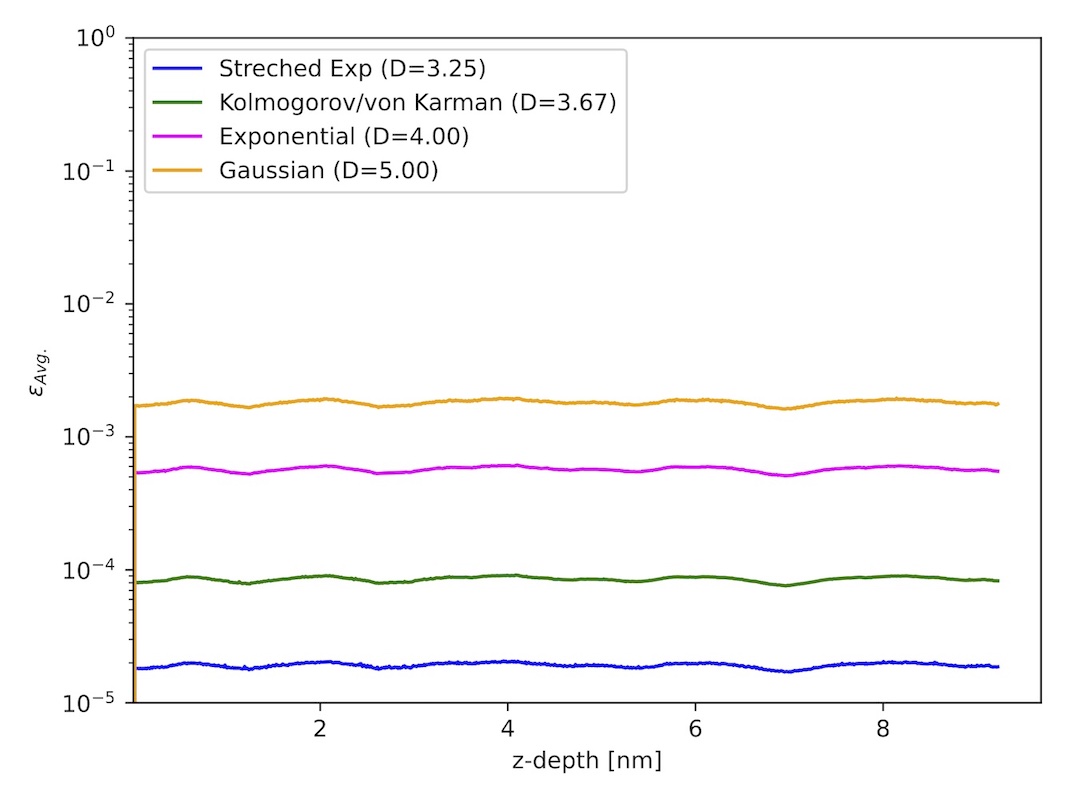}
\includegraphics[width=0.46\linewidth]{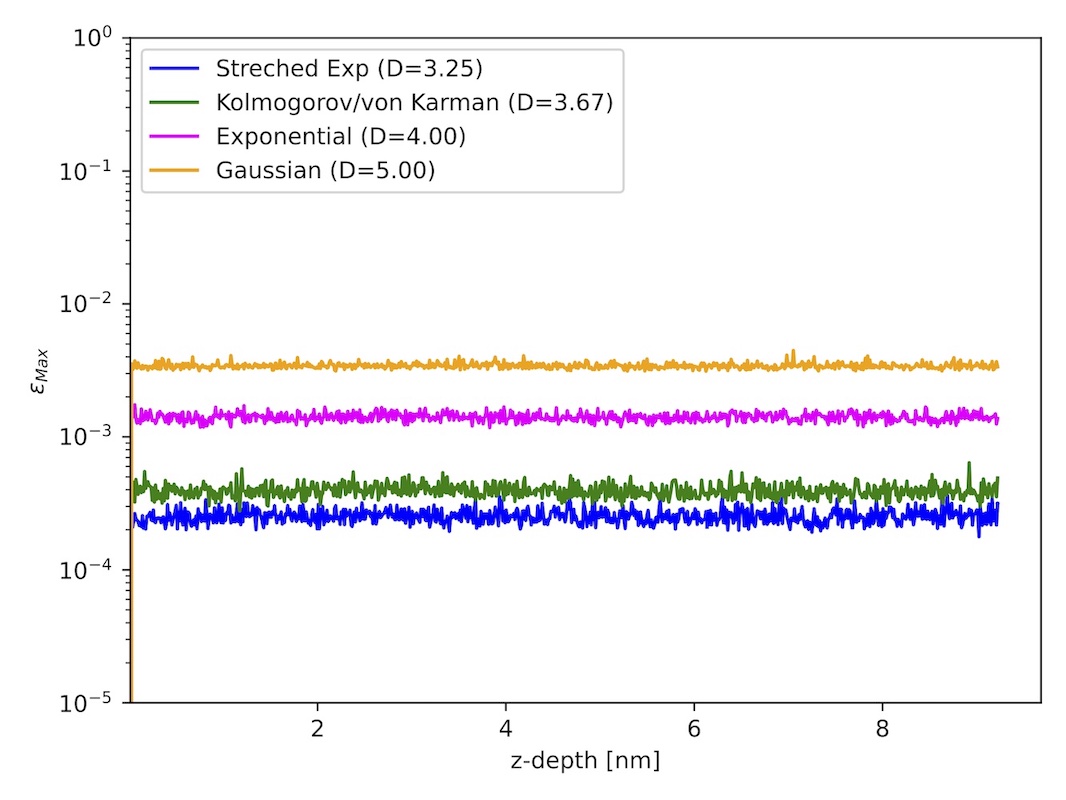}

\caption{Simulation results for propagation of the uniform intensity distributions [$I(r_{\bot},z_{initial}) = 1,000\ {\rm counts/pixel}$] through the refractive index distributions along axial direction. (a) Reduction of maximal intensity as a function of z axis. Dashed lines represent intensity reduction by Beer-Lambert law $I(z) = I(z=0)\ e^{-\mu(kn_0) z}$. (b) Intensity dispersion (i.e., RMS) as a function of z axis. (c) and (d) Averaged and maximal fractional errors as a function of z axis. Solid colored lines denote four different models of fractal media.}
\label{figB1;results02}
\end{figure*}

\vspace*{\fill}

\newpage

\vspace*{\fill}

%
%
%

\begin{figure*}[!h]
\leftline{\bf \hspace{0.02\linewidth} (a) $\lambda = 500\ {\rm nm}$ (visible light)}
\centering
\includegraphics[width=0.22\linewidth]{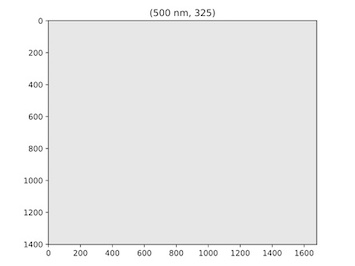}
\includegraphics[width=0.22\linewidth]{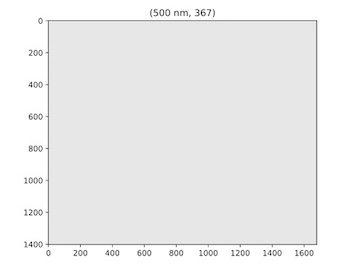}
\includegraphics[width=0.22\linewidth]{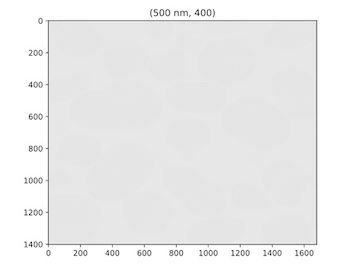}
\includegraphics[width=0.22\linewidth]{SI_figures/figure_B1-03-500nm400.jpg}

\leftline{\bf \hspace{0.02\linewidth} (b) $\lambda = 50.0\ {\rm nm}$}
\centering
\includegraphics[width=0.22\linewidth]{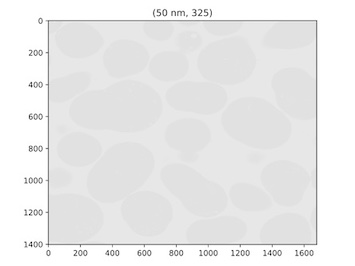}
\includegraphics[width=0.22\linewidth]{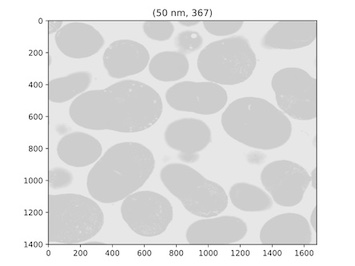}
\includegraphics[width=0.22\linewidth]{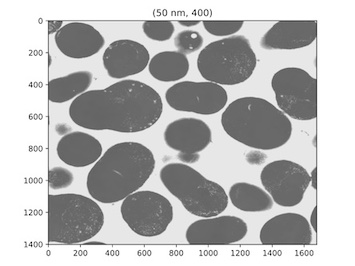}
\includegraphics[width=0.22\linewidth]{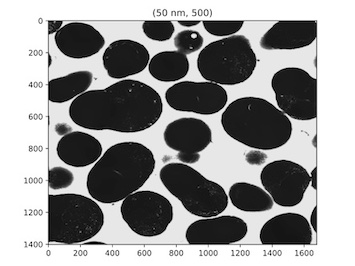}

\leftline{\bf \hspace{0.02\linewidth} (c) $\lambda = 5.00\ {\rm nm}$}
\centering
\includegraphics[width=0.22\linewidth]{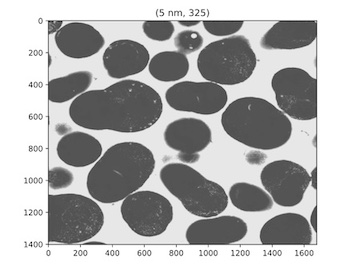}
\includegraphics[width=0.22\linewidth]{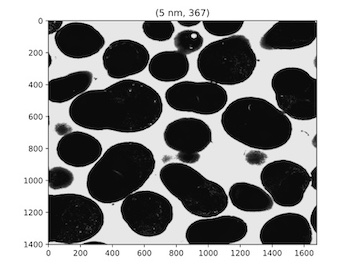}
\includegraphics[width=0.22\linewidth]{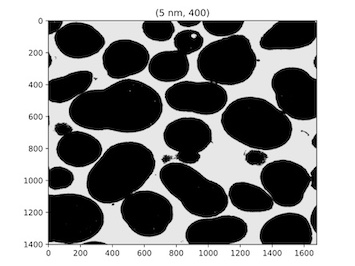}
\includegraphics[width=0.22\linewidth]{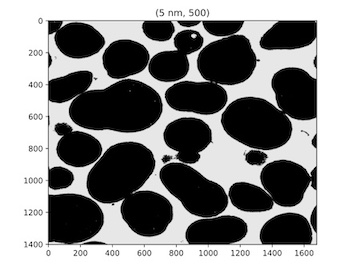}

\leftline{\bf \hspace{0.02\linewidth} (d) $\lambda = 500\ {\rm pm}$ (electron beam)}
\centering
\includegraphics[width=0.22\linewidth]{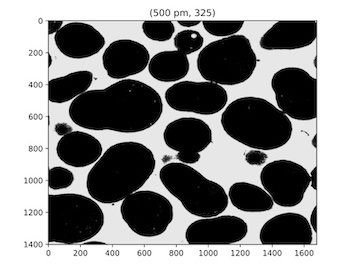}
\includegraphics[width=0.22\linewidth]{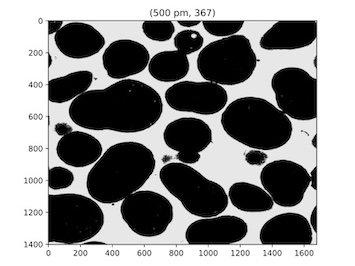}
\includegraphics[width=0.22\linewidth]{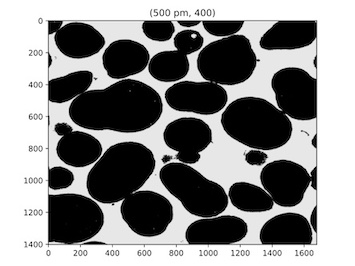}
\includegraphics[width=0.22\linewidth]{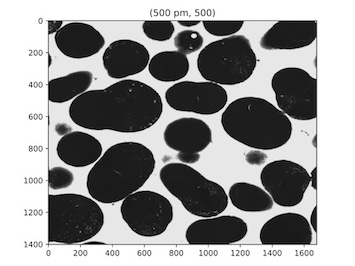}

\leftline{\bf \hspace{0.02\linewidth} (e) $\lambda = 50.0\ {\rm pm}$ (electron beam)}
\centering
\includegraphics[width=0.22\linewidth]{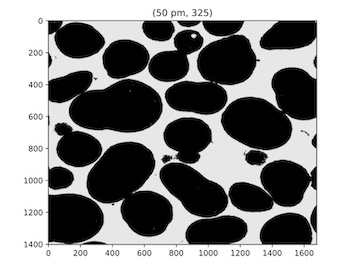}
\includegraphics[width=0.22\linewidth]{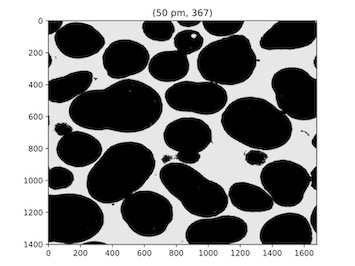}
\includegraphics[width=0.22\linewidth]{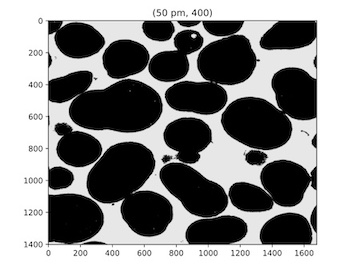}
\includegraphics[width=0.22\linewidth]{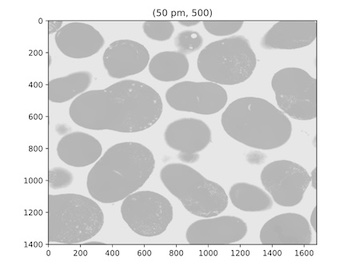}

\leftline{\bf \hspace{0.02\linewidth} (f) $\lambda = 5.00\ {\rm pm}$ (electron beam)}
\centering
\includegraphics[width=0.22\linewidth]{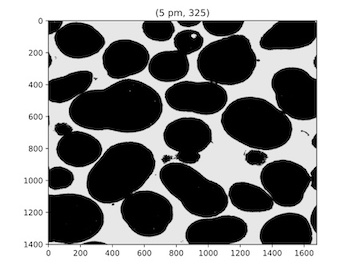}
\includegraphics[width=0.22\linewidth]{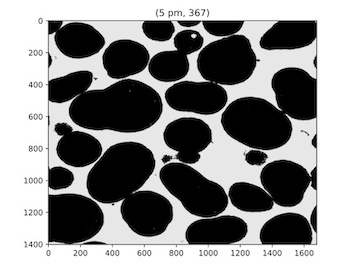}
\includegraphics[width=0.22\linewidth]{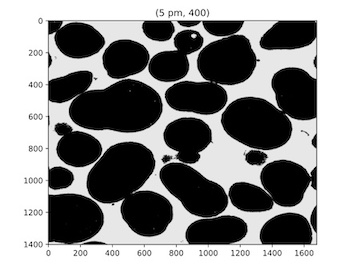}
\includegraphics[width=0.22\linewidth]{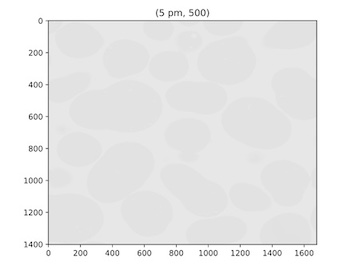}

\leftline{\bf \hspace{0.10\linewidth} Medium I \hspace{0.12\linewidth} Medium II \hspace{0.12\linewidth} Medium III \hspace{0.10\linewidth} Medium IV}

\caption{For sample thickness $L = 110\ {\rm nm}$, final images of the uniform intensity distributions are represented over a wide range of wavelength. Image intensities are  scaled from $0$ to $1,100$ counts.}
\label{figB2;results03}
\end{figure*}

\vspace*{\fill}

\newpage

\subsection{Standard image}

\begin{figure*}[!h]
\leftline{\bf \hspace{0.02\linewidth} (a) Medium I}
\centering
\includegraphics[width= 0.30\linewidth]{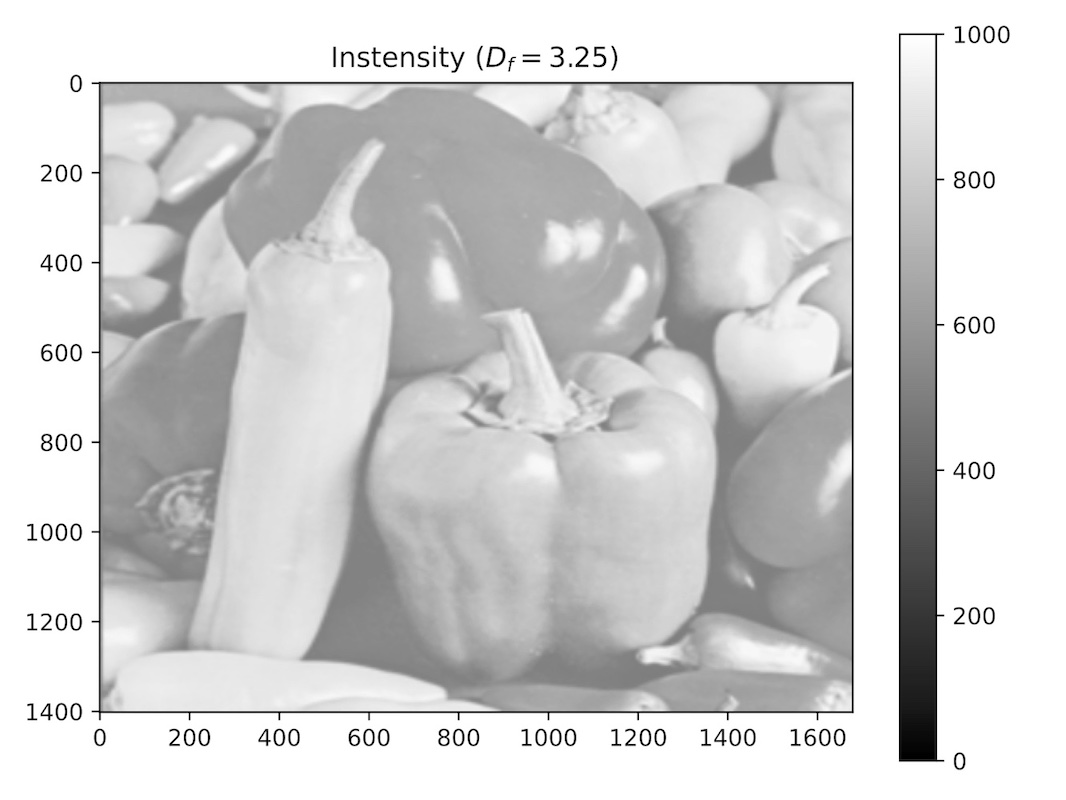}
\includegraphics[width= 0.30\linewidth]{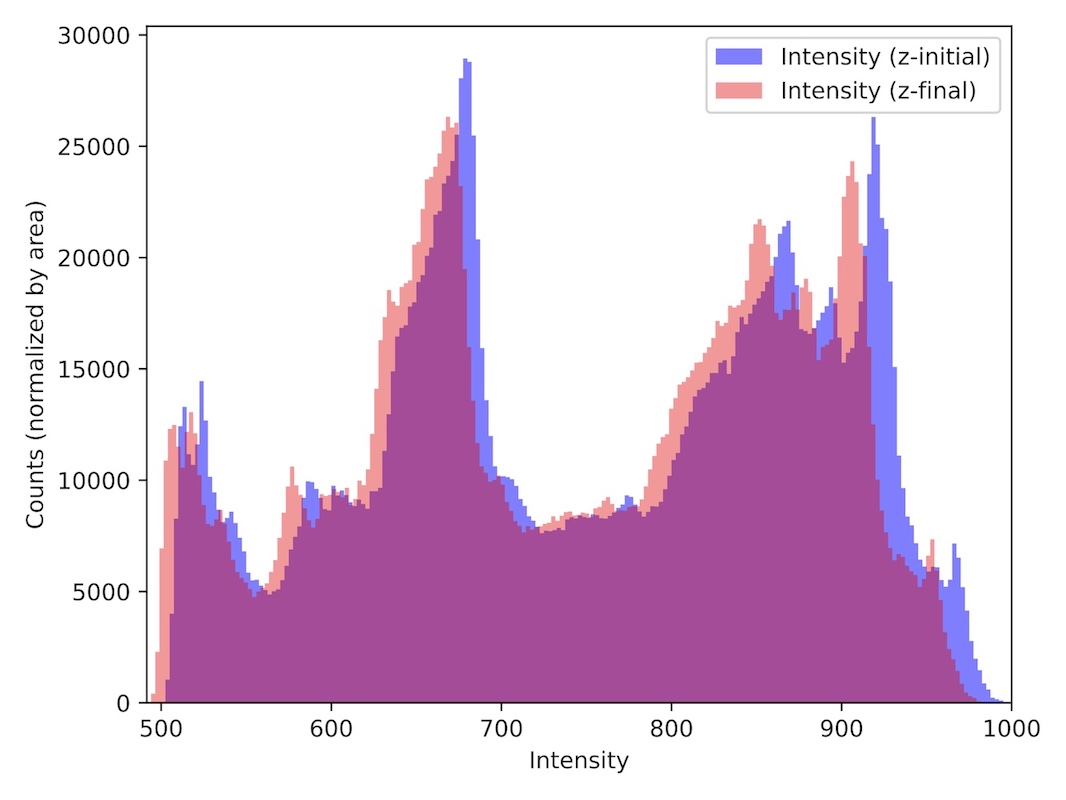}
\includegraphics[width= 0.30\linewidth]{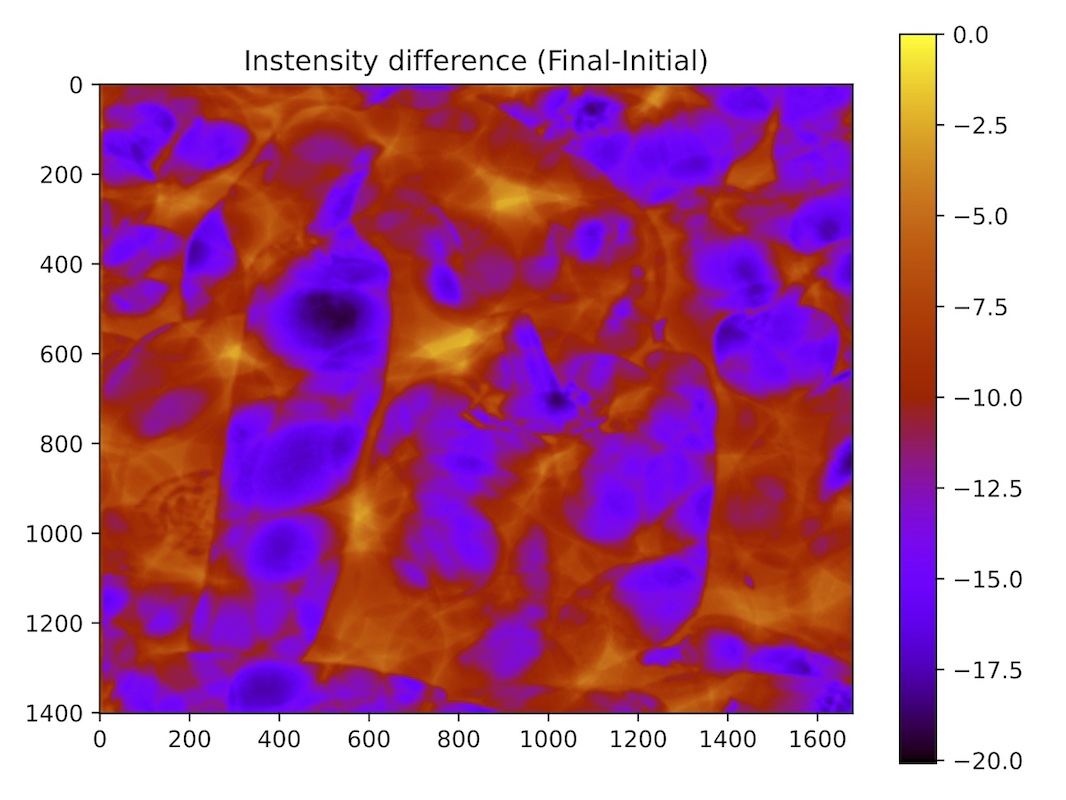}

\leftline{\bf \hspace{0.02\linewidth} (b) Medium II}
\centering
\includegraphics[width= 0.30\linewidth]{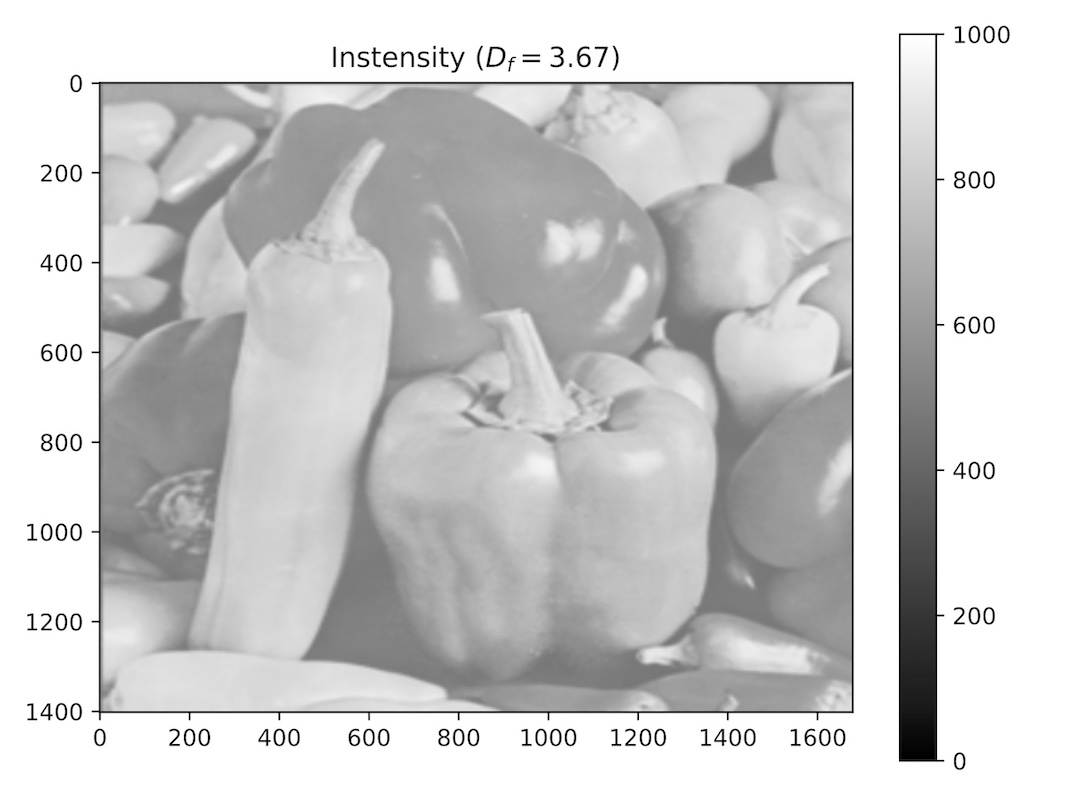}
\includegraphics[width= 0.30\linewidth]{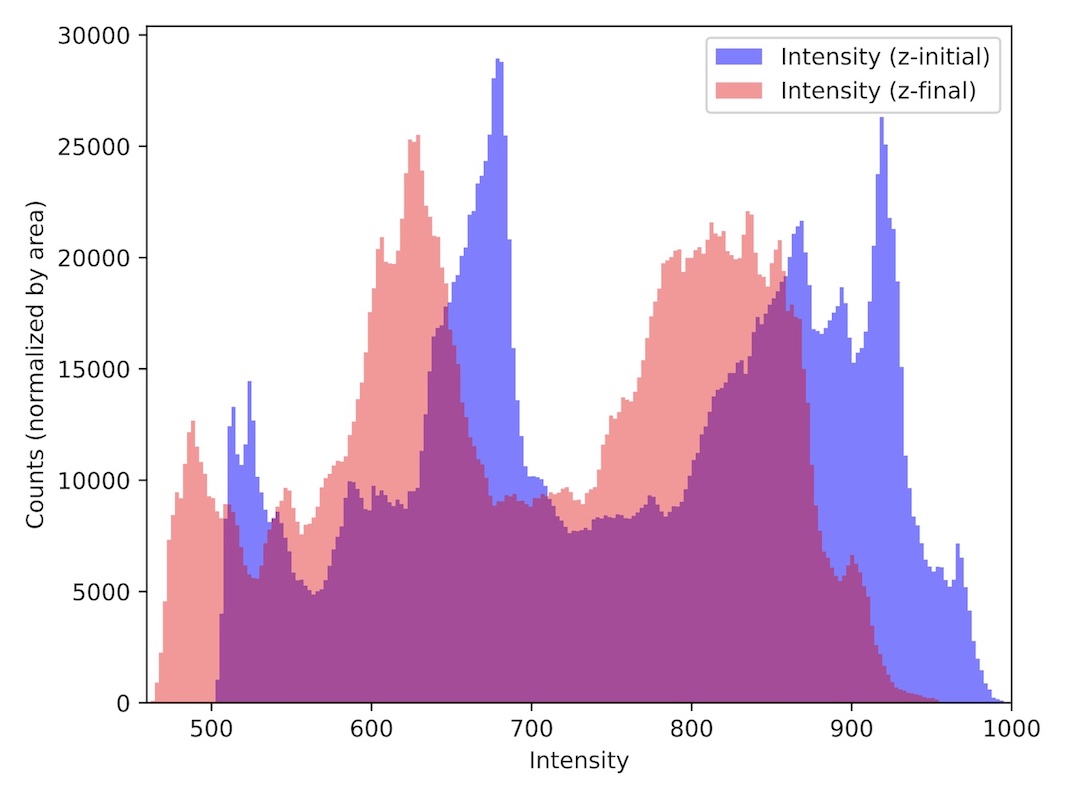}
\includegraphics[width= 0.30\linewidth]{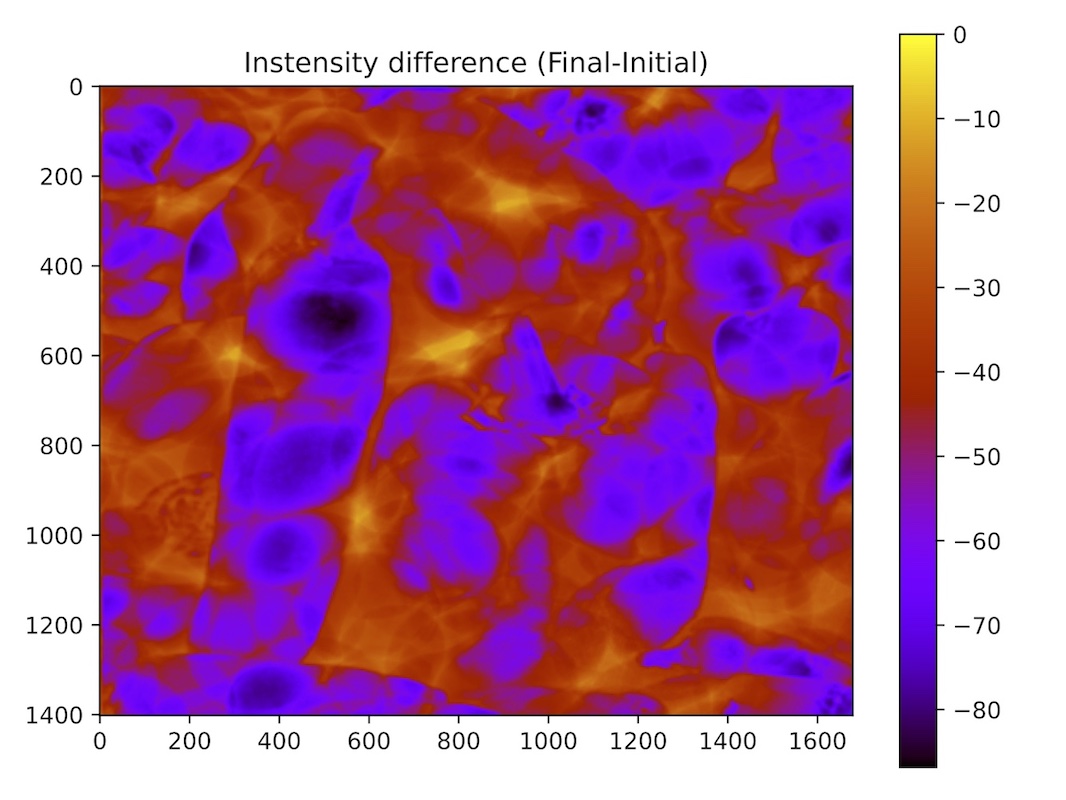}

\leftline{\bf \hspace{0.02\linewidth} (c) Medium III}
\centering
\includegraphics[width= 0.30\linewidth]{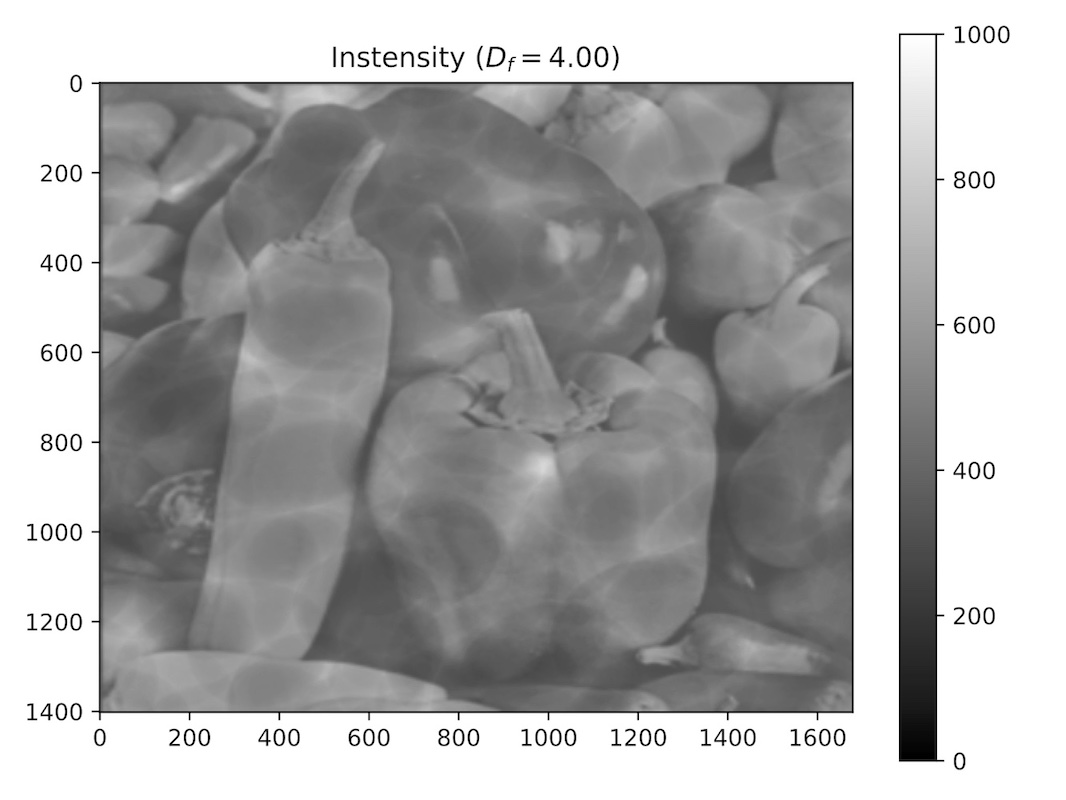}
\includegraphics[width= 0.30\linewidth]{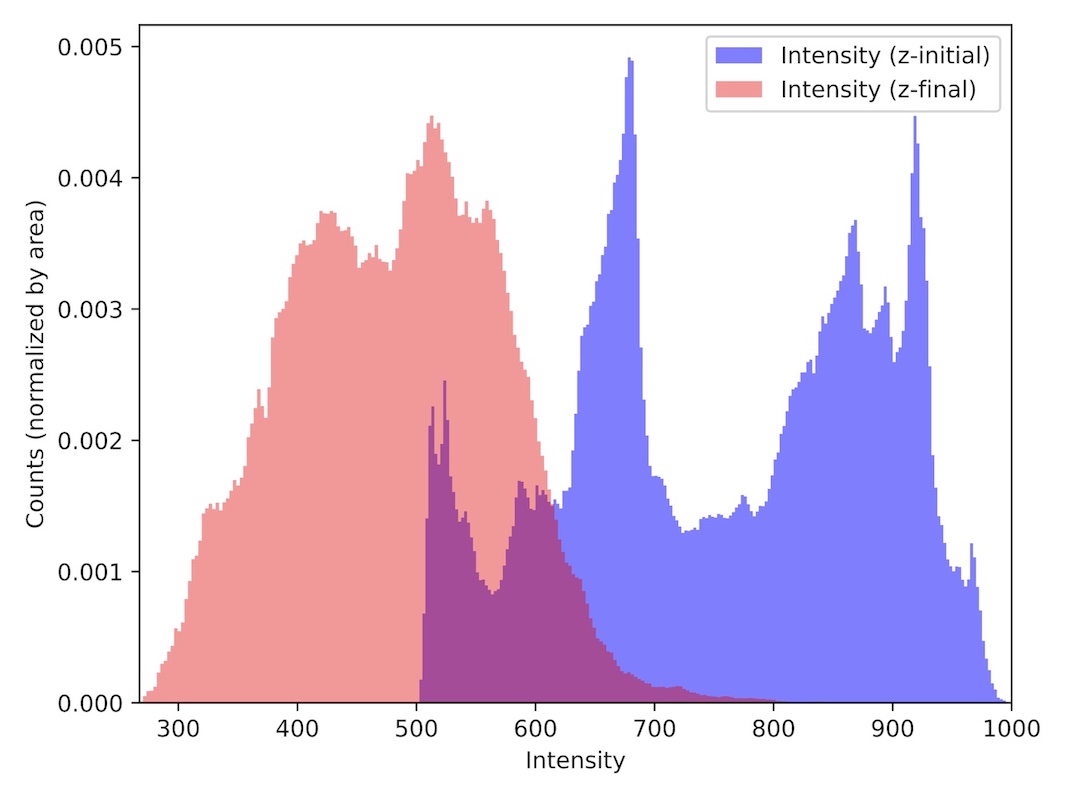}
\includegraphics[width= 0.30\linewidth]{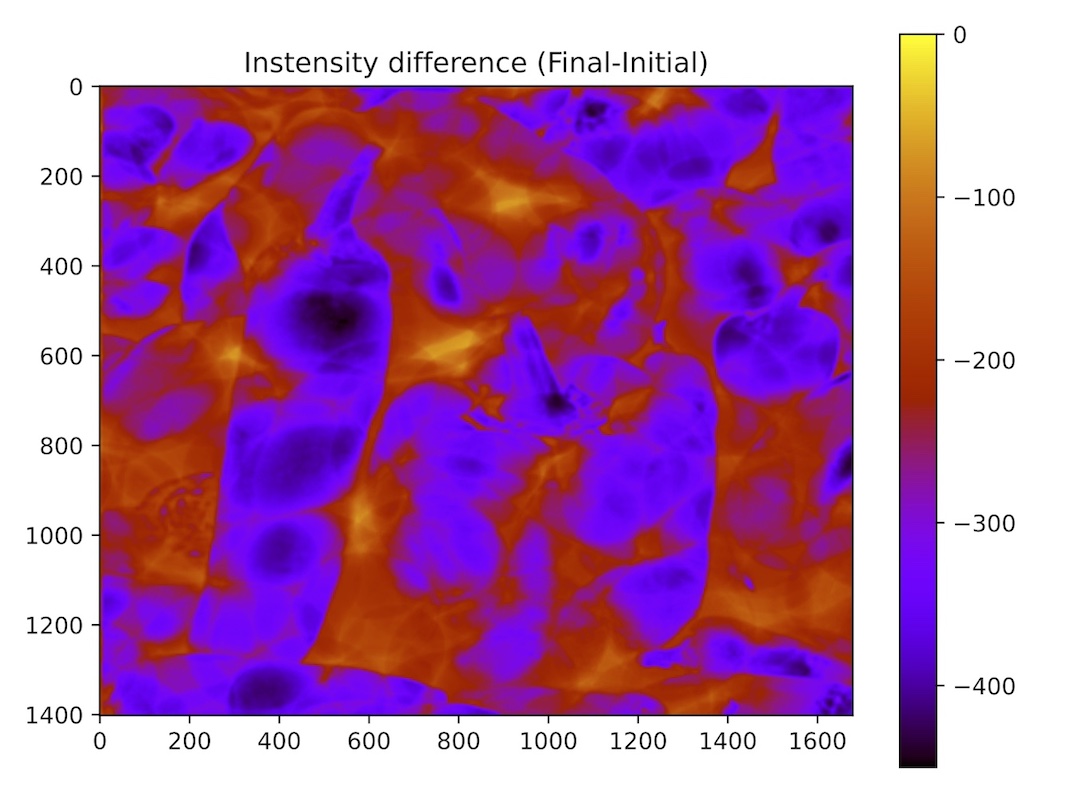}

\leftline{\bf \hspace{0.02\linewidth} (d) Medium IV}
\centering
\includegraphics[width= 0.30\linewidth]{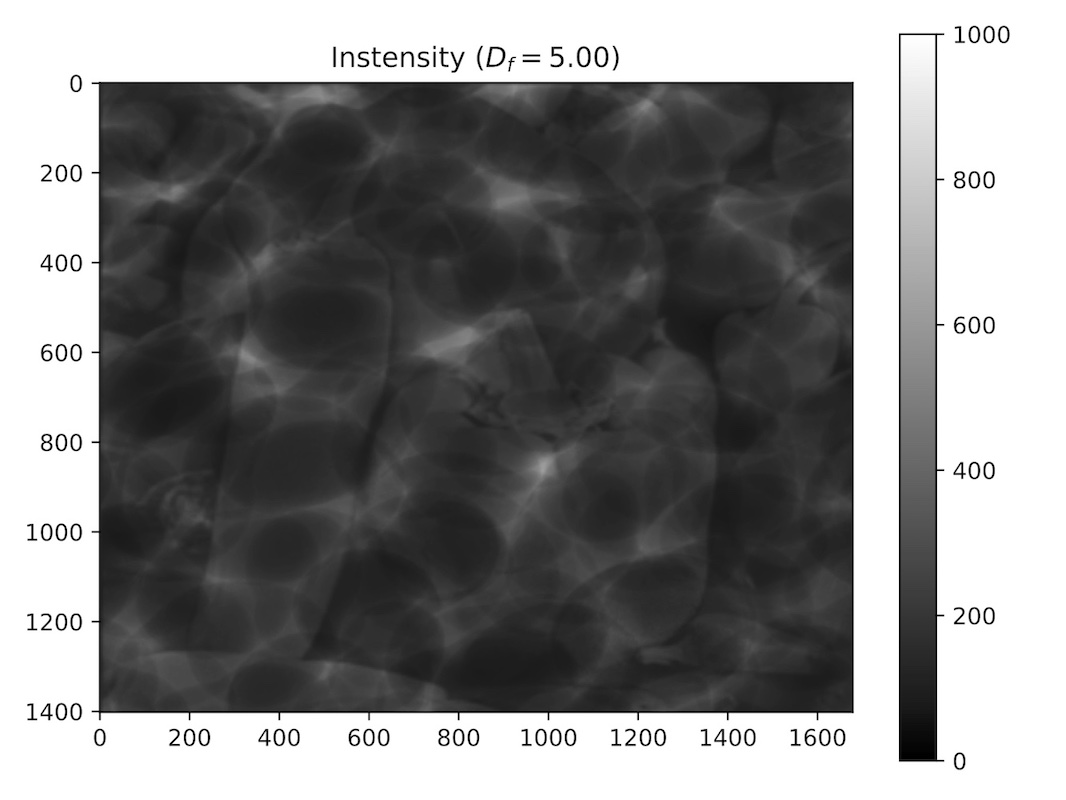}
\includegraphics[width= 0.30\linewidth]{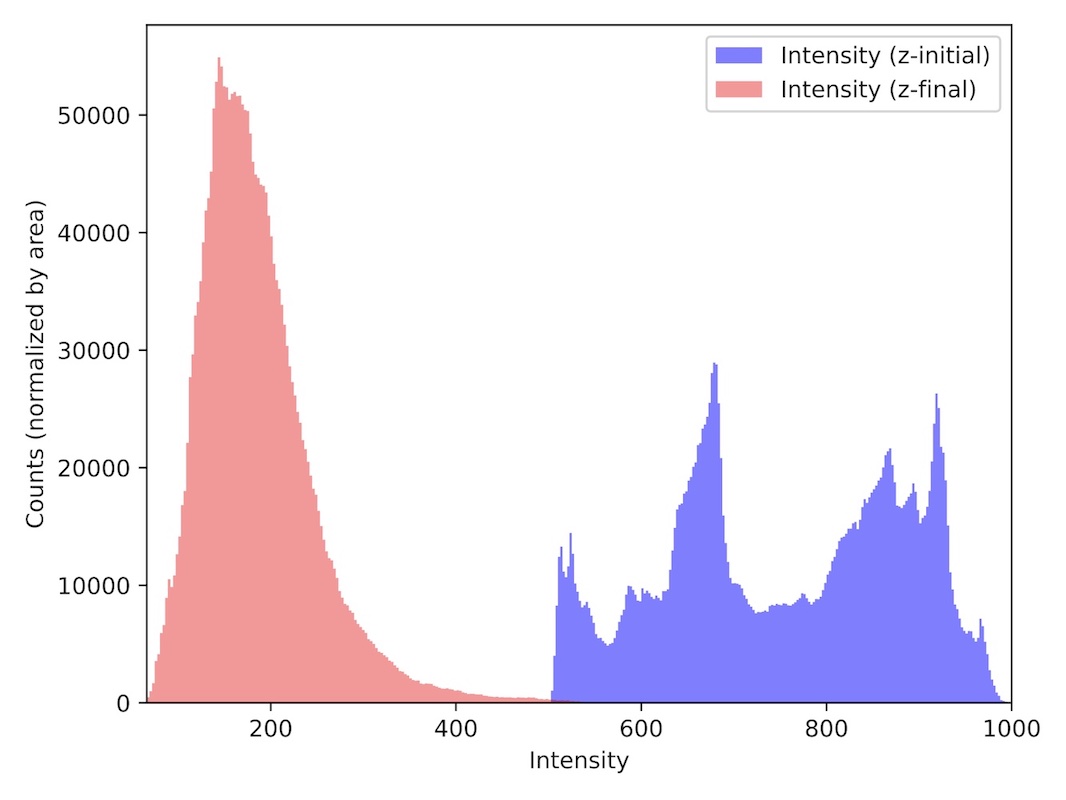}
\includegraphics[width= 0.30\linewidth]{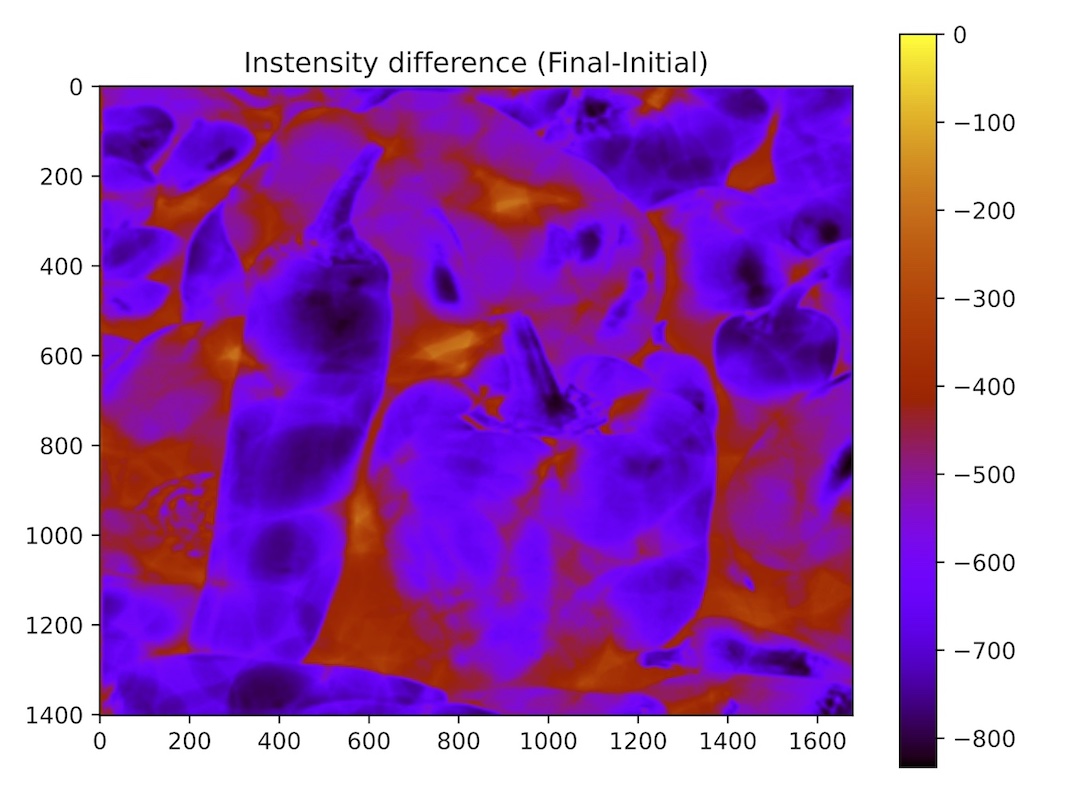}

\caption{Simulation results for intensity propagation of a standard image [$I(r_{\bot},z_{initial}) = 500 \sim 1,000\ {\rm counts/pixel}$] through the four different fractal media. Left, middle and right column figures represent final intensity distribution, histogram of final intensity distribution, and intensity difference $I(r_{\bot}, z_{initial}) - I(r_{\bot}, z_{final})$, respectively.}
\label{figB2;results01}
\end{figure*}

\vspace*{\fill}

\newpage

\vspace*{\fill}


\begin{table*}[!h]
    \centering
    \begin{tabular}{|l|c|c|c|}
    \hline
     & \hspace{2.0cm} RMS \hspace{2.0cm} &\hspace{2.0cm} Relative RMS ($\%$) \hspace{2.0cm} \\ \hline
    \hspace{1.0cm} Medium I\ \ \ (blue) \hspace{1.0cm} & $11.3470$ & $1.48$ \\ \hline
    \hspace{1.0cm} Medium II\ \ (green) & $50.7099$ & $6.62$ \\ \hline
    \hspace{1.0cm} Medium III (magenta) & $281.5880$ & $36.76$ \\ \hline
    \hspace{1.0cm} Medium IV (orange) & $581.6046$ & $75.92$ \\ \hline
    \end{tabular}
    \caption{RMS and its relative values between initial and final intensity distributions for wavelength $507\ {\rm nm}$.}
    \label{tabB2;rms}
\end{table*}

\begin{figure*}[!h]
\leftline{\bf \hspace{0.06\linewidth} (a) \hspace{0.42\linewidth} (b)}
\centering
\includegraphics[width=0.46\linewidth]{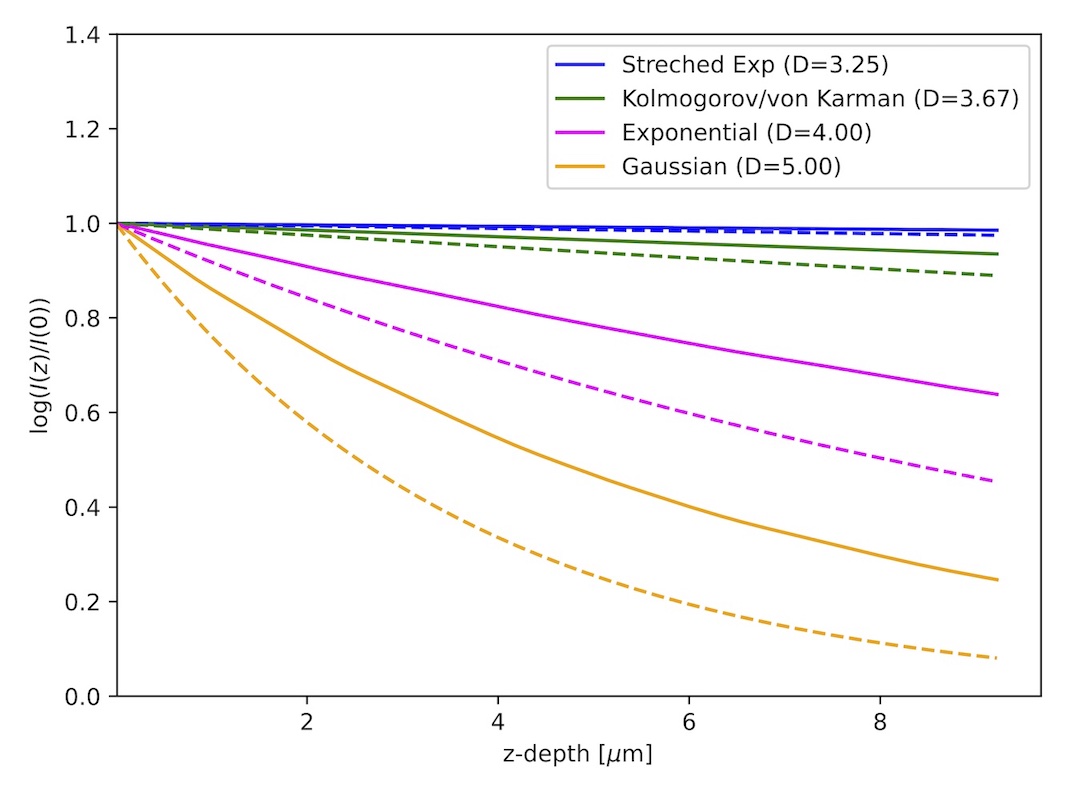}
\includegraphics[width=0.46\linewidth]{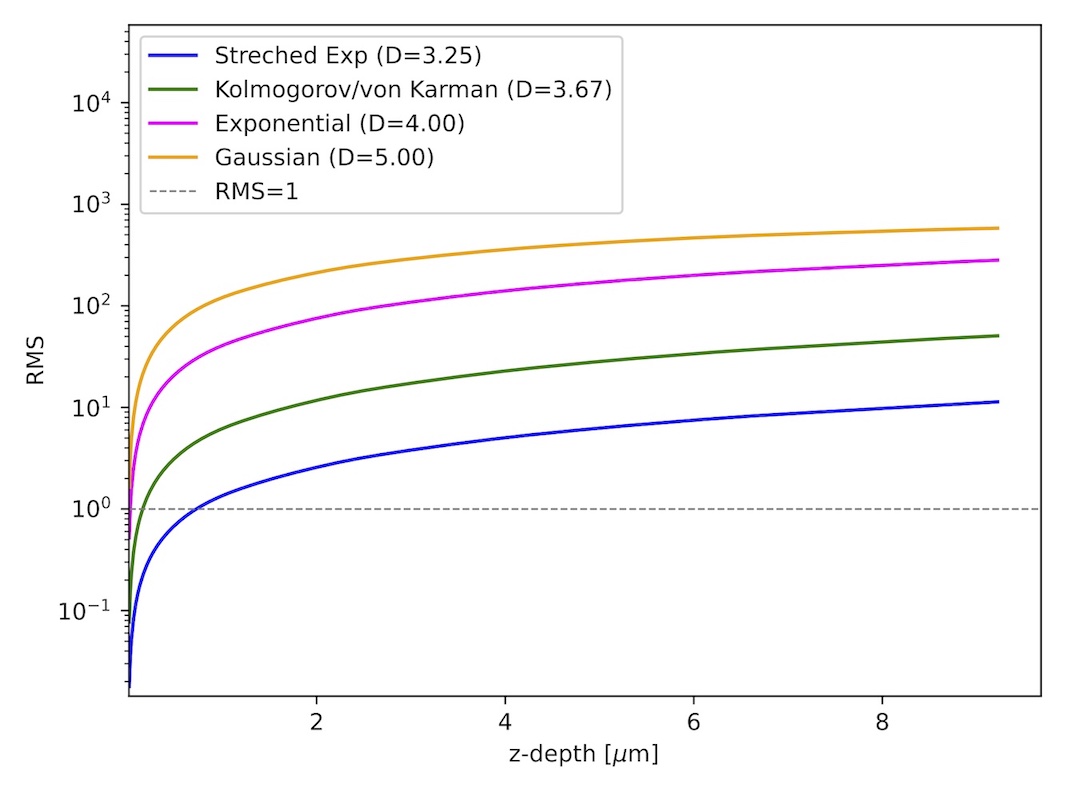}

\leftline{\bf \hspace{0.06\linewidth} (c) \hspace{0.42\linewidth} (d)}
\centering
\includegraphics[width=0.46\linewidth]{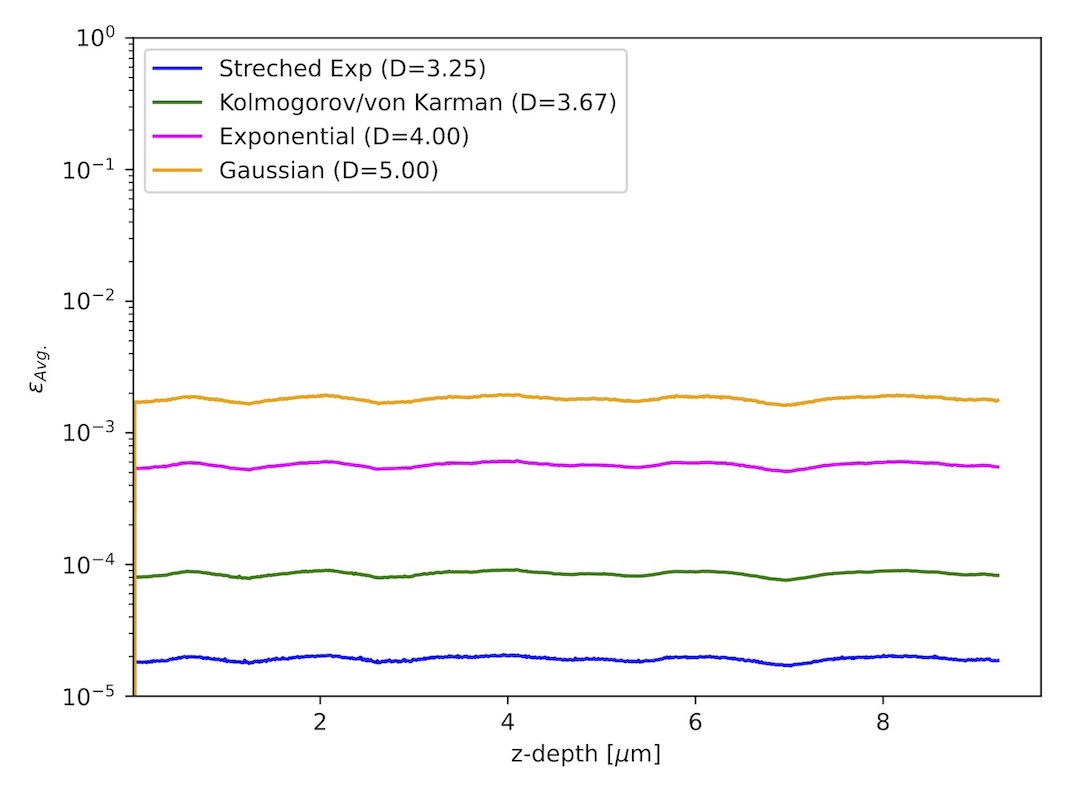}
\includegraphics[width=0.46\linewidth]{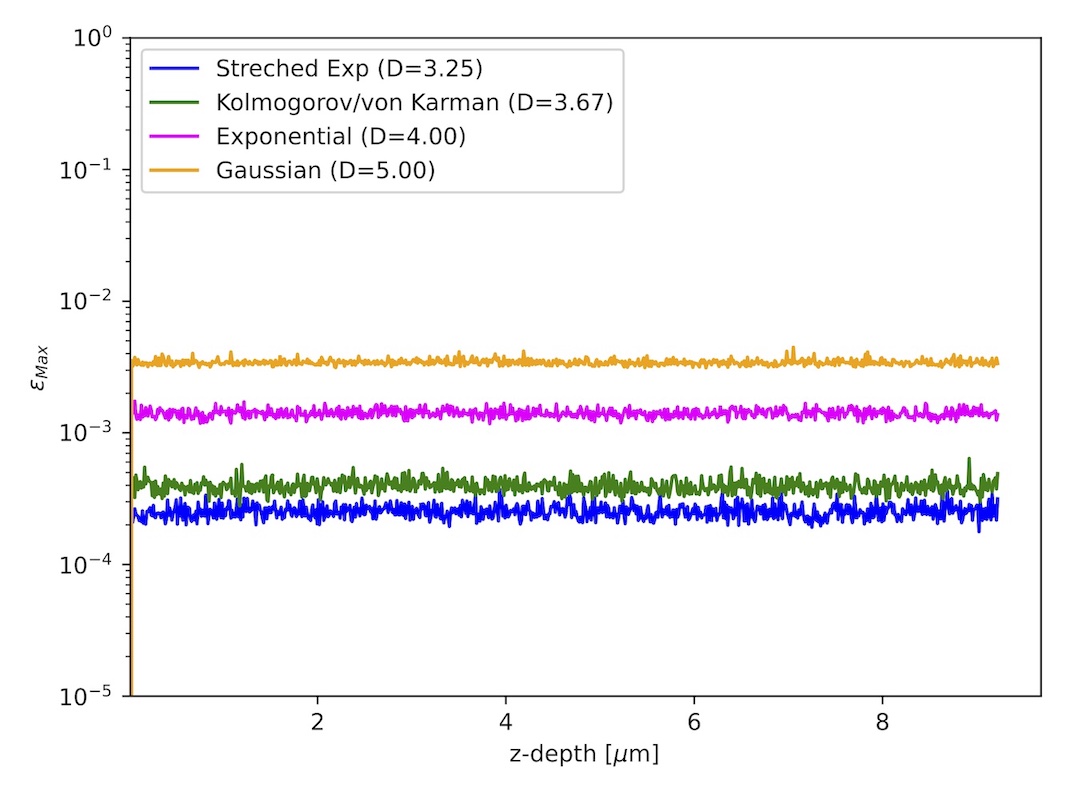}

\caption{Simulation results for intensity propagation of the standard image [$I(r_{\bot},z_{initial}) = 500 \sim 1,000\ {\rm counts/pixel}$] through the refractive index distributions along axial direction. (a) Reduction of maximal intensity as a function of z axis. Dashed lines represent intensity reduction by Beer-Lambert law $I(z) = I(z=0)\ e^{-\mu(kn_0) z}$. (b) Intensity dispersion (i.e., RMS) as a function of z axis. (c) and (d) Averaged and maximal fractional errors as a function of z axis. Solid colored lines denote four different models of fractal media.}
\label{figB2;results02}
\end{figure*}

\vspace*{\fill}

\newpage

\vspace*{\fill}

%
%
%

\begin{figure*}[!h]
\leftline{\bf \hspace{0.02\linewidth} (a) $\lambda = 500\ {\rm nm}$ (visible light)}
\centering
\includegraphics[width=0.22\linewidth]{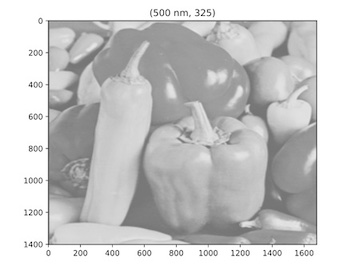}
\includegraphics[width=0.22\linewidth]{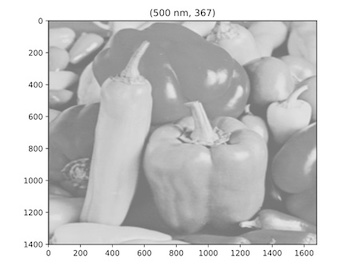}
\includegraphics[width=0.22\linewidth]{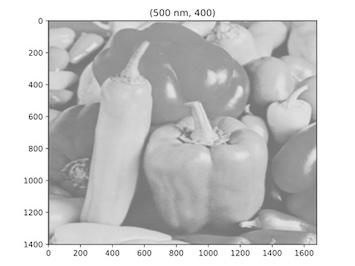}
\includegraphics[width=0.22\linewidth]{SI_figures/figure_B2-03-500nm400.jpg}

\leftline{\bf \hspace{0.02\linewidth} (b) $\lambda = 50.0\ {\rm nm}$}
\centering
\includegraphics[width=0.22\linewidth]{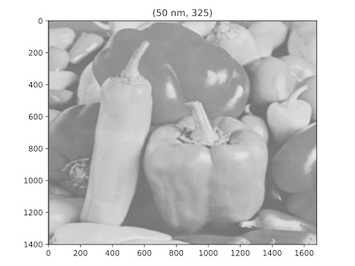}
\includegraphics[width=0.22\linewidth]{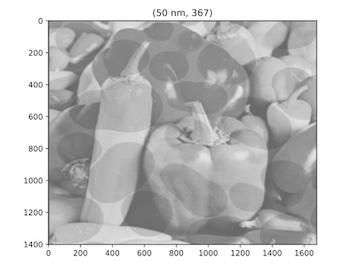}
\includegraphics[width=0.22\linewidth]{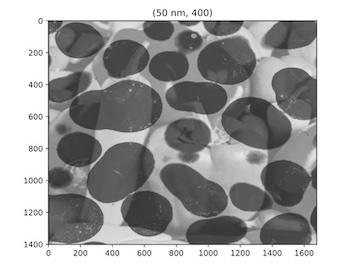}
\includegraphics[width=0.22\linewidth]{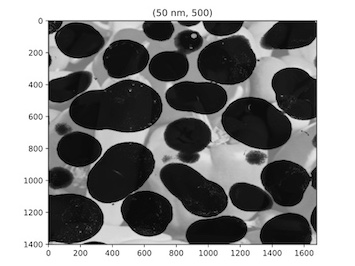}

\leftline{\bf \hspace{0.02\linewidth} (c) $\lambda = 5.00\ {\rm nm}$}
\centering
\includegraphics[width=0.22\linewidth]{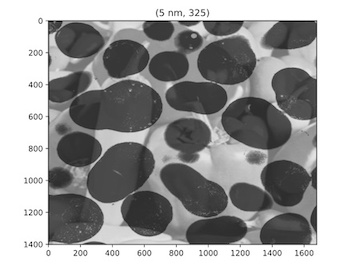}
\includegraphics[width=0.22\linewidth]{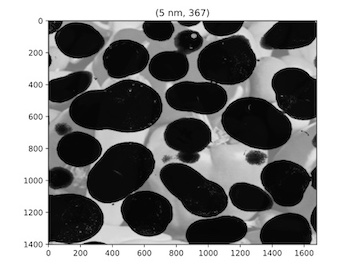}
\includegraphics[width=0.22\linewidth]{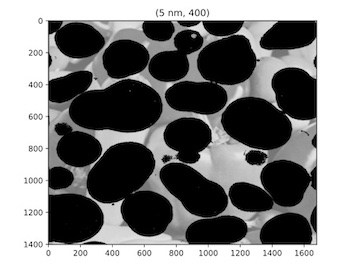}
\includegraphics[width=0.22\linewidth]{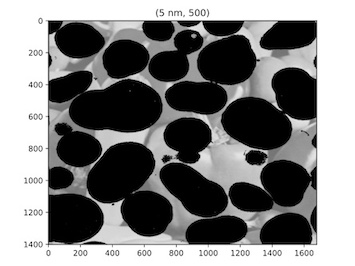}

\leftline{\bf \hspace{0.02\linewidth} (d) $\lambda = 500\ {\rm pm}$ (electron beam)}
\centering
\includegraphics[width=0.22\linewidth]{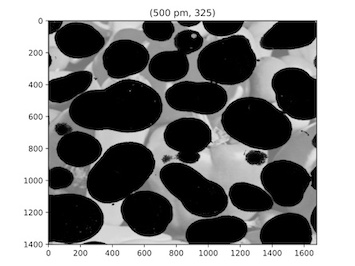}
\includegraphics[width=0.22\linewidth]{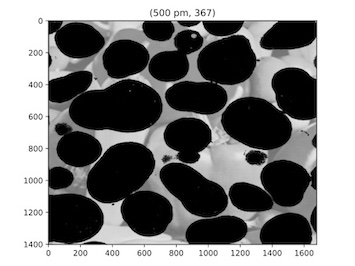}
\includegraphics[width=0.22\linewidth]{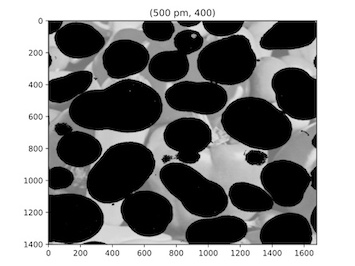}
\includegraphics[width=0.22\linewidth]{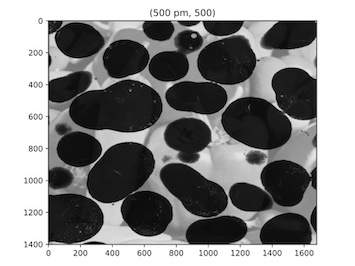}

\leftline{\bf \hspace{0.02\linewidth} (e) $\lambda = 50.0\ {\rm pm}$ (electron beam)}
\centering
\includegraphics[width=0.22\linewidth]{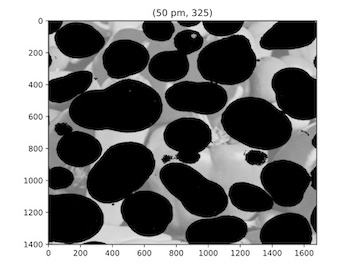}
\includegraphics[width=0.22\linewidth]{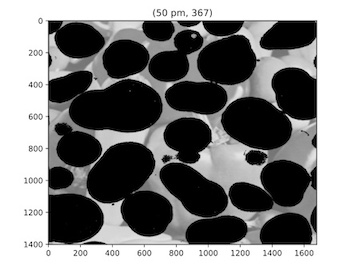}
\includegraphics[width=0.22\linewidth]{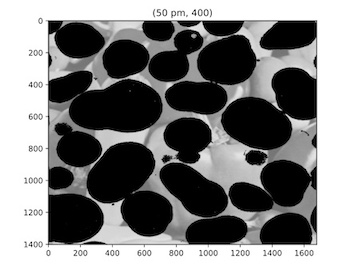}
\includegraphics[width=0.22\linewidth]{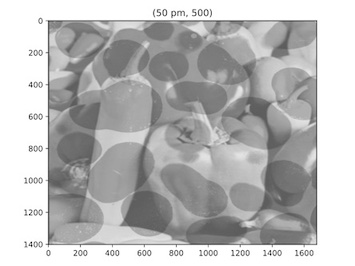}

\leftline{\bf \hspace{0.02\linewidth} (f) $\lambda = 5.00\ {\rm pm}$ (electron beam)}
\centering
\includegraphics[width=0.22\linewidth]{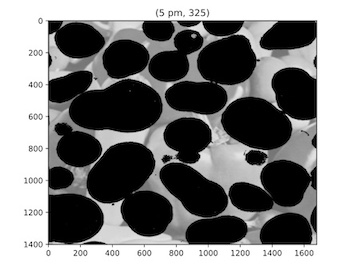}
\includegraphics[width=0.22\linewidth]{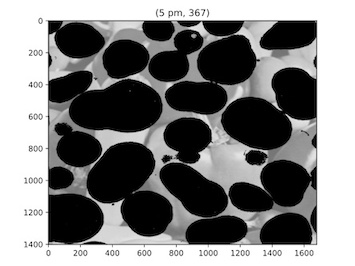}
\includegraphics[width=0.22\linewidth]{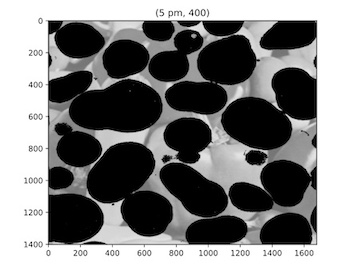}
\includegraphics[width=0.22\linewidth]{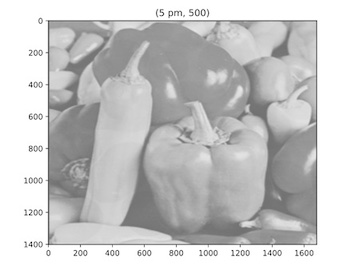}

\leftline{\bf \hspace{0.10\linewidth} Medium I \hspace{0.12\linewidth} Medium II \hspace{0.12\linewidth} Medium III \hspace{0.10\linewidth} Medium IV}

\caption{For sample thickness $L = 110\ {\rm nm}$, final intensity distributions of standard images are plotted over a wide range of wavelength. The image intensities are scaled from $0$ to $1,100$ counts.}
\label{figB2;results03}
\end{figure*}

\vspace*{\fill}

\newpage

\section{Application}
\subsection{Derivation of the autocorrelation function in differential-phase factors}
Using Eq. (5), (6) and (7) in the main text, the Eq. (10) is derived as follows.
\begin{eqnarray}
&& \left<\frac{\partial \beta(r)}{\partial z} \frac{\partial \beta(r + \rho)}{\partial z} \right>\ =\ \left< (k n_0)^2 \left( \Delta n(r) - \frac{1}{2} \kappa^2(r) \right) \left( \Delta n(r + \rho) - \frac{1}{2} \kappa^2(r + \rho) \right) \right> \nonumber\\
\nonumber\\
&& \hspace{0.10\linewidth} =\ (k n_0)^2 \bigg[ \left< \Delta n(r) \Delta n(r + \rho) \right> - \frac{1}{2} \left< \Delta n(r + \rho) \kappa^2(r) \right> - \frac{1}{2} \left< \Delta n(r) \kappa^2(r + \rho) \right> +  \frac{1}{4} \left< \kappa^2(r) \kappa^2(r + \rho) \right> \bigg] \nonumber\\
\nonumber\\
&& \hspace{0.10\linewidth} =\ (k n_0)^2 \bigg[ \left< \Delta n(r) \Delta n(r + \rho) \right> - \frac{1}{2} \left< \Delta n(r + \rho) \right> \left( \frac{\mu(k n_0)}{k n_0} \right)^2 - \frac{1}{2} \left< \Delta n(r) \right> \left( \frac{\mu(k n_0)}{k n_0} \right)^2 +  \frac{1}{4} \left( \frac{\mu(k n_0)}{k n_0} \right)^4 \bigg] \nonumber\\
\nonumber\\
&& \hspace{0.10\linewidth} =\ (k n_0)^2 \bigg[ B_n(\rho) +  \frac{1}{4} \left( \frac{\mu(k n_0)}{k n_0} \right)^4 \bigg] \nonumber\\
\nonumber\\
&& \hspace{0.10\linewidth} =\ (k n_0)^2 B_n(\rho) + \left( \frac{\mu(k n_0)}{\sqrt{2 k n_0}} \right)^4
\label{eqn;acorr}
\end{eqnarray}
\
\\
where $\left<\Delta n(r) \right> = 0$ and $\left<\Delta n(r + \rho) \right> = 0$.

\vspace*{\fill}

\begin{figure}[!h]
\centering
\includegraphics[width=0.65\linewidth]{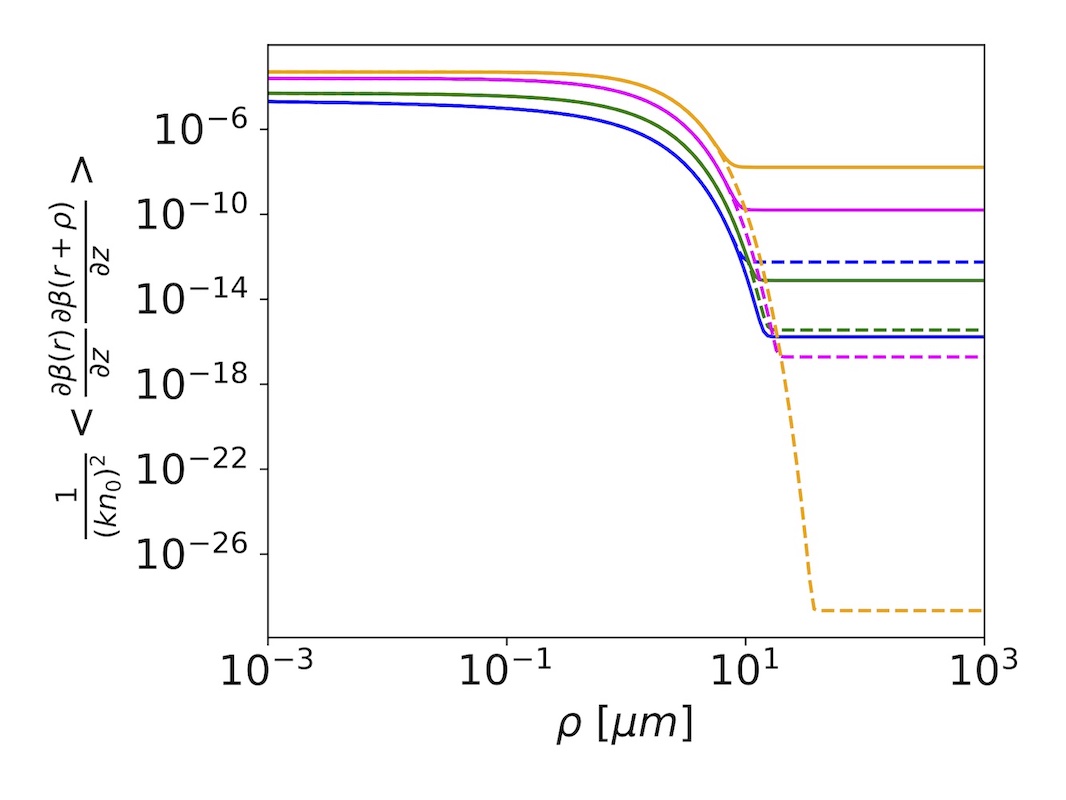}

\caption{The spatial-autocorrelation of phase derivatives normalized by $(k n_0)^2$ is shown as a function relative pixel distance $\rho$. Solid and dashed colored lines represent wavelength ranges for visible light ($\lambda = 507\ {\rm nm}$) and electron beams ($\lambda = 4.00\ {\rm pm}$), respectively. Each colors denote the four different fractal models.}
\label{fig05;relation}
\end{figure}

\vspace*{\fill}

\newpage

\subsection{Direct measurement of intracellular fractal profiles}
The Eq. (10) in the main text is applicable to optical as well as electron microscopy imaging. A direct measurement of the WM covariance model parameters in fluorescent cell imaging is demonstrated as follows.

\begin{itemize}
\item[(1)] Fluorescent cell imaging: 

\hspace*{1mm} Fluorescence microscopy system has been configured to visualize intracellular distribution of $\alpha$-tubulin and histone H2B in HeLa cells. The HeLa cells stably expressing histone H2B-GFP were a kind gift from Dr. Hiroshi Kimura (Tokyo Institute of Technology, Japan) \cite{kimura2006sm}. The HeLa cells were fixed with $4\%$ paraformaldehyde in 250 mM Hepes-NaOH (pH $7.2$) for $10$ min at room temperature (RT) (approximately 26$^{\circ}C$). After permeabilization and blocking by blocking buffer ($10\%$ Blocking One (Nacalai Tesque, Kyoto, Japan; Cat No. 03953-95) and $0.5\%$ TritionX-100 (FUJIFILM Wako Pure Chemical Corp., Tokyo, Japan; Cat No. 168-11805) in phosphate buffer saline) for $20$ min at RT, the cells were probed with anti-$\alpha$-tubulin antibody (1:1,000 dilution; DM1A; Merck, Darmstadt, Germany, Cat No. R6199) for $2$ h, followed by Alexa488-labeled anti-mouse IgG antibody (1:250 dilution; Cat No. R37120; ThermoFisher Scientific, Waltham, MA, USA) for $1$ h at RT. The cells were mounted with Prolong Glass Antifade mountant (ThermoFisher Scientific, Cat No. P36980) overnight at RT. The cells were observed using DeltaVison Elite system (GE Healthcare Inc., Chicago, USA) equipped with 60x PlanApo N OSC oil-immersion objective lens ($NA=1.4$, Olympus, Tokyo, Japan), and pco.edge $4.2$ sCMOS camera (PCO, Kelheim, Germany). Parameter configurations of the microscopy system are as follows: wavelength of emission light, $\lambda = 525\ {\rm nm}$; defocus distance, $\delta z = 200\ {\rm nm}$; image pixel length, $\delta x = \delta y = 107.50\ {\rm nm}$; total number of z-stack images, $31$ frames; and image size, $2040 \times 2040\ {\rm pixels}$. Figures \ref{figC2;images} show example snapshots of the intensity distribution for fluorescent cell images. 

\begin{figure}[b]
\centering
\includegraphics[width=0.32\linewidth]{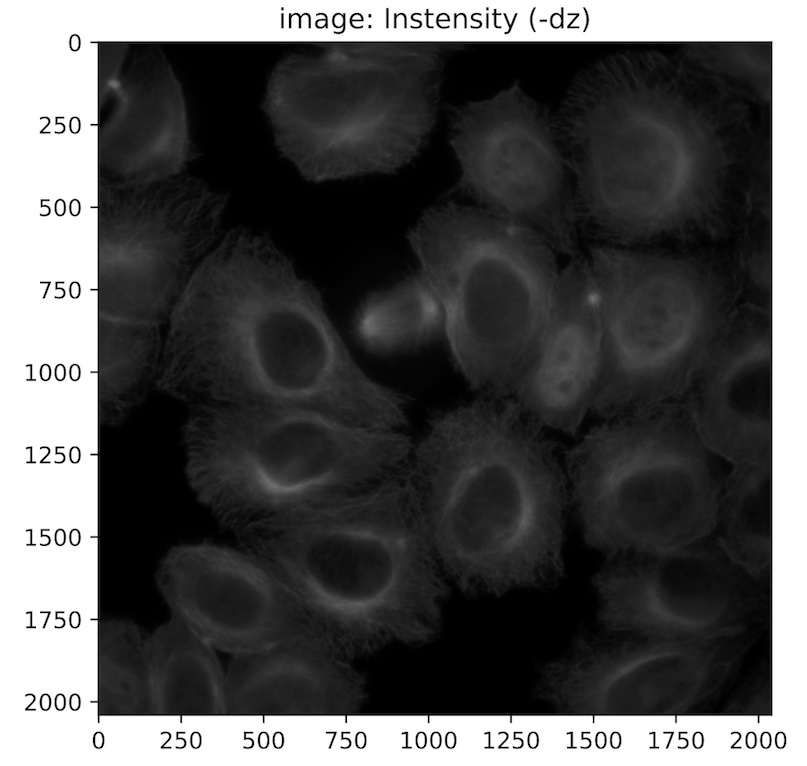}
\includegraphics[width=0.32\linewidth]{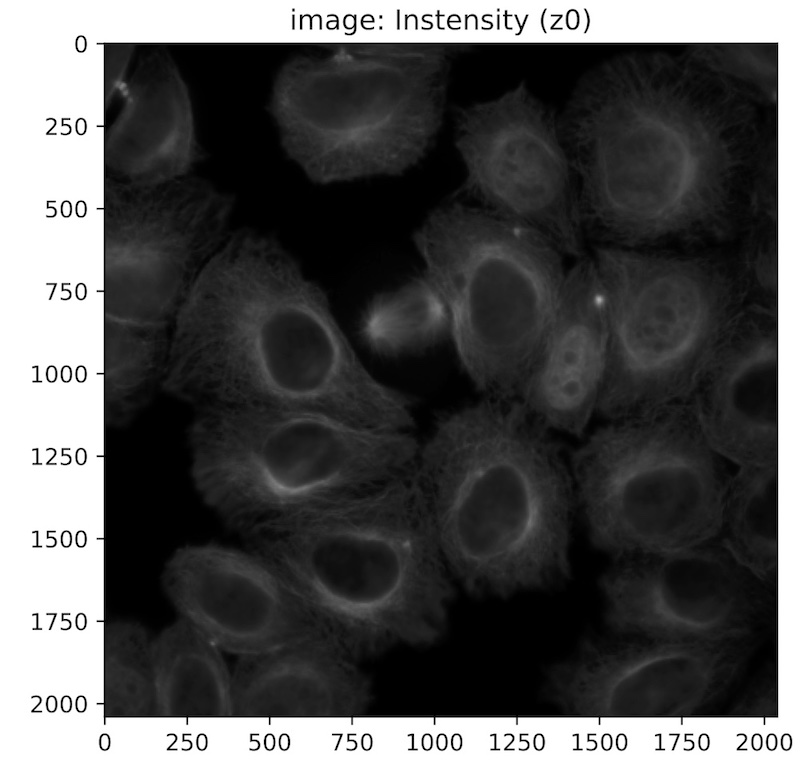}
\includegraphics[width=0.32\linewidth]{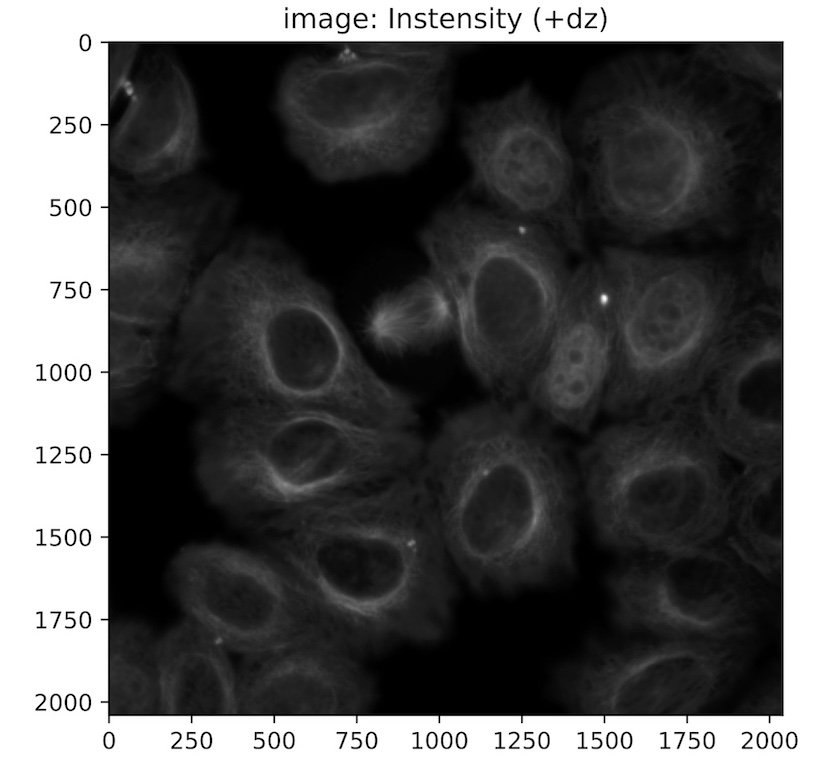}

\caption{Example snapshots of fluorescent cell images for the $5$-th, $10$-th and $15$-th index.}
\label{figC2;images}
\end{figure}

\item[(2)] Phase retrieval based on original TIE \cite{mitome2021sm, zuo2020sm}: 

\hspace*{1mm} We use the geometric multigrid (GMG) method to reconstruct the phase distribution from fluorescent cell images \cite{watabe2022sm, mazumder2016sm, xue2011sm, pinhasi2010sm}. The lateral phase distribution of the $p$-th axial index $\phi(r_{\bot},z_{p})$ can be reconstructed from three different images along the axial direction [see Figure \ref{figC2;gmgm}(a)]. In numerical analysis, the GMG method aims to solve original TIE by using a hierarchy of discretization as well as approximating derivatives with finite differences. This hierarchical discretization enables acceleration of the convergence of a basic iterative relaxation, efficiently reducing errors in high spatial-frequency components by global correction of the fine grid solution approximation in a step--by--step manner. The original equation can be rewritten in the approximate form of finite differences, following, in particular, the assumption that intensity differences between the two images $I(z_{p+5}) - I(z_{p-5})$ is approximately linear with an equal axial distance $5\delta z$ ($1.00\ {\rm \mu m}$). The finite difference approximation of the phase distribution for a specific axial position of the $p$-th index is given by
\begin{eqnarray}
\phi^{p}_{i,j} & = & \frac{1}{16 I^{p}_{i,j}} \Bigg[ \left(I^{p}_{i+1,j} - I^{p}_{i-1,j}\right) \left(\phi^{p}_{i+1,j} -  \phi^{p}_{i-1,j}\right) 
 + \left(I^{p}_{i,j+1} - I^{p}_{i,j-1}\right)\left(\phi^{p}_{i,j+1} - \phi^{p}_{i,j-1}\right) \nonumber\\
& & \hspace{1.5cm} +\ 4\ I^{p}_{i,j} \left( \phi^{p}_{i+1,j} + \phi^{p}_{i-1,j} + \phi^{p}_{i,j+1} + \phi^{p}_{i,j-1} \right) +\ 4\ \delta x \delta y k\ \left( \frac{I^{p+5}_{i,j} - I^{p-5}_{i,j}}{10\delta z} \right)
\Bigg]
\end{eqnarray}
where $i$ and $j$ represent the index of lateral intensity and phase distributions along the x- and y-axes; $\delta x$ and $\delta y$ are lateral resolutions; $\delta z$ is axial distance.

\hspace*{1mm} The entire procedure for the GMG algorithm [see Figure \ref{figC2;gmgm}(b)] can be summarized in the following five steps: presmoothing, reducing errors in high frequency components by iterative solvers such as Jacobi, Gauss-Seidel and conjugate gradient methods; residual estimations, computing residual errors after the iterative solvers; restriction, downsampling the residual errors to a coarser grid; prolongation, interpolating a correction estimated on a coarser grid into a finer grid; and, correction, modifying the interpolated coarser grid solution onto the finer grid. These five steps are implemented into the simplest multigrid process, termed the V-cycle, reconstructing the phase distributions from the $31$ fluorescent cell images. Figures \ref{figC2;gmgm}(c)-(e) show an example snapshot of the reconstructed phase distribution of the $10$-th axial index, as well as error convergence of intensity distributions in the V-cycle process.

\begin{figure}[t]
\leftline{\bf \hspace{0.02\linewidth} (a) \hspace{0.22\linewidth} (b)}
\centering
\includegraphics[width=0.28\linewidth]{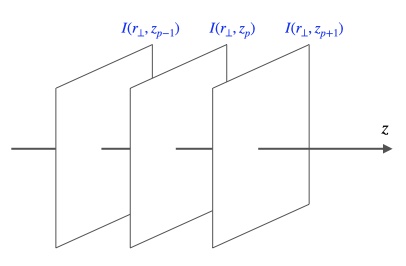}
\includegraphics[width=0.70\linewidth]{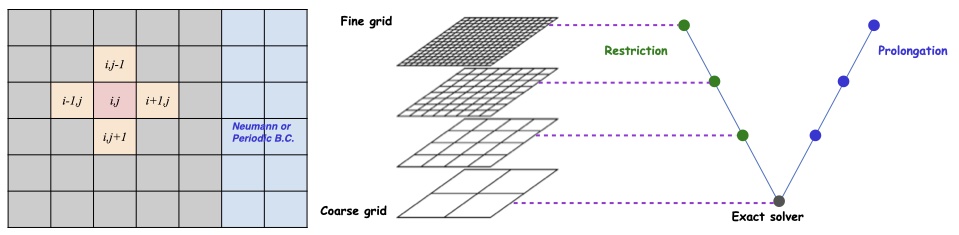}

\ 

\leftline{\bf \hspace{0.02\linewidth} (c) \hspace{0.28\linewidth} (d) \hspace{0.28\linewidth} (e)}
\centering
\includegraphics[width=0.32\linewidth]{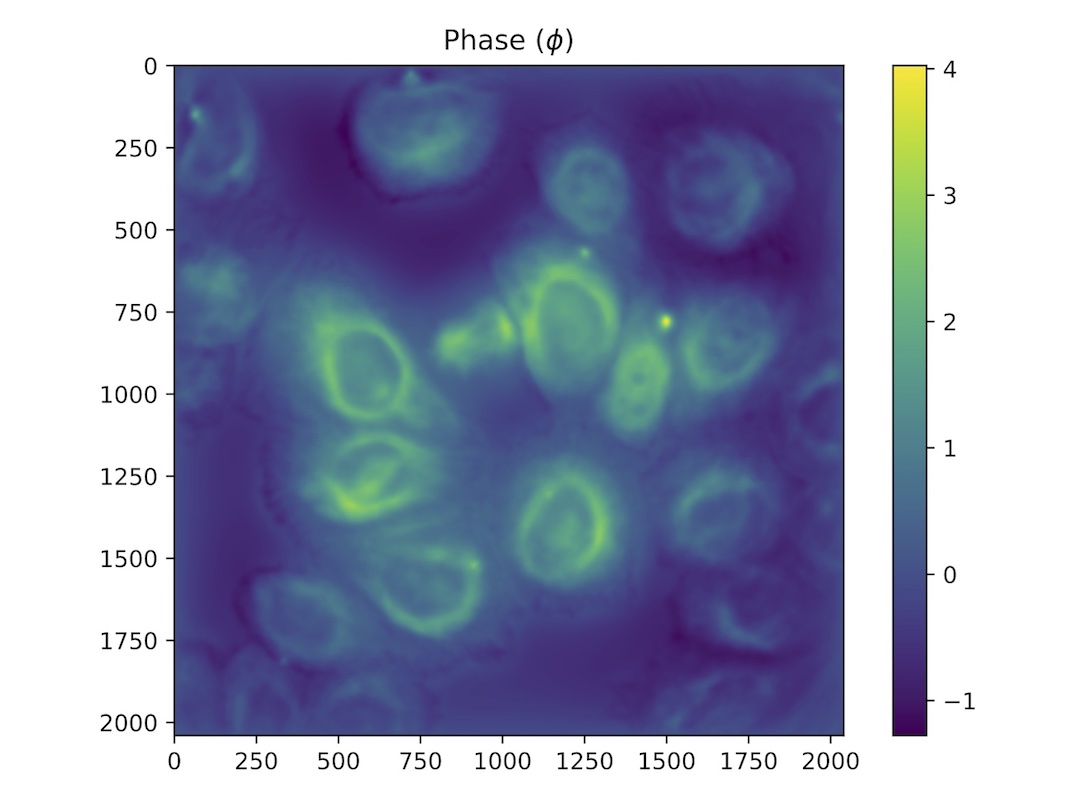}
\includegraphics[width=0.32\linewidth]{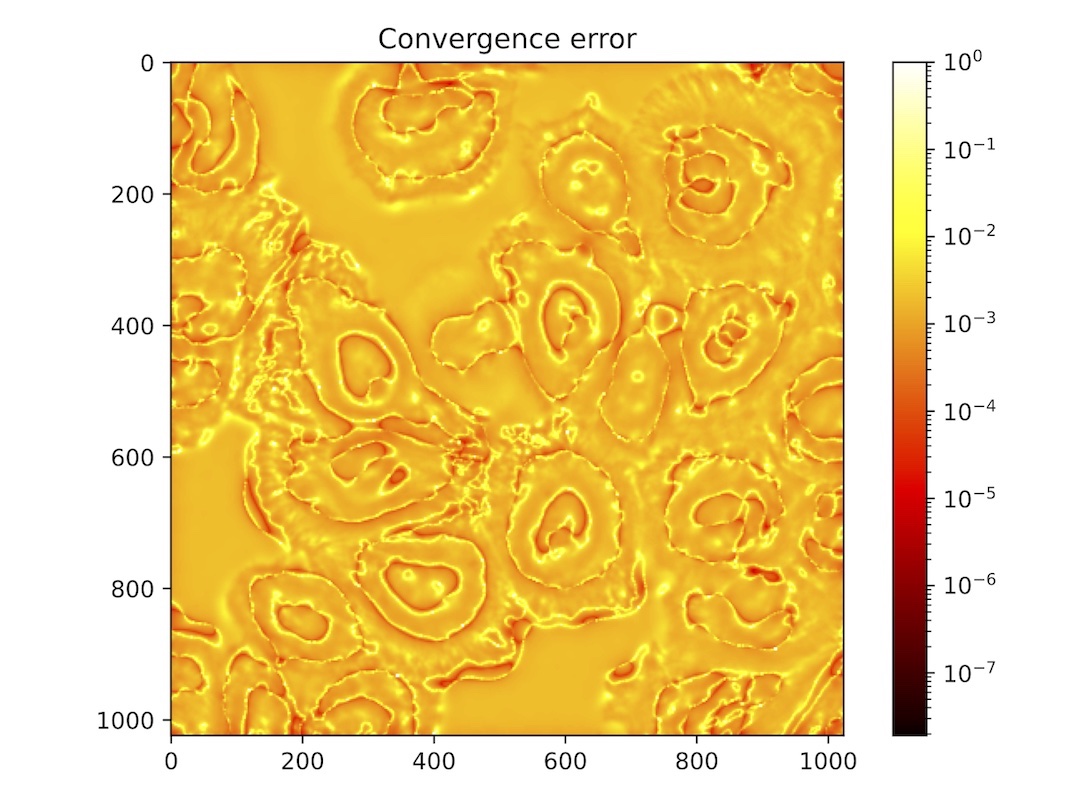}
\includegraphics[width=0.32\linewidth]{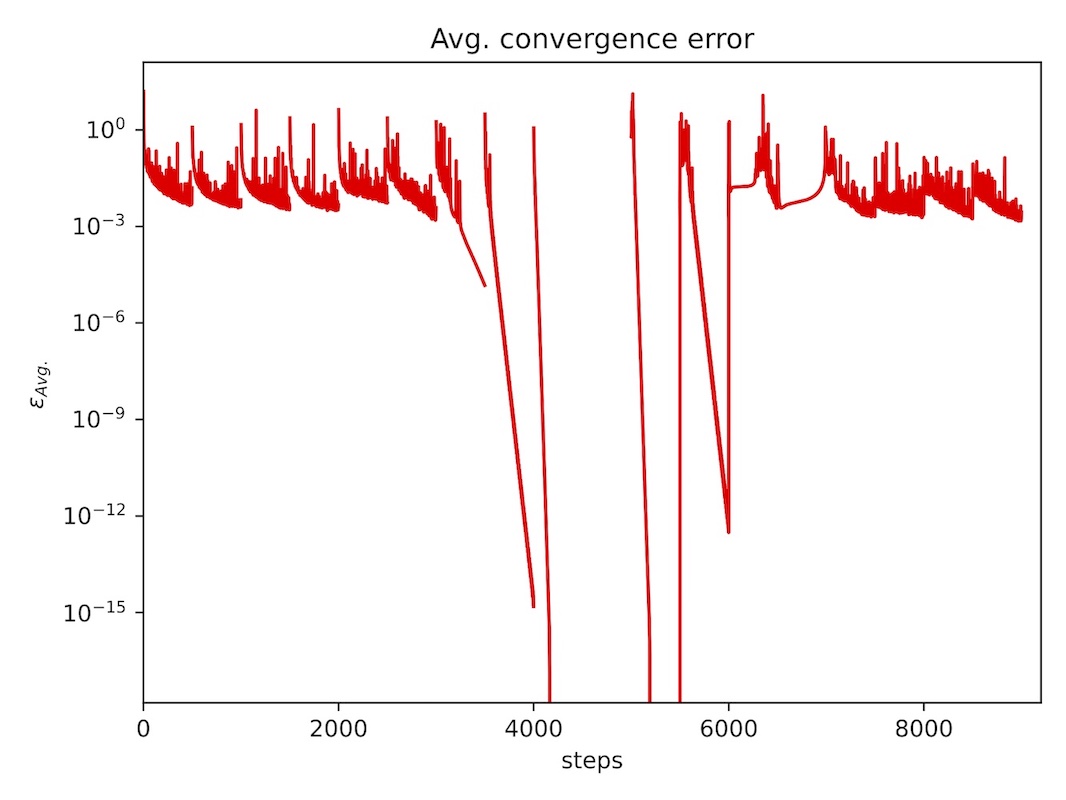}

\caption{Phase retrieval. (a) Lateral phase distribution for a given axial position of the $p$-th index can be reconstructed from three different intensity distributions, $I(r_{\bot},z_{p-5})$, $I(r_{\bot},z_{p})$, and $I(r_{\bot},z_{p+5})$. (b) Schematic overview for the V-cycle process of GMG algorithm. (c) Phase distributions reconstructed from the three fluorescent cell images. (d) Fractional error of intensity distribution converges to roughly $10^{-2}$. (e) Averaged fractional errors are represented as a function of iteration steps.}
\label{figC2;gmgm}
\end{figure}

\item[(3)] Parameter fitting:

\hspace*{1mm} For the sake of analysis simplicity, we first omit the offset of Eq. (\ref{eqn;acorr}) and then rewrite the spatial-autocorrelation equation in the form of
\begin{eqnarray}
\frac{1}{k^2}\left< \frac{\partial \phi(r)}{\partial z}\frac{\partial \phi(r + \rho)}{\partial z} \right> & = & \frac{(n_0\sigma_n)^2\ 2^{1-\nu}}{\Gamma(\nu)} \left(\frac{\rho}{l_{c}}\right)^{\nu} K_{\nu}\left(\frac{\rho}{l_c}\right)
\label{eqn;fitacorr}
\end{eqnarray}
where phase derivative $\partial \phi/\partial z$ is given by taking a linear difference of the two phase distributions in axial defocus distance $\delta z$, 
\begin{eqnarray}
\frac{\partial \phi(r_{\bot},z)}{\partial z} & \approx & \frac{\phi(r_{\bot}, z_{p+1}) -  \phi(r_{\bot},z_{p})}{\delta z}
\label{eqn;diff}
\end{eqnarray}
Figure \ref{figC2;fitting}(a) shows an example output of the phase derivative in fluorescent cell imaging.

\hspace*{1mm} Fractal properties of cellular interiors can be found in the shape and offset of the spatial-autocorrelation curve of phase derivatives. In our analysis, Eq. (\ref{eqn;acorr}) can be fitted to the observed spatial-autocorrelation curves of the derivatives obtained from the $5$-th to $25$-th index of phase distributions in fluorescent cell imaging. The best fit parameters are these which minimize $\chi^2 = -2\ln\mathcal{L}$ where $\mathcal{L}$ is the likelihood function,
\begin{eqnarray}
\chi^2 & = & \sum^{N {\rm bins}}_{i = 0} \frac{\left(E_i - O_i\right)^2}{\sigma_i^2}
\end{eqnarray}
where $E_i$ and $O_i$ are the expected and the observed spatial-autocorrelation at the $i$-th input. $\sigma_i$ is statistical error in $O_i$. Furthermore, no penalty term for nuisance parameters is introduced in this minimization function. Table \ref{tabC2;fitting} and Figure \ref{figC2;fitting} show the results from model parameter fitting without analyzing systematic uncertainties (or errors) arising from various biophysical sources in fluorescent cell imaging. In Figure \ref{figC2;fitting}(b), the best fit (red solid line) of spatial-autocorrelation function is directly compared with the observed spatial-autocorrelation curve (blue bars). We also estimate the statistical uncertainties in the model parameters to indicate numerically the validity and confidence of our fitting results [see Figures \ref{figC2;fitting}(c)-(f)].

\end{itemize}

\begin{table*}[!h]
\centering
\begin{tabular}[b]{|p{8cm}|c|}
\hline
\hspace{2.8cm} {\bf Model parameters} & \hspace{2.8cm} {\bf Fitting results} \hspace{2.8cm} \\\hline
\hline
Fluctuation of refractive index $(n_0 \sigma_n)^2$ & $3.841 \pm 0.399 \times 10^{-4}$ \\\hline
Correlation length $l_c$ & $11.159 \pm 1.02\ {\rm \mu m}$ \\\hline
Fractal dimension $D_f$ & $3.900 \pm 0.219$ \\\hline
Scattering coefficient $\mu_s$ & $1.073 \pm 0.373\ {\rm \mu m^{-1}}$ \\\hline
Minimization value $\chi^2_0\ (d.o.f)$ & $1605.616\ (2878)$ \\\hline
\end{tabular}
\caption{The fitting results. The best fit values and statistical uncertainties ($1\sigma$) of model parameters are listed in this table. $d.o.f$ stands for degree of freedom.}
\label{tabC2;fitting}
\end{table*}

\begin{figure*}[!h]
\leftline{\bf \hspace{0.02\linewidth} (a) \hspace{0.28\linewidth} (b) \hspace{0.28\linewidth} (c)}
\centering
\includegraphics[width=0.32\linewidth]{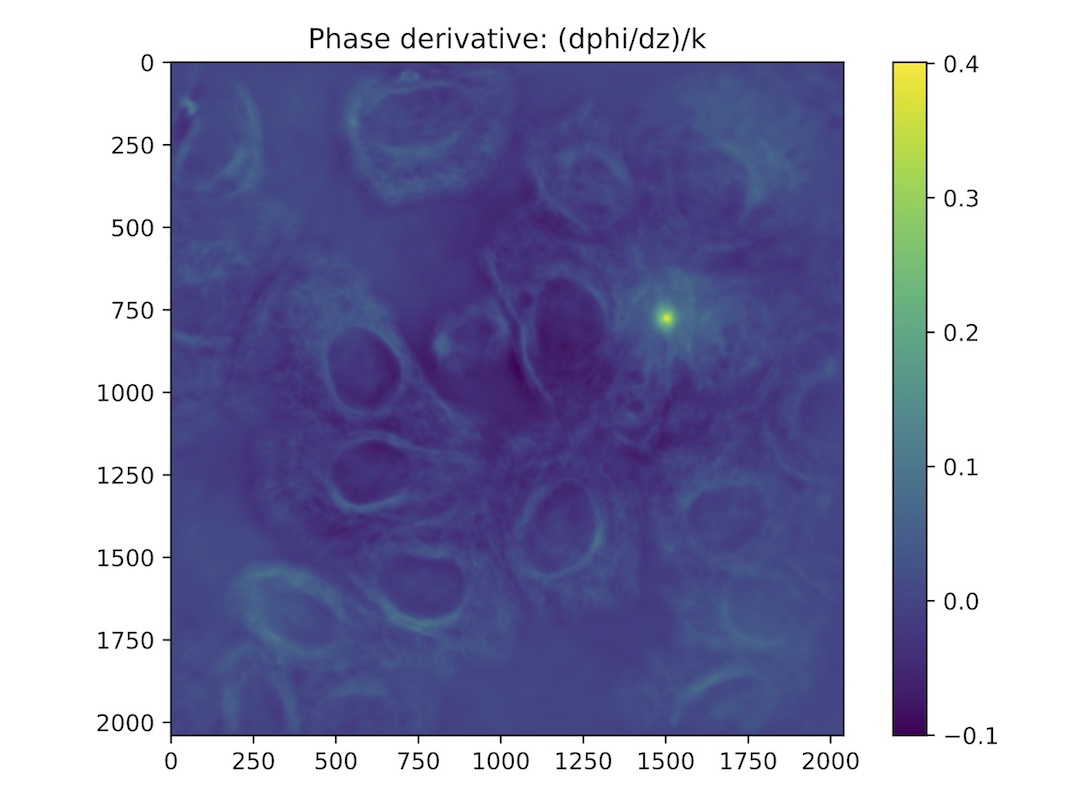}
\includegraphics[width=0.32\linewidth]{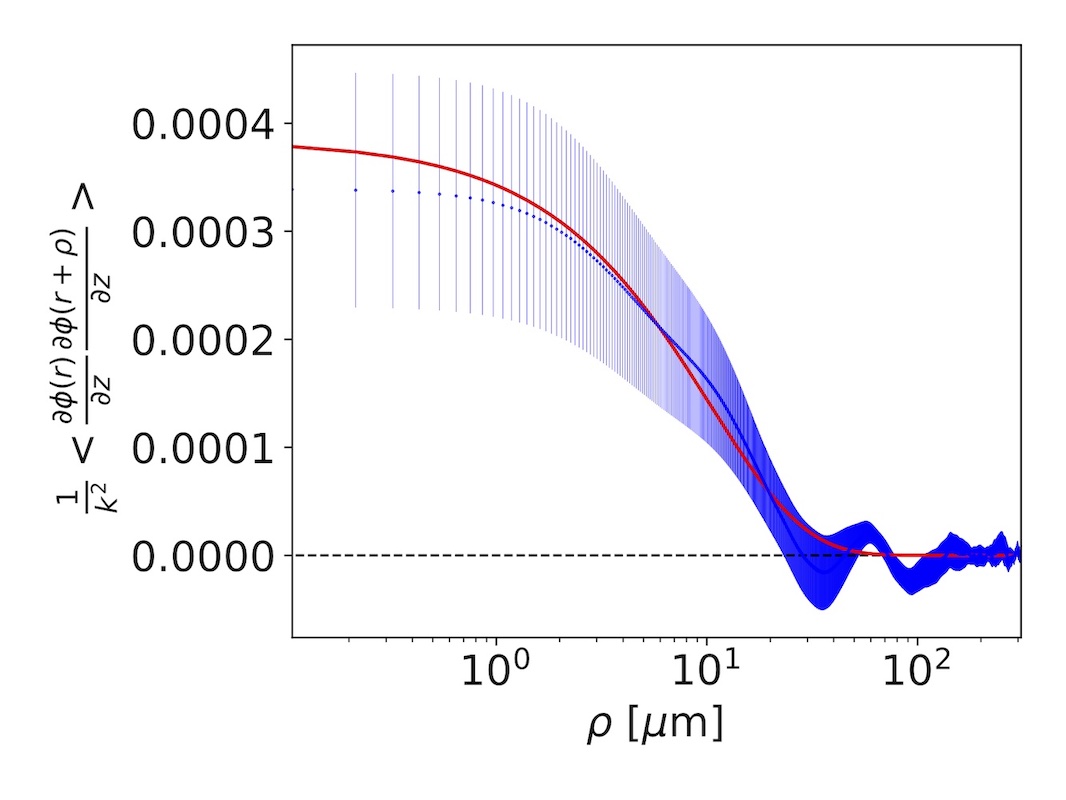}
\includegraphics[width=0.32\linewidth]{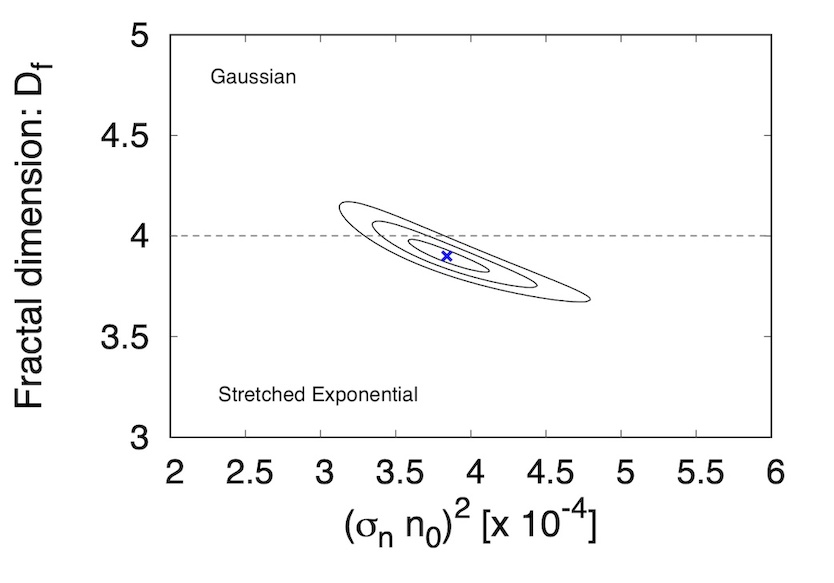}

\ 

\leftline{\bf \hspace{0.02\linewidth} (d) \hspace{0.28\linewidth} (e) \hspace{0.28\linewidth} (f)}
\centering
\includegraphics[width=0.32\linewidth]{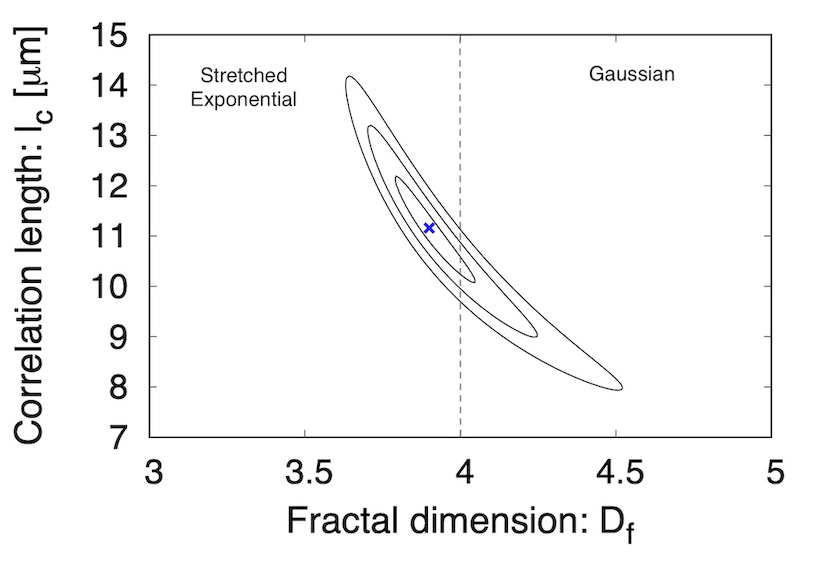}
\includegraphics[width=0.32\linewidth]{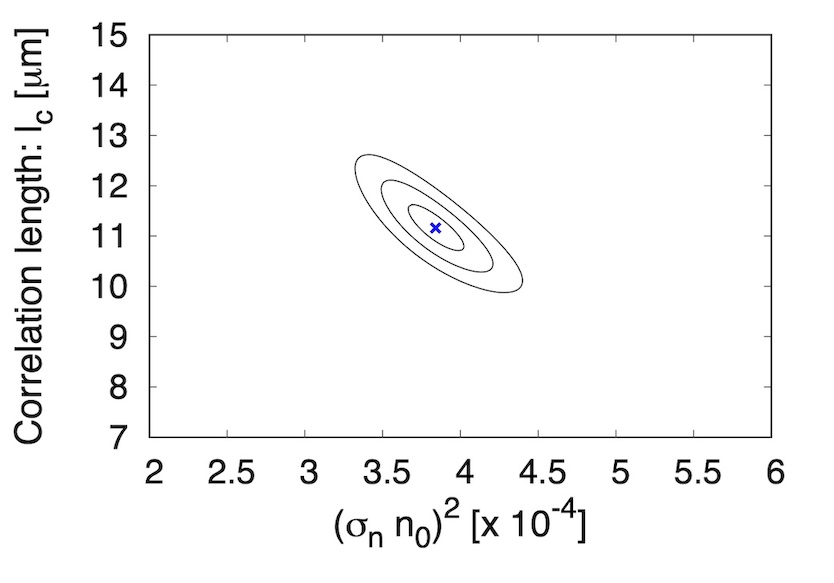}
\includegraphics[width=0.32\linewidth]{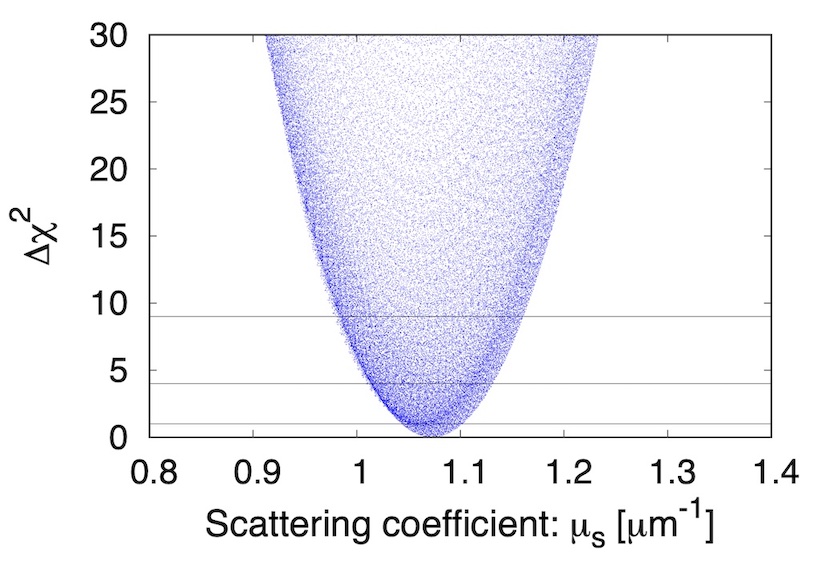}

\caption{Fitting results. (a) An example output of differential-phases along axial direction. (b) The spatial-autocorrelation curves of differential-phases in axial direction: blue bars and red line represent the observed spatial-autocorrelation and the best fit curves. (c-e) $\Delta \chi^2$-contour plots for fractal dimension $D_f$, fluctuation $(n_0 \sigma_n)^2$ and the correlation length $l_c$. Blue points indicate the best estimates. Black solid lines around the best fit values are allows at the $1\sigma$ ($68\%$), $2\sigma$ ($95\%$) and $3\sigma$ ($99\%$) confidence levels of the fitting results. (f) The hyperbolic curve of $\Delta \chi^2$ is represented as a function the mean scattering coefficients $\mu_s$ converted from the fitting results (assuming $n_0 = 1.38$).}
\label{figC2;fitting}
\end{figure*}

\newpage

\end{document}